\documentclass[a4paper,manyauthors,nocleardouble,COMPASS]{cernphprep}
\usepackage[a4paper, total={17 cm, 20cm}]{geometry}

\RequirePackage[T1]{fontenc}

\RequirePackage{graphicx}
\RequirePackage{newtxtext}
\RequirePackage{newtxmath}
\RequirePackage{flushend}
\RequirePackage[numbers,sort&compress]{natbib}

\RequirePackage[colorlinks=true,citecolor=blue,urlcolor=blue,linkcolor=blue]{hyperref}
\usepackage{orcidlink}
\usepackage{tikz}
\usepackage{lineno}
\usepackage[normalem]{ulem} % added by Anatolii to easily cross out text
\usepackage{amsmath}
\usepackage{balance}
\usepackage{lastpage}
\usepackage{array}
\usepackage{makecell}
\usepackage{adjustbox}

\renewcommand\vec{\mathbf}
\begin{document}
%\linenumbers
\begin{titlepage}
\PHnumber{2022--231}
%\PHdate{\today}
\PHdate{28 Oct 2022}
\title{Spin Density Matrix Elements in Exclusive $\rho ^0$ Meson Muoproduction}

\Collaboration{The COMPASS Collaboration}
\ShortAuthor{The COMPASS Collaboration}

%\ch@ckobl{author}{Name(s) and initial(s) of author(s) should be given}
%\ch@ckobl{institute}{Address(es) of author(s) should be given}

%\Collaboration{The COMPASS Collaboration}
%\ShortAuthor{The COMPASS Collaboration

%
%\author{
%{Draft version 2.0}
%}
%
%
%\institute{
% %First Address, Street, City, Country\label{addr1}
%}
\date\today
\maketitle
\begin{abstract}
We report on a measurement of Spin Density Matrix Elements  (SDMEs) in hard exclusive $\rho ^0$ meson muoproduction at COMPASS using 160~GeV/$c$ polarised $ \mu ^{+}$ and $ \mu ^{-}$ beams impinging on a liquid hydrogen target. The measurement covers the kinematic range 5.0~GeV/$c^2$ $< W <$ 17.0~GeV/$c^2$, 1.0 (GeV/$c$)$^2$ $< Q^2 <$ 10.0 (GeV/$c$)$^2$ and 0.01 (GeV/$c$)$^2$ $<  p_{\rm{T}}^2 <$  0.5 (GeV/$c$)$^2$.
Here, $W$ denotes the mass of the final hadronic system, $Q^2$ the virtuality of the exchanged photon, and $p_{\rm{T}}$ the transverse momentum of the $\rho ^0$ meson  with respect to the virtual-photon direction. The measured non-zero SDMEs for the transitions of transversely polarised virtual photons to longitudinally polarised vector mesons ($\gamma^*_T \to V^{ }_L$) indicate a violation of $s$-channel helicity conservation. Additionally, we observe a dominant contribution of natural-parity-exchange transitions and a very small contribution of unnatural-parity-exchange transitions, which is compatible with zero within experimental uncertainties. The results provide important input for modelling Generalised Parton Distributions (GPDs). In particular, they may allow one to evaluate in a model-dependent way the role of parton helicity-flip GPDs in exclusive $\rho ^0$ production.
\end{abstract}

\vfill
\Submitted{(accepted at Eur. Phys. J. C)}
\end{titlepage}
{
\pagestyle{empty}
%\input{Authors2020}
%Authors list to be provided by SPs
\clearpage
}
\setcounter{page}{1}

\section{Introduction}
\label{intro}

Exclusive vector meson production in lepton-nucleon scattering provides a convenient tool
for studying the  production mechanism and, in a model-dependent way,
the structure of the nucleon. In this paper, exclusive  $\rho ^0$ meson muoproduction on the proton is studied:
\begin{equation}
\mu + p \rightarrow \mu' + p' + \rho ^0.
\label{eq:introduction:01}
\end{equation}
In the one-photon-exchange approximation, this process
 is described by the interaction of a virtual photon $\gamma^{*}$
%emitted from the incoming muon with one of the partons from
with the target proton $p$:
\begin{equation}
\gamma^{*} + p \rightarrow p' + \rho ^0.
\label{eq:introduction:02}
\end{equation}

At high virtuality $Q^2$ of the photon, this process is known as Hard Exclusive Meson Production (HEMP). A wealth of information is contained in the spin density matrix elements (SDMEs), which are the observables describing how the spin components of the virtual photon are transferred to those of the created  vector meson~\cite{Schill,Diehl}.
%Comparing our new results on $\rho^0$ vector-meson production to our earlier published ones on $\omega$ vector-meson production~\cite{COMPASS-omega}, where different quantum numbers and quark-flavor and gluon contents are involved, will provide insight into details of their production mechanism(s).
%As different quantum numbers and quark-flavour and gluon contents are involved in $\rho^0$ and $\omega$ vector mesons,
The comparison of the new $\rho^0$ results presented in this paper to our previous $\omega$ results~\cite{COMPASS-omega} will provide insight into details of their respective production mechanism, because $\rho^0$ and $\omega$ vector mesons have different quantum numbers and hence different quark-flavour and gluon contributions to the cross section.

The colour dipole model describes HEMP as a fluctuation of the virtual photon into a quark-antiquark ($q \bar q$) pair that scatters off  the nucleon and then hadronises into the final vector meson~\cite{Devenish}.
Regge phenomenology and perturbative QCD (pQCD) provide complementary approaches to describe
the scattering of the $q \bar q$ pair off the nucleon.
The interaction of the $q \bar q$ pair with the nucleon depends on the transverse
separation between $q$ and $ \bar q$. A pair with large transverse separation is
thought to interact primarily softly, which is described in Regge phenomenology~\cite{Irving}
by the exchange of a pomeron or a secondary reggeon. The interaction of a $q \bar q$ pair with
%large transverse momentum and
small transverse separation is calculable in pQCD.
In lowest order of the strong coupling constant $\alpha_s$, this hard interaction is mediated
by the exchange of a gluon-gluon or quark-antiquark system.
%The model was able to predict
In this approach,
%\blue{[WDN 25.04.]
it is possible to calculate not only transitions without spin-flip induced by both longitudinally ($L$) and transversely ($T$) polarised
virtual photons,
$\gamma^{*}_{L} \rightarrow \rho^{0}_{L}$ and $\gamma^{*}_{T} \rightarrow \rho^0_{T}$,
%can be calculated.
but also to estimate single and double spin-flip transitions.
%{\sout{Moreover, the existence of single and double spin-flip transitions is predicted and their sizes can be estimated.}}
%\textcolor{red}{NOTE 2022-04-29: clarify last sentence - Witold / Andrzej}

In an alternative approach, % to describe HEMP is provided  by
the framework of General Parton Distributions (GPDs) \cite{gpd1, gpd2, gpd3, gpd4,
Radyushkin:1997ki} can be used to describe HEMP. These distribution functions
%, which
contain a wealth of new information on the parton structure of the nucleon.
For HEMP by
longitudinally polarised virtual photons, the amplitude was proven to factorise into a hard-scattering part
and a soft part~\cite{gpd4, Collins:1996fb}. %The soft part contains
While the former is calculable in pQCD, %perturbative QCD (pQCD),
the latter contains GPDs that describe the structure of the probed nucleon and a
distribution amplitude %, which GPDs
that accounts for the structure
of the produced meson. This factorisation is referred to as collinear
because parton transverse momenta are neglected.
For HEMP by transversely polarised virtual photons no
corresponding proof of
factorisation exists. Instead, phenomenological pQCD-inspired models
%were proposed
~\cite{Martin-1997,Goloskokov:2005,Goloskokov:2008,Goloskokov:2009}
%go beyond the collinear factorisation by
postulate the so-called $k_{\perp}$ factorisation, where
$k_{\perp}$ denotes parton transverse momentum.
In particular, the Goloskokov-Kroll (GK) model~\cite{Goloskokov:2005,Goloskokov:2008,
Goloskokov:2009,GK:epjC-2014,GK:epjA-2014}
%hereafter referred to as GK model, cross sections
%, Spin Density Matrix Elements (SDMEs)
allows for a simultaneous description of SDMEs as well as target
and beam-spin asymmetries for HEMP induced by both longitudinally and transversely polarised
virtual photons. % can be described simultaneously.

% At leading twist, vector-meson production by longitudinal virtual photons is described by the
% chiral-even GPDs
% $H^{f}$ and $E^{f}$,
% where $f$ denotes a quark of a given flavour or a gluon. When higher-twist
% effects are included in the three-dimensional {\blue {[PK:] light-cone}} meson wave function,
% in addition to chiral-even GPDs $H^{f}$, $E^{f}$, $\widetilde{H}^{f}$ and $\widetilde{E}^{f}$, the chiral-odd GPDs $H_T^{f}$ and $\bar{E}_T^{f}$ appear, which describe process
% amplitudes with helicity flip of the `active' quark.

The chiral-even GPDs $H^{f}$ and $E^{f}$ are used to describe at leading twist the production of
longitudinally polarised vector mesons by longitudinally polarised virtual
photons. Here, $f$ denotes a quark of a given flavour or a gluon. In the GK model, the chiral-odd GPDs $H_T^{f}$ and $\bar{E}_T^{f}$ are used together with higher-twist effects in the three-dimensional light-cone wave function to describe $\gamma^{*}_{T} \rightarrow \rho^0_{L}$ transitions.
%longitudinally polarised vector mesons by transversely polarised virtual photons.
These GPDs account for a helicity flip of the ``active'' quark and are hence related to the violation of s-channel helicity conservation (SCHC).
The GPDs $\widetilde{H}^{f}$ and $\widetilde{E}^{f}$\!\!, and also the pion-pole exchange mechanism treated in the GK model as one-boson exchange contribution, provide unnatural parity
exchange (UPE) contributions to the transitions $\gamma^{*}_{T} \rightarrow \rho^0_{T}$ and $\gamma^{*}_{L} \rightarrow \rho^0_{T}$.
%These UPE contributions are different by an order of magnitude when comparing exclusive production of $\rho^0$ and  $\omega$ mesons.
With this ansatz the GK model offers an explanation for the contrast between a substantial UPE contribution in exclusive $\omega$ production and a small UPE contribution
in exclusive $\rho^0$ production.

Spin density matrix elements are related to helicity amplitudes that describe transitions between specified spin states of virtual photon, target proton, produced vector meson, and recoil proton. In the case of an unpolarised nucleon target,
%after summing over initial and final spin states of the proton,
 SDMEs depend only on the helicities of virtual photon and produced meson, if the initial and
final spin states of the proton are summed over.
The interpretation of the measured SDME values is a rich field and in this paper we will address the following: the test of SCHC, the evaluation of %contribution of
UPE contributions, the determination of the phase difference between helicity amplitudes, and the calculation of the longitudinal-to-trans\-ver\-se cross-section ratio.
%\textcolor{blue}{\emph{CKR 2022-04-29 The other three items, which are not covered in this paper, are hierarchy, role of chiral-odd GPDs, and GPD constraints, and I integrated them to the end of this section. }}

There exist numerous measurements of hard exclusive $\rho ^0$ production in lepton scattering off hydrogen, deuterium and $^3$He targets. At small values of $W$, measurements were performed at CORNELL~\cite{cornell} and by CLAS~\cite{clas1,clas2}.
%{\color{red} Indicate energy ranges (?), may be also Q2 (?)}
For intermediate values of $W$, results were obtained by HERMES~\cite{HERMES:EPJC-17, HERMES:EPJC-18, DC-24}, NMC~\cite{NMC} and Fermilab experiment E665~\cite{E665}.
At highest values of $W$, results were obtained by the H1~\cite{h1-2000,h1-2010} and ZEUS~\cite{zeus-1999, zeus-2000,zeus-2007} Collaborations.

However, among the quoted publications only three~\cite{h1-2000,zeus-2000,DC-24} are providing extensive sets of SDME values that were obtained through an analysis of three-dimensional angular distributions of $\rho ^0$ production and decay. Such an analysis allows for the determination of all 15 SDMEs that are not coupled to the beam polarisation (``unpolarised SDMEs''), as in Refs.~\cite{h1-2000,zeus-2000}.
The complete set of 23 SDMEs, which includes the 8 SDMEs coupled to the beam polarisation (``polarised SDMEs''), was obtained for the first time by HERMES~\cite{DC-24}.
The published results confirm
the dominance of amplitudes for NPE transitions and the violation of the SCHC hypothesis that is observed for the transitions $\gamma^*_T \to \rho^0_L$.
 % The SDMEs describe the spin structure of the reaction shown in Eq.~(\ref{eq:introduction:01}). They are

The present COMPASS results on SDMEs for exclusive $\rho ^0$ muoproduction have the potential to further constrain GPDs, in particular in conjunction with the published COMPASS results on SDMEs for exclusive $\omega$ production~\cite{COMPASS-omega}.
These additional constraints on GPD parameterisations are  beyond those obtained from measurements of cross sections and spin asymmetries in HEMP.
The COMPASS SDME results provide input to asses the role of chiral-odd, i.e., parton helicity-flip GPDs in exclusive vector-meson production, which are related
%and how the chiral-odd GPDs relate
%in the framework of the GK model
to the mechanism of SCHC violation.
%\sout{In the framework of the GK model it may become possible to assess the role of the chiral-odd GPDs in the mechanism of SCHC violation.}
%\textcolor{red}{ Do we need the hierarchy at all? \\
%i) a hierarchy of helicity amplitudes can be established; }

\section{ Theoretical formalism}
\label{formalism}

Throughout this article, the theoretical formalism of SDMEs and helicity amplitudes introduced by K. Schilling and G. Wolf~\cite{Schill} is used, thereby following the notation from Refs.~\cite{COMPASS-omega,HERMES:2014}.
%Adopting the notation from Refs.~\cite{COMPASS-omega,HERMES:2014}, the theoretical formalism of SDMEs and helicity amplitudes introduced by K. Schilling and
%G. Wolf~\cite{Schill} is used throughout this paper.
\subsection{Definition of Spin Density Matrix Elements}

In the hard exclusive process of vector-meson production on a nucleon $N$ with helicity $\lambda^{}_N$ ($\lambda^\prime_N$) in the initial (final) state (Eq.~\ref{eq:introduction:02}), the transition of a virtual photon $\gamma^*$ with helicity $\lambda_{\gamma}$  to a vector meson $V$ with helicity $\lambda _V$ is described by helicity amplitudes {$F_{\lambda_V^{ } \lambda^\prime_N\lambda_{\gamma}^{}\lambda_N^{}}$}, which depend on the three kinematic variables $W$, $Q^{2}$, and $t^\prime$ with $t'\equiv|t| - t_{0}\approx p^{2}_{\rm T}$. Here $t$ is the squared four-momentum transfer to the
proton, $t_0$ represents the smallest kinematically allowed value of $|t|$ for given $Q^2$ and meson mass and  $p^{2}_{\rm T}$ the squared transverse momentum of the vector meson with respect to the virtual-photon direction. In the $\gamma^*$-$N$ centre-of-mass (CM) system, the vector-meson spin density matrix $\rho_{\lambda_V^{}\lambda_V^\prime}$ is related to the helicity amplitude {$F_{\lambda_V^{} \lambda^\prime_N\lambda _{\gamma}^{} \lambda_N^{}}$} as
%The  helicity amplitudes {$F_{\lambda _{V} \lambda '_{N}\lambda %_{\gamma} \lambda
%_{N}}$} describe the transition of a
%virtual photon with helicity $\lambda _{\gamma}$ to a vector meson with
%helicity $ \lambda _{V}$,
%where $\lambda_{N}$ ($\lambda '_{N}$) is the
%nucleon helicity  in the initial (final) state.
%The helicity amplitudes depend on $W$, $Q^{2}$, and $t'\equiv|t| %- t_{0}\approx p^{2}_{\rm T}$, where $t_{0}$ represents the %smallest kinematically allowed value of $|t|$ for given
%$Q^{2}$ and meson mass, and $p^{2}_{\rm T}$ is the square of the %vector-meson transverse
%momentum with respect to the direction of the virtual photon. In %the centre-of-mass (CM) system of virtual photon and nucleon, %the vector-meson spin density matrix %$\rho_{\lambda_{V}\lambda_{V}^{'}}$ is related to the helicity %amplitude {$F_{\lambda _{V} \lambda '_{N}\lambda _{\gamma} %\lambda_{N}}$} as
%given by equation~\ref{eq:rhotohel}
\cite{Schill}
\begin{eqnarray}
\rho_{\lambda_V^{} \lambda '_V}=
\frac{1}{2 \mathcal{N} }
  \sum_{\lambda_{\gamma}^{}
\lambda '_{\gamma}\lambda_N^{} \lambda '_N}
   F^{}_{\lambda_{V}^{}\lambda '_N\lambda_{\gamma}^{}\lambda _N^{}}~
 \varrho^{U+L}_{\lambda_{\gamma}^{} \lambda '_{\gamma}}~
  F_{\lambda '_{V} \lambda '_N\lambda '_{\gamma}\lambda
 _N^{}}^{*},\,\,
 \label{eq:rhotohel}
\end{eqnarray}
with the normalisation factor $\mathcal{N}$~\cite{Schill,DC-24}. The virtual-photon spin density matrix
$\varrho^{U+L}_{\lambda_{\gamma} \lambda^\prime_{\gamma}}$~\cite{DC-24} describes the QED-calculable sub-process
$ \mu \rightarrow \mu^\prime + \gamma^{*}$. It can be decomposed into elements coupled to the longitudinal beam polarisation $P_\text{b}$ (indicated by a superscript $L$) and elements not coupled to $P_\text{b}$ (superscript $U$):
%where $\mathcal{N}$ is a normalisation %factor~\cite{Schill,DC-24}.
%\sout{the $\varrho^{U+L}_{\lambda_{\gamma} \lambda'_{\gamma}} $ %is the virtual photon spin density matrix~\cite{DC-24}.}
%The virtual-photon spin density matrix
%$\varrho^{U+L}_{\lambda_{\gamma} \lambda'_{\gamma}}$  %~\cite{DC-24} describes the subprocess
%$ \mu \rightarrow \mu' + \gamma^{*}$, which is
%calculable in quantum electrodynamics. It can be decomposed as
\begin{equation}
\varrho^{U+L}_{\lambda_{\gamma} \lambda '_{\gamma }}
 = \varrho^{U}_{\lambda_{\gamma} \lambda '_{\gamma }} +
P_{\text{b}}~\varrho^{L}_{\lambda_{\gamma} \lambda '_{\gamma }}.
\label{phspden}
\end{equation}
The vector-meson spin density matrix elements (SDMEs) discussed below are related to either $U$ or $L$ elements in Eq.~\ref{phspden} and will correspondingly be referred to as ``unpolarised'' or ``polarised'' in the following.
%where the matrix with superscript $L$ ($U$) contains elements %that
%are coupled (not coupled) to the beam polarisation %and
%$P_{b}$. % is the value of the beam polarisation.
%In the following the corresponding vector-meson SDMEs, which are %related to these elements, will be referred to as ``polarised'' %(``unpolarised'').

 %\sout{After decomposition of %$\varrho^{U+L}_{\lambda_{\gamma} \lambda
 %'_{\gamma}} $ into the standard  set of $3\times 3$ %Hermitian
 %matrices $ \Sigma^{\alpha}$,
 %the vector-meson spin density matrix
 %is expressed in
 %terms of a set of nine matrices %$\rho^{\alpha}_{\lambda_V\lambda'_V}$, which are %related to various
 %photon-polarisation states: transversely polarised %photon ($\alpha$=0, ... ,3),
 %longitudinally  polarised photon ($\alpha$=4), and  %terms describing
 %their interference  ($\alpha$=5, ..., %8)~\cite{Schill}.}
After the decomposition of   $\varrho^{U+L}_{\lambda_{\gamma} \lambda
 '_{\gamma}}$ into the set of $3\times 3$ Hermitian matrices~\cite{Schill}, the vector-meson spin density matrix can be expressed in terms of a set of nine matrices
 $\rho^{\alpha}_{\lambda_V^{}\lambda'_V}$  corresponding to different virtual-photon polarisation states. Here $\alpha$= 0 corresponds to unpolarised transverse photons, $\alpha$= 1, 2 to the two directions of linear polarisation, $\alpha$ = 3 to circular photons and
 $\alpha$ = 4 represents longitudinal virtual photons. The terms with $\alpha$ = 5 ... 8 correspond to the interference of transverse and longitudinal amplitudes.

Lacking the possibility to separate contributions from longitudinally and transversely polarised virtual photons, as is the case for this experiment, one usually defines SDMEs as follows:
%In case it is
%If it is experimentally %]
%not possible to separate the contributions from longitudinally and transversely polarised photons,
%%it is usual to define the SDMEs as
%SDMEs are usually defined as follows:
\begin{equation}
r^{04}_{\lambda_V^{}\lambda '_V} = (\rho^{0}_{\lambda_V^{}\lambda '_V}
+ \epsilon R \rho^{4}_{\lambda_V^{}\lambda '_V})( 1 + \epsilon R )^{-1},
\nonumber
\end{equation}
\begin{equation}
r^{\alpha}_{\lambda_V^{}\lambda'_V} =
\begin{cases}
{  \rho^{\alpha}_{\lambda_V^{}\lambda'_V}}{(  1 + \epsilon R )^{-1}},
\; \alpha = 1,2,3,\\
{ \sqrt{R} \rho^{\alpha}_{\lambda_V^{}\lambda '_V}}
{(1 + \epsilon R )^{-1}},\; \alpha = 5,6,7,8.
\end{cases}
  \hspace*{0.25cm}
\label{rmatr}
\end{equation}

The quantity $R= \sigma_{L}/ \sigma_{T}$ is the
longitudinal-to-transverse cross-section ratio of virtual photons and $\epsilon$ the
virtual-photon polarisation parameter given in Eq.~(\ref{expreps}). There are in total 23 SDMEs defined in Eq.~(\ref{rmatr}). The relations between these SDMEs and the corresponding helicity amplitudes are provided in Appendix A of Ref.~\cite{DC-24}.
%Here, $R= d\sigma_{L}/ d\sigma_{T}$ is the differential longitudinal-to-transverse cross-section ratio of virtual photons and $\epsilon$ is the
%virtual-photon polarisation parameter, see Eq.~(\ref{expreps}).
%The relations between the 23 SDMEs defined in Eq.~(\ref{rmatr})
% \magenta{[WDN: we better use all indices here: %$r^{\alpha}_{\lambda_{V}\lambda'_{V}}$}]
% $r$
%and the helicity amplitudes are given in Appendix A of Ref.~\cite{DC-24}.

\subsection{Properties of Helicity Amplitudes}
\label{hel_ampli}

Each helicity amplitude $F\equiv F_{\lambda_V^{}\lambda^\prime_N\lambda_{\gamma}^{}\lambda_N^{}}$ can be decomposed linearly into a natural-parity-exchange (NPE) amplitude $T$ and an unnatural-parity-exchange (UPE) amplitude $U$~\cite{Schill,DC-24}: $F=T+U$.
%, where the set of indices for $T$ and $U$ are the same as for $F$ and are skipped for simplicity {\color{red} it is not skipped in the text below}.
The NPE and UPE amplitudes are related to helicity amplitudes as follows~\cite{Schill}:
 %\indent As detailed in Refs.~\cite{Schill,DC-24}, each helicity amplitude can be linearly decomposed into a natural-parity-exchange (NPE) amplitude $T$ and an unnatural-parity-exchange (UPE) amplitude $U$,
%\begin{equation}
%F_{\lambda_{V} \lambda '_{N} \lambda_{\gamma}  \lambda_{N} } =
%T_{\lambda_{V} \lambda '_{N} \lambda_{\gamma}  \lambda_{N} }+
%U_{\lambda_{V}\lambda '_{N} \lambda_{\gamma}  \lambda_{N}},
%\label{nu}
%\end{equation}
%with the following relations~\cite{Schill}:
\begin{eqnarray}
T_{\lambda_V^{} \lambda'_N \lambda_{\gamma}^{} \lambda_N^{}}
&=&
\frac{1}{2}[
F_{\lambda_V^{} \lambda'_N \lambda_{\gamma}^{} \lambda_N^{}}
\nonumber ~~~~~~~~~~~~~~~~~~~~~~~~~~~\\
& & \hspace{0.3cm}
+(-1)^{\lambda_V-\lambda_{\gamma}}F_{-\lambda_V^{} \lambda'_N -\lambda_{\gamma}^{}\lambda_N^{}}],
\label{fnat}\\
U_{\lambda_V^{} \lambda'_N \lambda_{\gamma}^{} \lambda_N^{}}&=&\frac{1}{2}[
F_{\lambda_V^{} \lambda'_N \lambda_{\gamma}^{} \lambda_N^{}}
\nonumber~~~~~~~~~~~~~~~~~~~~~~~~~~~ \\
& & \hspace{0.3cm}
-(-1)^{\lambda_V^{}-\lambda_{\gamma}^{}}F_{-\lambda_V^{} \lambda'_N -\lambda_{\gamma}^{} \lambda_N^{}}].
\label{funnnat}
\end{eqnarray}
The asymptotic behaviour of amplitudes $F$
 at small $t'$ was derived from angular-momentum conservation~\cite{Wang} and reads~\cite{Diehl}
\begin{eqnarray}
 F_{\lambda_V^{} \lambda'_N \lambda_{\gamma}^{} \lambda_N^{}} \propto \Bigl (\frac{\sqrt{t'}}{M}\Bigr
 )^{|(\lambda_V^{}-\lambda'_N)-(\lambda_{\gamma}^{}-\lambda_N^{})|}.
 \label{asytpr}
 \end{eqnarray}
%and it follows from  angular-momentum conservation.
Here and in the following $M$ denotes
the proton mass.
Equations (\ref{fnat}-\ref{asytpr}) show that double-helicity-flip amplitudes with
 $|\lambda _V-\lambda_{\gamma}|=2$ are suppressed at least by a factor of $\sqrt{t'}/M$, and that their contributions to SDMEs are suppressed by $t'/M^2$.

% \blue{There is no interference between NPE and UPE amplitudes and no linear contribution from nucleon-helicity-flip amplitudes to SDMEs in case of an unpolarised target.}\blue{\bf [$\leftarrow$ This sentence is a bit left alone and it is not clear to me what use it has here?]}

Introducing %Using
the notation
\begin{eqnarray}
\widetilde{\sum}T^{}_{\lambda_V^{} \lambda_{\gamma}^{}} T^*_{\lambda'_V
\lambda'_{\gamma}}\equiv
 \frac{1}{2} \sum_{\lambda_N^{} \lambda'_N}
T^{}_{\lambda_V^{} \lambda'_N\lambda_{\gamma}^{}\lambda_N^{}}
T^*_{\lambda'_V \lambda'_N\lambda'_{\gamma}\lambda_N^{}}
\label{tilde-sum}
\end{eqnarray}
and the symmetry properties~\cite{Schill,DC-24} of the amplitudes
$T$, Eq.~(\ref{tilde-sum}) becomes
\begin{eqnarray}
\widetilde{\sum}T^{}_{\lambda_V^{} \lambda_{\gamma}^{}} T^*_{\lambda'_V
\lambda'_{\gamma}} &=&
T^{}_{\lambda_V^{} \frac{1}{2}\lambda_{\gamma}^{}\frac{1}{2}}
T^*_{\lambda'_V \frac{1}{2}\lambda'_{\gamma}\frac{1}{2}}
\nonumber \\
& + &
%\hspace{1.5cm} + \hspace{0.1cm}
T^{}_{\lambda_V^{} -\frac{1}{2}\lambda_{\gamma}^{}\frac{1}{2}}
T^*_{\lambda'_V -\frac{1}{2}\lambda'_{\gamma}\frac{1}{2}}.~~~~~
\label{sum-two}
\end{eqnarray}
Note that the first product $TT^*$ on the right-hand side represents contributions from NPE amplitudes without nucleon-helicity flip, while the second product of NPE amplitudes $TT^*$ includes a nucleon-helicity flip. The relations for the UPE amplitudes can be written in an analogous way. For brevity, the nucleon-helicity indices will be omitted for amplitudes with $\lambda_N^{}=\lambda^\prime_N$ in the following, i.e.,
%Here, both products on the right-hand side represent the contribution of
%NPE amplitudes, the first without and the second with nucleon-helicity flip. The relations for the UPE amplitudes can be written in an analogous way.
%In the abbreviated notation used in the text, the nucleon-helicity indices will be omitted
%for amplitudes with $\lambda_N=\lambda'_N$, i.e.
\begin{eqnarray}
T_{\lambda_V^{} \lambda_{\gamma}^{}}
&\equiv T_{\lambda_V^{} \frac{1}{2}\lambda_{\gamma}^{}\frac{1}{2}}&=
\phantom{-}T_{\lambda_V^{} -\frac{1}{2}\lambda_{\gamma}^{}-\frac{1}{2}},\nonumber\\
 U_{\lambda_V^{} \lambda_{\gamma}^{}}
&\equiv U_{\lambda_V^{} \frac{1}{2}\lambda_{\gamma}^{}\frac{1}{2}}
&=-U_{\lambda_V^{} -\frac{1}{2}\lambda_{\gamma}^{}-\frac{1}{2}}.
\label{abrr}
\end{eqnarray}
The assumption that there exist only diagonal $\gamma^{*} \to V$ transitions ($\lambda_V=\lambda_{\gamma}$) is usually referred to as  hypothesis of $s$-channel helicity conservation.
%The hypothesis of $s$-channel helicity conservation implies that there exist only
%, there exist only
%The dominance of
%diagonal $\gamma^{*} \to V$ transitions ($\lambda_{V}=\lambda_{\gamma}$).
%is referred to as $s$-channel helicity conservation.

%\newpage
\section{Experimental access to SDMEs}
\label{access}

\begin{figure}[hbt!]
\begin{center}
\includegraphics[width=8.0cm]{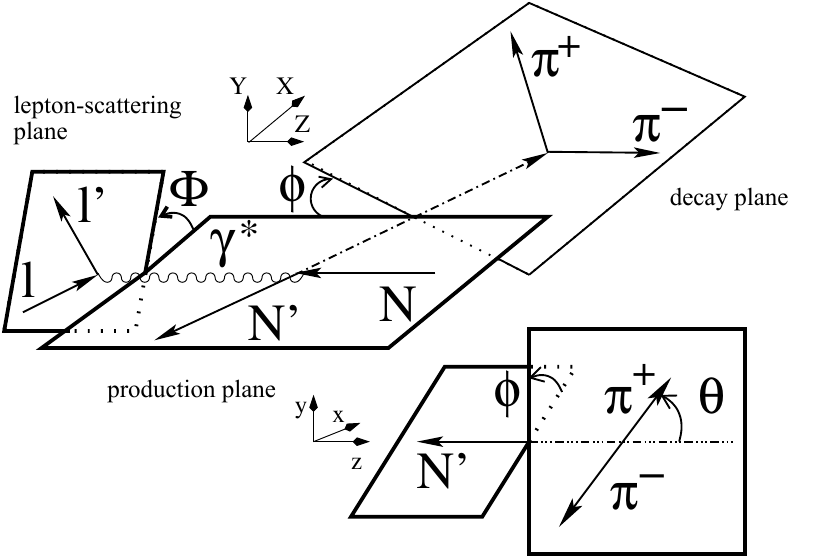}
%\vspace{0.5cm}
\caption{
Definition of angles in the process
$\mu N \to \mu' N'  \rho ^0$ with  $\rho ^0
 \to \pi^+ \pi^- $.
Here, $\Phi$ is the angle between the $\rho ^0$ production plane
and the lepton scattering plane in the
centre-of-mass system
of the virtual photon and the target nucleon, while $\phi$ is the angle between the $\rho ^0$ production and decay planes.
The variable $\Theta$ is
the polar angle of the decay $\pi ^+$ in the
$\rho ^0$ meson rest frame.}
\label{defang}
\end{center}
\end{figure}

Spin density matrix elements are extracted from COMPASS data in exclusive muoproduction of $\rho ^0$ mesons (Eq.~(\ref{eq:introduction:01})). The SDMEs are fitted as parameters of the three-di\-men\-sio\-nal angular distribution
$\mathcal{W}^{U+L}(\Phi,\phi,$ $\cos\Theta)$, which is defined below,
to the corresponding experimental distribution.
The angles and reference frames for the production and decay process $\mu N\to\mu N^\prime \rho^0\;\;(\rho^0\to\pi^+\pi^-)$ are shown in Fig.~\ref{defang}.
\vspace{1cm}

The right-handed ``hadronic CM system'' of virtual photon and target nucleon is identical to that used in Ref.~\cite{COMPASS-omega}.
The directions of axes of the hadronic CM system and  the $\rho ^0$-meson rest frame coincide with the directions of axes of the helicity frame~\cite{Schill,DC-24,joos}.
The angles $\Phi$, $\phi$, and $\Theta$ involved in %with
$\rho^0$-meson production and decay are defined as follows~\cite{DC-24}.
The azimuthal angle $\Phi$ between $\rho^0$-meson production plane and lepton scattering plane in the hadronic CM system is given by:
\begin{eqnarray}
\cos \Phi = \frac{ (\vec{q} \times \vec{v}) \cdot (\vec{k} \times \vec{k}')}
{ | \vec{q} \times \vec{v} | \cdot |\vec{k} \times \vec{k}'| } ,
\label{phicap1-def}
\end{eqnarray}
\begin{eqnarray}
\sin \Phi =
 \frac{ [ (\vec{q} \times \vec{v} )\times (\vec{k} \times \vec{k}' )] \cdot \vec{q} }
 { |\vec{q} \times \vec{v}| \cdot |\vec{k} \times \vec{k}'| \cdot |\vec{q}|
 }.
\label{phicap2-def}
\end{eqnarray}
Here,  $\vec{k}$, $\vec{k'}$, $\vec{q}=\vec{k}-\vec{k'}$, and $\vec{v}$  are the three-momenta of the incoming  and outgoing lepton, the virtual photon, and the $\rho ^0 $ meson respectively.
The azimuthal angle $\phi $ between $\rho ^0$-meson decay and production planes is defined by:
\begin{eqnarray}
\cos \phi =
\frac{ (\vec{q} \times \vec{v} )\cdot (\vec{v} \times \vec{p_{\pi ^+}} ) }
{ | (\vec{q} \times \vec{v} ) |
%{ | (\vec{q} \times \vec{v} )\times \vec{v} |
\cdot |\vec{v} \times \vec{p_{\pi ^+} |}} ,
\label{phismall1-def}
\end{eqnarray}
\begin{eqnarray}
 \sin \phi =
 \frac{[ (\vec{q} \times \vec{v} )\times \vec{v} ] \cdot ( \vec{p_{\pi ^+}}  \times \vec{v} ) }
{ | (\vec{q} \times \vec{v} )\times \vec{v} | \cdot |\vec{p_{\pi ^+}} \times \vec{v} |
 } ,
\label{phismall2-def}
\end{eqnarray}
where $\vec{p_{\pi ^+}}$ is the three-momentum of the positively charged decay pion $\pi^+$ in the hadronic CM system.\\
The polar angle  $\Theta$ of the decay $\pi ^+$ in the vector-meson rest frame, with the $z$-axis aligned opposite to the outgoing nucleon three-momentum $\vec{P}^\prime$ and the $y$-axis directed along $\vec{P}^\prime \times \vec{q}$, is given by:
\begin{eqnarray}
\cos \Theta  = -\frac{ \vec{P}^\prime \cdot \vec{P_{\pi^+}}}{| \vec{P}^\prime| \cdot |\vec{P_{\pi^+}}|},
\label{theta-def}
\end{eqnarray}
where $\vec{P}_{\pi^+}$ is the three-momentum of the
positively charged decay pion in the vector-meson rest frame.

%Spin density matrix elements are extracted from COMPASS data in exclusive muoproduction of $\rho ^0$ mesons (Eq.~\ref{eq:introduction:01}). The SDMEs are fitted as parameters of the three-di\-men\-sio\-nal angular distribution
%$\mathcal{W}^{\text{U+L}}(\Phi,\phi,$ $\cos\Theta)$
%to the corresponding experimental distribution.
The angular distribution $\mathcal{W}^{U+L}$ is decomposed
into contributions that are not coupled ($\mathcal{W}^{U}$ - unpolarised) or coupled ($\mathcal{W}^{L}$ - polarised) to the longitudinal beam polarisation $P_\text{b}$:
\begin{eqnarray}
\mathcal{W}^{U+L}(\Phi,\phi,\cos{\Theta}) =
\nonumber \\
\mathcal{W}^{U}(\Phi,\phi,\cos{\Theta}) & + &
P_{\text{b}}\mathcal{W}^{L}(\Phi,\phi,\cos{\Theta}).%,
\label{eqang1}
\end{eqnarray}

Since the data were collected  with longitudinally polarised muon beams, both unpolarised and polarised SDMEs can be accessed, allowing the extraction of 15 unpolarised SDMEs from $\mathcal{W}^{U}$:
\begin{equation}
\mathcal{W}^{U}(\Phi,\phi,\cos{\Theta}) \nonumber
\end{equation}
\vspace{-0.6cm}
\begin{eqnarray}
%\mathcal{W}^{U}(\Phi,\phi,\cos{\Theta})
&=& \frac{3} {8 \pi^{2}} \Bigg[
         \frac{1}{2} (1 - r^{04}_{00}) + \frac{1}{2} (3 r^{04}_{00}-1) \cos^2{\Theta}
\nonumber \\
& & \hspace{0.5cm} -\sqrt{2} \mathrm{Re} \{ r^{04}_{10} \} \sin 2\Theta
\cos \phi
- r^{04}_{1-1}  \sin ^{2} \Theta\cos 2 \phi
\nonumber \\
&-& \epsilon \cos 2 \Phi \Big( r^{1}_{11} \sin ^{2} \Theta  + r^{1}_{00} \cos^{2}{\Theta} \nonumber \\
& &  \hspace{0.5cm}
  - \sqrt{ 2}  \mathrm{Re} \{r^{1}_{10}\} \sin 2  \Theta  \cos  \phi
   - r^{1}_{1-1} \sin ^{2} \Theta \cos 2 \phi   \Big)   \nonumber  \\
&-& \epsilon \sin 2 \Phi \Big( \sqrt{2} \mathrm{Im} \{r^{2}_{10}\} \sin 2 \Theta \sin \phi \nonumber \\
& &  \hspace{3.5cm}
 +\mathrm{Im} \{ r^{2}_{1-1} \} \sin ^{2} \Theta \sin 2 \phi  \Big)  \nonumber \\
&+& \sqrt{ 2 \epsilon (1+ \epsilon)}  \cos \Phi
\Big(  r^{5}_{11} \sin ^2 {\Theta} +
 r^{5}_{00} \cos ^{2} \Theta \nonumber \\
& & \hspace{0.5cm}
 - \sqrt{2} \mathrm{Re} \{r^{5}_{10}\} \sin 2 \Theta \cos \phi
 - r^{5}_{1-1} \sin ^{2} \Theta \cos 2 \phi  \Big)  \nonumber \\
&+& \sqrt{ 2 \epsilon (1+ \epsilon)}  \sin \Phi
\Big( \sqrt{ 2} \mathrm{Im} \{ r^{6}_{10} \} \sin 2 \Theta \sin \phi \nonumber \\
& & \hspace{3.5cm}
+ \mathrm{Im} \{r^{6}_{1-1} \} \sin ^{2} \Theta \sin 2 \phi \Big) \Bigg],
\label{eqang2}%\\
%\hspace*{-3.0cm}
\end{eqnarray}
and the extraction of 8 polarised SDMEs from $\mathcal{W}^{L}$:
\begin{equation}
\mathcal{W}^{L}(\Phi,\phi,\cos \Theta) \nonumber
\end{equation}
\vspace{-0.6cm}
\begin{eqnarray}
%\mathcal{W}^{L}(\Phi,\phi,\cos \Theta) & &\\
 & =&  \frac{3}{8 \pi^{2}}  \Bigg[
  \sqrt{ 1 - \epsilon ^{2} }  \Big(  \sqrt{ 2}  \mathrm{Im} \{ r^{3}_{10} \}
  \sin 2 \Theta \sin \phi \nonumber  \\
& & \hspace{3.5cm} +   \mathrm{Im} \{ r^{3}_{1-1}\} \sin ^{2} \Theta \sin 2 \phi  \Big)  \nonumber  \\
&+& \sqrt{ 2 \epsilon (1 - \epsilon)} \cos \Phi
\Big( \sqrt{2} \mathrm{Im} \{r^{7}_{10}\} \sin 2 \Theta \sin \phi \nonumber  \\
& & \hspace{3.5cm} +  \mathrm{Im} \{ r^{7}_{1-1} \}  \sin ^{2} \Theta \sin 2 \phi   \Big)  \nonumber \\
&+& \sqrt{ 2 \epsilon (1 - \epsilon)} \sin \Phi
\Big( r^{8}_{11} \sin ^{2} \Theta + r^{8}_{00} \cos ^{2}
\Theta \nonumber  \\
& & \hspace{0.5cm} -  \sqrt{2} \mathrm{Re}\{ r^{8}_{10}\} \sin 2 \Theta \cos \phi    - r^{8}_{1-1} \sin ^{2} \Theta \cos 2\phi \Big)  \Bigg].\hspace*{-20.0cm}
\label{eqang3}
\end{eqnarray}

The virtual-photon polarisation parameter $\epsilon$, which
represents the  ratio of fluxes of
longitudinal and transverse virtual photons, is given by:
\begin{eqnarray}\label{expreps}
\epsilon &=& \frac{1-y - y^2\frac{Q^2}{4\nu^2}}{1-y+ \frac{1}{4}y^2
(\frac{Q^2}{\nu^2} + 2)},
\end{eqnarray}
where $y = p\cdot q / p\cdot k   \stackrel{\text{lab}}{=} \nu / E$.
%Here
The symbols $p$, $q$ and $k$ denote the four-momenta of target proton, virtual photon and incident lepton, respectively. The energy of virtual photon and incident lepton in the target rest frame is denoted by $\nu$ and $E$, respectively.

\section{Experimental setup and data selection}
\label{sec:exp} % CKR added

The fixed-target experiment COMPASS is located at CERN in the M2-beamline of the Super Proton Synchrotron (SPS). The experiment consists of a versatile setup that can use variety of targets, detectors and make use of different beams. It uses a two-stage spectrometer with a number of tracking and particle identification detectors placed over a length of approximately 60~m. Each stage of the spectrometer is built around one of the two dipole magnets (SM1 and SM2). The first stage covers large scattering angles up to 180~mrad, and the second stage smaller
scattering angles down to 0.5~mrad. More detailed descriptions of the COMPASS experiment can be found in Refs.~\cite{comp1,comp2,comp3}.
%The main component of the COMPASS setup is the two-stage magnetic
% spectrometer. Each spectrometer stage comprises a dipole magnet
% complemented by a variety of tracking detectors, a muon filter for muon
% identification and an electromagnetic as well as a hadron
% calorimeter. A detailed description of the setup can be found in Refs.
% \cite{comp1,comp2,comp3}.

In this paper, we analyse data collected during 4 weeks in 2012 that were dedicated to the pilot run of a program designed to study GPDs and hadron tomography through Deeply Virtual Compton Scattering (DVCS) and HEMP processes. The experiment made use of the 160 GeV$/c$ polarised muon beam and an unpolarised liquid-hydrogen target. The target was surrounded by a time-of-flight (TOF) system for the detection of recoil protons. The use of the recoil-proton detector (RPD) is important in the studies of the DVCS process, but for HEMP processes, like $\rho ^0$-meson production, it restricts the kinematic coverage, so that the RPD information is not used in the present analysis.

%{\color{red} The effect of a different amount of non-exclusive background onto the extracted values of SDMEs was studied in the earlier COMPASS analysis of exclusive $\omega$ production~\cite{COMPASS-omega}. Two event samples were used for the extraction of SDMEs. The first one was obtained by applying selections similar to those described in the present $\rho ^0$ analysis, while for the second one the more restrictive selections using the information from the RPD were added. The SDME values extracted from the two data samples were found to be consistent within statistical uncertainties.
%Motivated by this observation, no attempt was made to include the RPD information into the $\rho ^0$ analysis.
%}
%The data used for this analysis were collected within four weeks in 2012. In this period
%the COMPASS spectrometer was complemented by a 2.5 m long
%liquid-hydrogen target surrounded by a time-of-flight (TOF) system for the detection of recoiling protons and by a
%third electromagnetic calorimeter placed directly downstream of
%the target.
%The recoil detector restricts the kinematic coverage towards the region of small squared transverse momentum of the $\rho ^0$-meson with respect to the virtual-photon direction. Hence it was used only for an additional check of the background correction procedure as explained in Sec.~\ref{sec:sys} and not for the determination of SDMEs.

The muon beam originates from in-flight decays of pions and kaons produced by SPS protons impinging on a primary
target. Due to the weak nature of the decay, the muon beam is naturally polarised. The beam is negatively polarised for $\mu^{+}$ and positively polarised for $\mu^{-}$, and the achieved absolute value of polarisation is $(80 \pm 4) \%$. The data has been taken using both $\mu^{+}$ and $\mu^{-}$ beams. The SM1 and SM2 spectrometer magnets polarities were changed accordingly with beam charge to ensure equal acceptance of the COMPASS spectrometer in both cases.

%Data  with $\mu^{+}$ and $\mu^{-}$ beams were
% taken separately.
 %The natural polarisation of the 160 GeV$/c$ muon beam provided by the
 %CERN SPS originates from the parity-violating decay-in-flight of the parent
 %meson, which implies opposite polarisations for $\mu^{+}$ and $\mu^{-}$
 %beams. Within regular time intervals during
%data taking, charge and
 %polarisation of the muon beam were swapped simultaneously. In order to equalise the spectrometer acceptance for the two beam charges,
 %also the polarities of the two spectrometer magnets were changed accordingly. For  both beams,
%the absolute value of the average beam polarisation is about 0.8 with an
% uncertainty of about 0.04.

The analysis is focused on the process
\begin{center}
\begin{tikzpicture}
\node [right] at (0., 1.2) {$\mu p \rightarrow \mu' p' \rho ^0$};
\node [right] at (3.0, 0.6) {$\pi^+ \pi^-$};
%\node [right] at (2.5, 0.6) {$\pi^+ \pi^-$};
%\draw [thick, ->] (2.1, 0.9) -- (2.1, 0.6) -- (2.7, 0.6);
\draw [thick, ->] (2.0, 0.85) -- (2.0, 0.55) -- (2.7, 0.55);
\node [right] at (6.7, 0.6) {$\mathrm{BR} \approx 99\%,$};
%\node [right] at (6.7, 0.0) {$\mathrm{BR} \approx 99\%.$};
\label{eq:selection:reaction}
\end{tikzpicture}
\end{center}
which defines the topology of the accepted events. The events are required to have two hadron tracks of opposite charge and one reconstructed vertex inside the target with incoming and outgoing muon associated. The outgoing muon is required to traverse more than 15 radiation lengths of material and to have the same charge as the incoming muon. Charged hadron tracks are identified by requiring to traverse less than 10 radiation lengths of material.
%The selected events should have one reconstructed vertex inside the
%liquid-hydrogen target  associated with the incoming and the outgoing muon, and
%two hadron tracks of oppo\-si\-te charge. The outgoing muon has to have the
%same charge as the incoming muon and is required to traverse more than 15 radiation lengths to be identified as a muon. The
%charged hadron tracks are selected by requiring the traversed path to be
%shorter than 10 radiation lengths.

\subsection{Kinematic selections}
\label{sec:kine-sel} % CKR added
In order to select exclusively produced $\rho ^0$ mesons, events are required to meet the following kinematic constraints:
%The following kinematic selections
%are applied to
 %\st{select} \blue{extract}  \st{the}
%select exclusively produced
%$\rho ^0$ mesons:

\begin{itemize}
\item 1.0~(GeV/$c)^2$ < $Q^{2} < 10.0$~(GeV/$c)^2$, which selects the region of perturbative QCD (lower limit) and suppresses background from hadrons produced in DIS, hereafter referred to as ``SIDIS background'' (upper limit);
%where the lower limit
%ensures applicability of pQCD and the upper one suppresses background due to hadrons produced in DIS, which hereafter is referred to as ``SIDIS background''.

\item $y$ < 0.9 to minimize the effect of radiative corrections;
%the lower limit suppresses events with  poorly reconstructed
%kinematics and the upper one removes events with large radiative corrections.

\item $W$ > 5.0~GeV/$c^2$ to avoid significant fluctuations in the cross section that appear in the lower $W$ region because of the production of resonances;
%to remove the kinematic region where the cross section changes rapidly due to the production of resonances.

\item $\nu$ > 20~GeV, which is the energy of the virtual photon in the laboratory frame;
%where in the laboratory system $\nu$ is equal
%to the energy of the virtual photon.

\item   0.01~(GeV/$c)^2$  < $ p_{\rm{T}}^2 <  0.5$~(GeV/$c)^2$ to remove events with poorly determined azimuthal angle (lower limit) and to suppress SIDIS background (upper limit);

%where $p_{\rm{T}}$ is the
%transverse momentum of the $\rho ^0$ meson  with
%respect to the virtual-photon direction.
%The lower limit removes
%events with a poorly determined
%azimuthal angle of the produced meson and the upper one suppresses SIDIS background.

\item $ P_{\rho ^0}$ > 15~GeV/$c$ to reduce the SIDIS background contribution, where $P_{\rho ^0}$ is the $\rho ^0$ momentum in the laboratory frame;

\item 0.5~GeV/$c^2$ < $M_{\pi^{+}\pi^{-}}$ <  1.1~GeV/$c^2$ to select $\rho ^0$ mesons, where $M_{\pi^{+}\pi^{-}}$ is the two-pion invariant mass.
\end{itemize}

In order to select exclusively produced $\rho ^0$ mesons, the missing energy
%is required to be $-$2.5~GeV $<  E_{\rm miss} <$ 2.5~GeV
%in order to count for experimental resolution. The missing energy
%is given as
%The information from the recoil proton detector is not used for the extraction of SDMEs. Instead, in order to enhance the fraction of events with exclusively produced $\rho ^0$ mesons, the missing energy
\begin{equation}
E_{\rm miss}  = \frac{ M^{2}_{\rm X} - M^{2}}{2 M}
\end{equation}
is used.
Here $M^{2}_{\rm X}=({p} + {q}- {p}_{\pi^+} - {p}_{\pi^-})^{2}$ is the missing mass squared, ${p}_{\pi^{+}(\pi^-)}$  the pion four-momenta and $M$ the mass of the proton. In order to account for experimental
resolution the selection $-$2.5~GeV $<  E_{\rm miss} <$ 2.5~GeV is applied.
The distribution of the missing energy is shown in Fig.~\ref{emiss}, where the exclusive peak in the experimental data appears within the selection limits.

% is constrained by $-$2.5~GeV $<  E_{\rm miss} <$ 2.5~GeV
%to take into account the experimental resolution. Here $M$ is
%the proton mass, $ M^{2}_{\rm X}=({p} + {q}- {p}_{\pi^+} - {p}_{\pi^-})^{2}$ is the missing mass squared, and ${p}_{\pi^+}$ and $
%{p}_{\pi^-}$ are the four-momenta of the two pions.
%The $E_{\rm miss}$ distribution for the experimental data is shown in
%Fig.~\ref{emiss}. The exclusive peak is apparent.

After having applied all the selection requirements, the data set for physics analysis consists of 23785 events taken with the $\mu ^+$ beam and 28472 events with the $\mu ^-$ beam.
%After applying all selection criteria, there are 23785 events for $\mu +$ and 28472 for $\mu ^-$ beam that are available for further analysis.

\subsection{Invariant mass distribution}
\label{invariantmass}
\begin{figure}[hbt!]\centering
\includegraphics[width=0.5\textwidth]{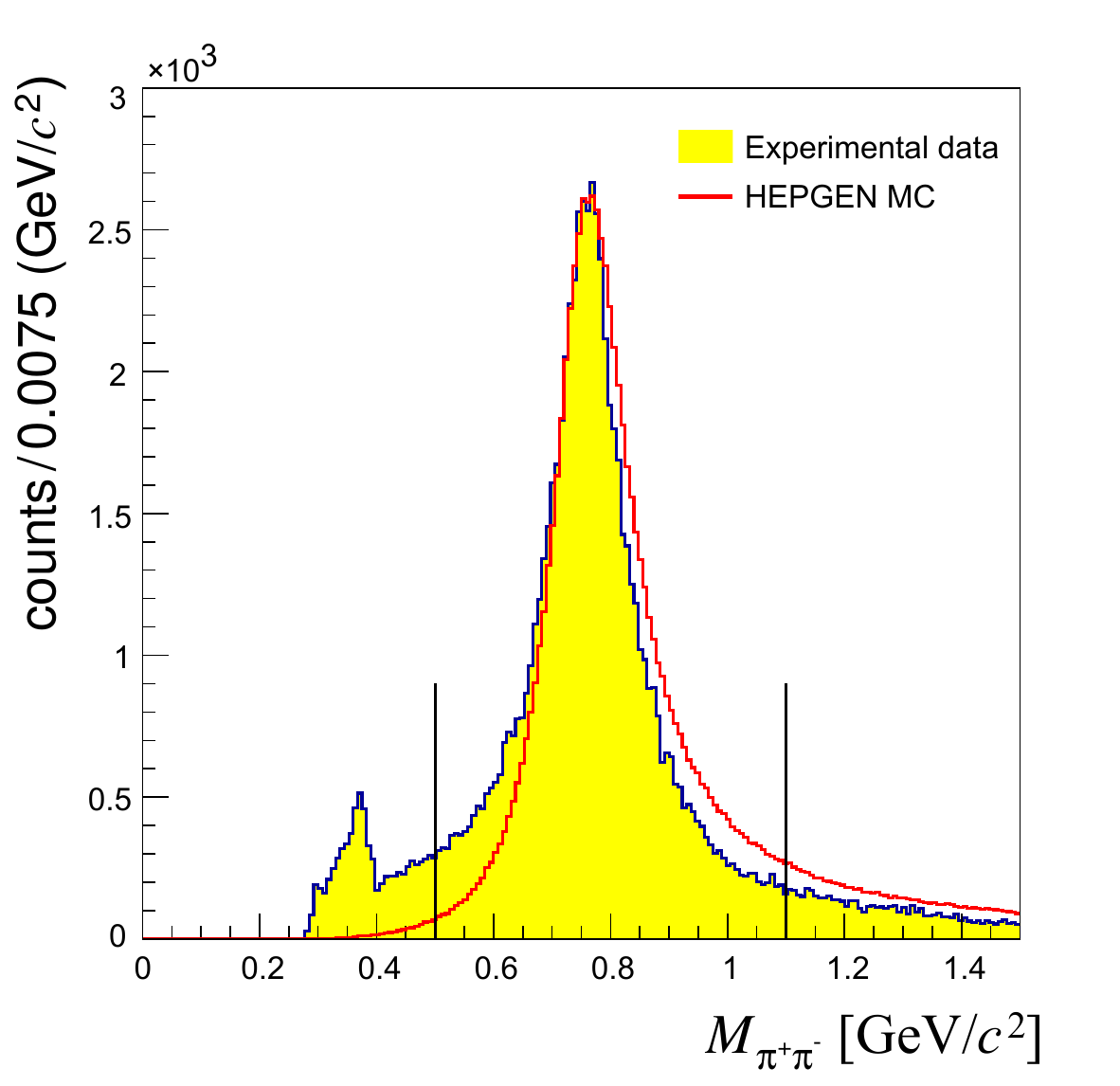}
\caption{\small{
Distribution of the $\pi ^+ \pi ^-$ invariant mass
for experimental data (shaded histogram) and HEPGEN MC (open histogram). The distributions are obtained applying all event selections except the selection on the invariant mass.
The invariant mass distribution from HEPGEN  is normalised to the data in the region
0.75 GeV/$c^2$ < $M_{\pi^{+}\pi^{-}}$ < 0.77 GeV/$c^2$.
The vertical lines indicate the applied limits.
 }}
\label{invmass}
\end{figure}

The two-pion invariant mass distribution is shown in Fig.~\ref{invmass}.
A clear $\rho ^0$ signal is observed.
Background coming from exclusive production of $\phi$ and its decay $\phi \to K^+ K^-$, where the kaons are misidentified as pions, is expected and seen at $M_{\pi^{+}\pi^{-}} <$ 0.4 GeV/$c ^2$.
Therefore, applying the selection 0.5~GeV/$c^2$ < $M_{\pi^{+}\pi^{-}}$ <  1.1~GeV/$c^2$ removes this background.
The distribution for the experimental data is compared to that for the reconstructed events
of Monte Carlo (MC) events
obtained with the HEPGEN++ $\rho ^0$ generator,
in the following denoted by HEPGEN~\cite{hepg1,hepg2}. As in HEPGEN only
exclusive $\rho ^0$ production is generated, while the data contains exclusive production of both resonant and non-resonant pi+pi- pairs, as well as their interference,
a difference in shapes between the experimental and simulated distributions is observed.
The effect is in agreement with the expectation from
the S\"oding model~\cite{Soeding}, which predicts a sizeable interference between the small amplitude for non-resonant $\pi^{+}\pi^{-}$ pair production and the large one for resonant
$\pi^{+}\pi^{-}$ production. The characteristic prediction of the model is the change of the sign of the interference term at the maximum of the $\rho ^0$ resonance from being positive at smaller values of $M_{\pi^{+}\pi^{-}}$ to negative at larger $M_{\pi^{+}\pi^{-}}$ values.

In order to evaluate the contribution of non-resonant $\pi^{+}\pi^{-}$ pair production, the following
procedure is used. The invariant mass distribution from HEPGEN is normalised to the data in the region
0.75 GeV/$c^2$ < $M_{\pi^{+}\pi^{-}}$ < 0.77 GeV/$c^2$. The difference  between the integrals of the distributions for the data and HEPGEN over the full range 0.5~GeV/$c^2$ < $M_{\pi^{+}\pi^{-}}$ <  1.1~GeV/$c^2$ is approximately equal to the global contribution of non-resonant
production. This contribution, which includes
the interference between the
amplitudes for resonant  and non-resonant production as well as the squared  amplitude for the latter one, is equal to about 3$\%$ and hence neglected.

%A possible background from exclusive $\omega$ production and its decay $\omega \to \pi^+ pi^- pi^0$, where $\pi^0$ remains undetected, that may contribute~\ref{kreisel} at $M_{\pi \pi}$ < 0.65 Gev/$c^2$ is included in the aforementioned
In addition, the $\omega \rightarrow \pi^{+}\pi^{-}$
channel, with a branching fraction of 1.5\%,
gives an irreducible background to the $\rho ^0$ channel.
As the branching fraction is small and the contribution of $\rho^0 - \omega$ interference was found to be very small~\cite{HERMES:EPJC-17}, the contribution of $\omega \rightarrow \pi^{+}\pi^{-}$ channel is neglected in this analysis.

\subsection{Backgrounds for exclusive $\rho^0$ production}
\label{sec:SIDIS-bg}

i) {\it SIDIS background}

\begin{figure}[hbt!]\centering
\includegraphics[width=0.5\textwidth]{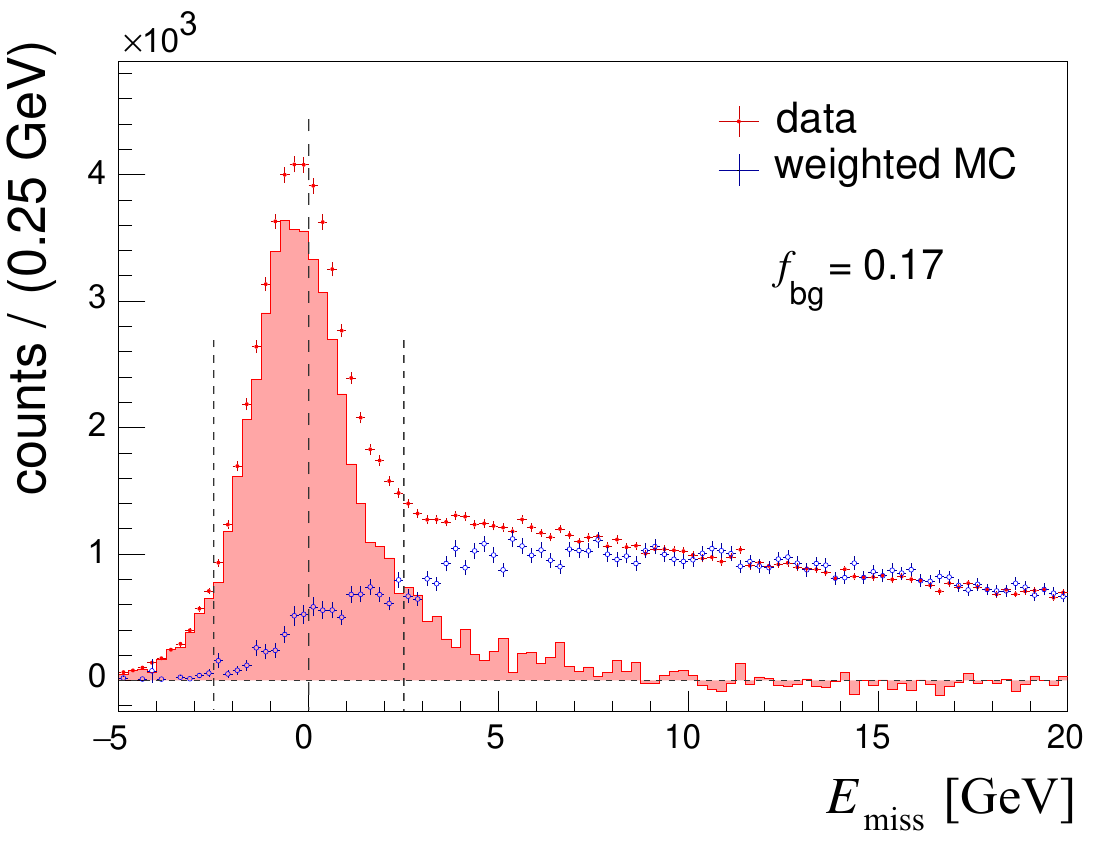}
\caption{ The missing-energy distribution from experimental data (red points) compared to the distribution of SIDIS events from a LEPTO MC
simulation
(blue points).
Each LEPTO MC event is
reweighted by a $E_{\rm miss}$-dependent weight that is calculated using both
experimental and simulated data
with same-charge hadron pairs. See text for a detailed
explanation.
The reweighted MC distribution is
normalised to the data in
the region 7~GeV  $< E_{\rm miss} < 20$~GeV.
The SIDIS-background-corrected distribution for the data is shown as shaded histogram. The vertical lines at $|E_{\rm miss}| =$ 2.5 GeV indicate
the limits of the exclusive region.
The shown distributions are obtained using all event selections except the selection on $E_{\rm miss}$.
}
%; to Kamil: do you need still to modfiy Caption}}

\label{emiss}
\end{figure}

The largest background contribution is due to the SIDIS background, i.e., events with hadrons produced in DIS. In order to determine the fraction of SIDIS background in the selected $\rho^0$ events, the $ E_{\rm miss} $ distribution is used as shown in Fig.~\ref{emiss}. The procedure is described in detail in Refs.~\cite{COMPASS-asy,psz}. The SIDIS background simulation is performed using the LEPTO~6.5.1 generator with the COMPASS tuning of parameters~\cite{Comptune} and processed with the simulation of the COMPASS setup~\cite{hepg3}.
The simulated events are selected using the same criteria as for the experimental data. In order to improve the agreement between LEPTO events and the data, the simulated events are reweighted. For this purpose, events with the same-sign hadron pairs are selected.
%Additionally, the simulated events are reweighted to ensure agreement with experimental data in the procedure for the background fraction estimation.
The reweighting is applied on a bin-by-bin basis to the $ E_{\rm miss} $ distribution with the following weight:
%Its treatment is detailed in the following.

%The $ E_{\rm miss} $ distribution shown in Fig.~\ref{emiss} is used to determine the fraction of
%SIDIS background under the exclusive peak, following the procedure described
%in Refs.~\cite{COMPASS-asy,psz}. For the simulation of background, the LEPTO 6.5.1 %Monte Carlo
%generator is used with the COMPASS tuning of parameters~\cite{Comptune}.
%In order to achieve the best possible agreement between experimental and simulated
%$E_{\rm miss}$ distributions, the
%simulated data
%are reweighted on a bin-by-bin basis using the weight
\begin{equation}
 w(E_{\rm miss}) =
 \frac{N^{\text{sc}}_{\text{D}}(E_{\rm miss})}{N^{\text{sc}}_{\text{MC}}(E_{\rm miss})}.
\end{equation}
Here $N^{\text{sc}}_{\text{D}}(E_{\rm miss})$ ($N^{\text{sc}}_{\text{MC}}(E_{\rm miss})$) is the number of events  with same-sign hadron pairs selected from experimental (D) or simulated (MC) data.

The distribution of the reweighted LEPTO  events is normalised to the experimental data in the background dominated region of 7~GeV $<E_{\rm miss}<$ 20~GeV.
It is shown in Fig.~\ref{emiss} as the blue points.
The procedure estimates the background fraction $f_{\rm bg}$  for the selected $\rho^0$ to be 0.17 in the signal region $-$2.5~GeV < $E_{\rm miss}$ < 2.5~GeV. However, it was found that the fraction of SIDIS background changes within the kinematic coverage of this measurement, in particular it is increasing with increasing $Q^2$ and $p^2_{\rm T}$ and with decreasing $W$. Therefore, the background fraction is estimated in each kinematic bin separately, resulting in values of $f_{\rm bg}$ from 0.10 to 0.32 for the determination of the SDME values as functions of kinematic variables.

%Here $N^{sc}_{rd}(E_{\rm miss})$ and $N^{sc}_{MC}(E_{\rm miss})$  are
%numbers of events containing same-charge hadron pairs,
%$h^{+} h^{+}$ and $h^{-} h^{-}$, in the %3
%two-pion system for experimental and simulated data, respectively.
%In order to improve the statistical significance, the
%constraint on the $\omega$ invariant mass is not used for the purpose of
%estimating the weight $w$.
%In Fig.~\ref{emiss} the blue points represent the simulated
%SIDIS background, which is generated by LEPTO
%and processed through the full simulation of the COMPASS setup~\cite{hepg3}, followed by the same event
%reconstruction and selection procedure as for the real data, and then reweighted in the way
%described above. The distribution is
%normalised to the
%experimental data  in the region
%7~GeV $<E_{\rm miss}<$ 20~GeV.
%The fraction of background in the signal window $-$2.5~GeV < $E_{\rm miss}$ < 2.5~GeV
%for the total kinematic range
%is found to be $f_{\rm bg}$= 0.17 for the total kinematic range.
%The fraction of SIDIS background increases with increasing
%$Q^2$ and
%$p^2_{\rm T}$, and it decreases with increasing $W$. For the results on kinematic
%dependences of SDMEs, which are presented in the following, the background
%fraction $f_{\rm bg}$
%is evaluated separately for each kinematic bin, the values
%ranging between 0.10 and 0.32.

ii) {\it Background from proton-dissociation processes}

This background is due to the processes
$\gamma^* + p \rightarrow \rho^0 + N^*$ with a baryon $N^*$
decaying
into a system of hadrons. As observed by HERA experiments,
the cross section for such process, when integrated over all $N^*$ states, is at the level of 20\% of the cross section for exclusive $\rho^0$ production~\cite{zeus-2007}.
In the present analysis such processes are suppressed by the applied selections, in particular on
$E_{\rm miss}$ and $ p_{\rm{T}}^2$,
%The residual contribution from proton-dissociation
%processes was estimated by analysing the shape
%of $E_{\rm miss}$ distribution after correction for SIDIS background ({\it eg.} see the shaded histogram in Fig.~\ref{emiss}).
%The amount of this background contribution is approximately {\color{red} XXX - to be evaluated}.
which reduce the contribution from proton-dissociation by a factor of 2.
As it is reported in Refs.~\cite{h1-2010,zeus-2000} that
the angular distributions for $\rho^0$ decay and production
in the exclusive and
proton-dissociation channels are compatible, no correction is applied for proton-dissociation events.

iii) {\it Other backgrounds}

In addition to the background contributions discussed above and in Sec.~\Ref{invariantmass}, several others were considered in Ref.~\cite{h1-2010}. They may originate from the processes $\omega \rightarrow \pi^{+} \pi^{-} \pi^0$, $\phi \rightarrow \rho \pi$, $\phi \rightarrow \pi^{+} \pi^{-} \pi^0$ and  $\rho' \rightarrow \pi^{+} \pi^{-} \pi^0 \pi^0$.
These events are expected to be strongly suppressed after applying the selections on
$E_{\rm miss}$ and $ p_{\rm{T}}^2$ and their contributions are neglected in the present analysis.

\section{Extraction of SDMEs}
\subsection{Unbinned Maximum Likelihood method}
\label{sec:uml} % CKR added

The method to determine SDMEs was described in Sec.~\ref{access}.  Equations~(\ref{eqang1}, \ref{eqang2}, \ref{eqang3}) relate the angular distribution ${\mathcal{W}}$ to the 23 SDMEs $r^{\alpha}_{\lambda_V^{}{\lambda'_V}}$. In order to extract  SDME values in this measurement, the Unbinned Maximum Likelihood (UML) method is used to fit the experimental three-dimensional angular distribution of $\rho ^0 $ production and decay to the function ${\mathcal{W}} ({\cal R};\Phi,\phi,\cos\Theta)$, where $\cal R$ is the set of the 23 SDMEs. In the fit, the negative log-likelihood function
\begin{eqnarray}
-\ln L({\cal R})=
-\sum_{i=1}^{N}\ln\frac{\mathcal{W}^{U+L}({\cal
R};\Phi_{i},\phi_{i},\cos{\Theta_{i}})}{\widetilde{\mathcal
N}({\cal R})}
\label{loglik-def}
\end{eqnarray}
is minimised. Here, $N$ represents the number of selected events and $ {\mathcal N}({\cal R})$ is the likelihood normalisation factor defined as
\begin{equation}
 \widetilde {\mathcal N}({\cal R})=
\sum_{j=1}^{N_{\text{MC}}}\mathcal{W}^{U+L}({\cal
 R};\Phi_{j},\phi_{j},\cos{\Theta_{j}}),
\label{loglik-def1}
\end{equation}
where $N_{\text{MC}}$ is the number of simulated $\rho ^0$ events generated by the HEPGEN generator~\cite{hepg1,hepg2}. In order to simulate exclusive $\rho ^0$ production, the option of an isotropic three-dimensional angular distribution of $\rho ^0$ production and decay was chosen. The generated events are further processed with the simulation of the COMPASS setup~\cite{hepg3}. Identical selection requirements are applied as for the experimental data.

%The SDMEs are determined by an Unbinned Maximum Likelihood fit of the
%function
%${\mathcal{W}} ({\cal R};\Phi,\phi,\cos\Theta)$ to the experimental
%three-dimensional
% angular distribution of
%$\rho ^0 $ production and decay. The explicit expression for the dependence
% of ${\mathcal{W}}$
%on SDMEs was given in Sec.~\ref{access} by Eqs.~(\ref{eqang1},
%\ref{eqang2}, \ref{eqang3}). Here $\cal R$ denotes the set of 23 SDMEs $r
%^{\alpha}_{\lambda_{V}{\lambda'_{V}}}$.
%The negative log-likelihood function to be minimised reads

%where $N$ is the number of selected events.The likelihood normalisation factor

%is calculated numerically using the sample of MC events generated with
%the HEPGEN $\rho ^0$ generator
%\cite{hepg1,hepg2}. This
%generator is used to model the kinematics of exclusive $\rho ^0$ production. For
%the purpose of the present analysis, the option with an isotropic
%three-dimensional angular distribution of $\rho ^0$ production and decay is
%chosen. The generated events are passed through a complete description of the
%COMPASS setup and the resulting  data are treated in the same
%way as it was done for experimental data. The number of HEPGEN events is denoted $N_{MC}$
%in Eq.~(\ref{loglik-def1}).

\subsection{Background-corrected SDMEs}
\label{sec:bg-corr-in}

The above described procedure of SDME extraction does not account for the SIDIS background contamination. In order to determine the background-corrected SDMEs, a two-step approach is applied.

In the first step, the 23 ``background SDMEs'' are determined using a parameterisation of the background angular distributions. The SIDIS background events simulated by the LEPTO generator are treated by the same method as described above. The UML fit is performed in the signal region, $-$2.5~GeV$<E_{\rm miss} <$ 2.5~GeV according to Eq.~(\ref{loglik-def}), resulting in the set ${\cal R}_{\rm bg}$ of background SDMEs.

%In order to determine SDMEs that are corrected for SIDIS background, a two-step procedure is used. First, the parameterisation of the background angular distributions
%is obtained by applying the above described
%maximum likelihood method to selected SIDIS events simulated with the LEPTO generator.
%These events are required to pass the same selection criteria as experimental data. Performing an unbinned likelihood fit according to Eq.~(\ref{loglik-def}) using simulated events in the range $-$2.5~GeV$<E_{\rm miss} <$ 2.5~GeV yields
%the set $\cal B$ of 23 ``background SDMEs''.
%\label{sec:bg-corr} % CKR added

In the second step, the set ${\cal R}_{\rm bg}$ and the background fraction $f_{\rm bg}$ determined in Sec.~\ref{sec:SIDIS-bg} are used to extract the set ${\cal R}_{\rm sig}$ of the  background-corrected SDMEs by fitting the negative log-likelihood function
%. The extraction is performed with fit to the following negative log-likelihood function:}
%In the second step, the set $\cal B$ of background SDMEs %${\cal B}$
%is used to extract the set $\cal R$ of background-corrected SDMEs %$\cal R$
%by applying the unbinned maximum likelihood fit to the experimental data.
%For this purpose the following negative log-likelihood function is
%fitted:
\begin{eqnarray}
 -\ln L({\cal R}_{\rm sig})= \nonumber
 ~~~~~~~~~~~~~~~~~~~~~~~~~~~~~~~~~~~~~~~~~~~~~~~~~~~~~~\\
\;\;\; - \sum_{i=1}^{N}\ln\Bigl[ \frac{ (1-f_{\rm bg})~\mathcal W^{U+L}({\cal R}_{\rm sig};\Phi_{i},\phi_{i},\cos\Theta_{i})
}{\widetilde {\mathcal N}({\cal R}_{\rm sig},{\cal R}_{\rm bg})} \nonumber\\
 \hspace{1cm}
+\frac{f_{\rm bg}~\mathcal W^{U+L}({\cal R}_{\rm bg}; \Phi_{i},\phi_{i}, \cos \Theta_{i})}
{\widetilde{\mathcal N}({\cal R}_{\rm sig},{\cal R}_{\rm bg})}\Bigr].
\label{logbac}
\end{eqnarray}
Here, $\widetilde{\mathcal N}$ is the likelihood normalisation factor defined as
%Here, $f_{\rm bg}$ is the fraction of background events in the selected experimental data as determined in Sec.~\ref{sec:kine-sel} and $\widetilde{\mathcal N}$ is the
%normalisation factor:
\begin{eqnarray}
\widetilde{\mathcal N}({\cal R}_{\rm sig},{\cal R}_{\rm bg})= \nonumber  ~~~~~~~~~~~~~~~~~~~~~~~~~~~~~~~~~~~~~~~~~~~~~~~~~~~~ \\
\;\;\sum_{j=1}^{N_{MC}}[(1-f_{\rm bg})~\mathcal W^{U+L}({\cal R}_{\rm sig}; \Phi_{j},
\phi_{j},\cos\Theta_{j}) \nonumber\\
\hspace{0.5cm}
+f_{\rm bg}~\mathcal W^{U+L}({\cal R}_{\rm bg}; \Phi_{j},
\phi_{j},\cos\Theta_{j})].
\label{logbacnor}
\end{eqnarray}

\subsection{Statistical uncertainties of the observables depending on SDMEs}
\label{sec:stat}
The statistical uncertainties of the observables, which depend on SDMEs and are discussed in Sec.~\ref{sec:discus}, were calculated by propagating the statistical uncertainties of SDMEs and using their covariance matrix as obtained from the fit. The corresponding correlation matrix for the data in the total kinematic range is presented in Table~\ref{corrtab2}.

\subsection {Systematic uncertainties}
\label{sec:sys} % CKR added
The systematic uncertainties of the measured SDME values are considered to arise from the following sources:
%The following sources of systematic uncertainties are considered:

\begin{itemize}
\item[i)]{\it Difference between $\mu ^+$ and $\mu ^-$ beam}\\
In the measurement, $\mu ^+$ and $\mu ^-$ beams were used, which were not identical in terms of  intensity. The intensity of the $\mu^{+}$  beam was higher by a factor of approximately 2.7 than that of the $\mu ^-$ beam. In order to account for a possible impact of this difference on the measured SDMEs, the latter are extracted separately for $\mu ^+$ and $\mu ^-$ data, and half of the difference between the two results is assigned as systematic uncertainty.

%The $\mu^{+}$  beam intensity was about 2.7 times higher than that of the $\mu ^-$ beam.
% A possible impact of this difference on the determination of SDMEs is  checked
%by comparing the SDMEs extracted
%separately for the
%$\mu^{+}$ beam (negative polarisation) and the $\mu^{-}$
%beam (positive polarisation).
%For each SDME, half of the difference between the SDME values determined  with  opposite beam polarisations
% is taken as systematic uncertainty.

\item[ii)]{\it Position of the $E_{\rm miss}$ peak}\\As observed in Fig.~\ref{emiss}, the signal peak in the $E_{\rm miss}$ distribution is not centred at zero, but rather slightly shifted towards negative values. The reason for this shift is a small imbalance between the measured incoming muon energy and energies of the measured final state particles in the spectrometer. Some SDME values depend on the position of this peak~\cite{comnote}, hence a systematic uncertainty is assigned based on the difference between the SDMEs  extracted without and with a correction of +0.25~GeV/$c$ to the beam momentum to centre the $ E_{\rm miss} $ peak at zero. It was checked that this method of the beam momentum correction yields a similar systematic uncertainty as in case of the method that rescales the measured momenta of the final-state particles to centre the $E_{\rm miss}$ peak position at zero.

%It was  observed in Ref.~\cite{comnote}  that certain SDME values depend on the position of the $E_{\rm miss}$ peak.
% The $E_{\rm miss}$ distribution shown
%  in Fig.~\ref{emiss} is not precisely centred at zero, but slightly shifted towards negative values.
%This results from an imbalance between the energy
%measured for
%  the incoming muon and the energies of the final-state particles measured in the forward spectrometer.
%The effect of this  shift  on the extracted SDMEs is investigated by applying the
% small kinematic correction (+0.25~GeV/$c$) to the beam momentum that is
% needed to centre the $ E_{\rm miss} $ peak at zero. The difference between the
% values of final SDMEs and those obtained with corrected kinematics is taken as
 %\blue{the}
% systematic uncertainty.

% A similar shift of the $E_{\rm miss}$ peak
% is obtained by rescaling the momenta of the final-state particles measured in the spectrometer. The differences between SDME values obtained without and with the rescaling are comparable to those obtained with corrected beam momentum. In order to avoid double counting, only the differences obtained with corrected beam momentum are taken as systematic uncertainties.

\item[iii)]{\it Dependence on the background angular distribution}\\
  The method to evaluate the background-corrected SDMEs, described in Sec.~\ref{sec:bg-corr-in}, uses the LEPTO generated events for the estimation of the SIDIS background SDMEs in the signal region $-$2.5~GeV $< E_{\rm{miss}} <$ 2.5~GeV.  Note that the LEPTO generator was not tuned to reproduce the experimental angular distributions in the specific phase space of this analysis.
 % data never experimentally verified for the event selection used in the presented analysis.
  In order to account for a  possible source of uncertainty, another procedure was applied to estimate the background SDMEs using the background-dominated region  7.0~GeV$< E_{\rm miss} <$ 20.0~GeV in the experimental data. The systematic uncertainty is assigned based on the difference between the two methods of evaluating the background SDMEs.

 %As detailed in Sec.~\ref{sec:bg-corr-in},
 %and\ref{sec:bg-corr}}
 % the background-corrected SDMEs are obtained with  background SDMEs  obtained from LEPTO events in the exclusive region
 % $-$2.5~GeV $< E_{\rm{miss}} <$ 2.5~GeV.
%As the angular distributions from LEPTO were never experimentally
%verified for the event selection used in the present analysis,
%as a check
%an alternative method
%is used, in which background SDMEs
%are estimated from experimental
% data in the
%region  7.0~GeV$< E_{\rm miss} <$ 20.0~GeV.
%The difference between SDMEs obtained by these two methods is taken as
% systematic uncertainty.

\item[iv)]{\it Uncertainty in the determination of the background fraction}\\
Another contribution to the systematic uncertainty of SDME values is related
to the uncertainty of the background fraction determination.
It is estimated to be about $1\%$ based
on the comparison of background fraction values
that were evaluated using two different methods to normalise LEPTO MC results with respect to experimental data.
The difference between the respective SDME values
%that were obtained using two methods of background fraction determination
is taken as systematic uncertainty.

\item[v)]{\it Sensitivity to the shapes of the kinematic distributions generated by HEPGEN}\\
  The SDME values can be sensitive to the shapes of the kinematic distributions generated by the HEPGEN generator. In order to check for such an effect, the SDMEs were extracted using modified HEPGEN weights so that the reconstructed MC $Q^2$ and $\nu $ distributions match those of the experimental data.
  %Despite fairly small effect on the measured SDMEs, the systematic uncertainty on SDMEs is assigned as the difference between the extraction from original simulated sample and the reweighted sample.
  Although the effect on the measured SDMEs is fairly small, the difference between the extraction using the original simulated sample and the one with reweighting is assigned as systematic uncertainty.
%In order to check the sensitivity of SDMEs to the shapes of kinematic distributions in the HEPGEN generator, the SDME extraction
%was repeated by reweighting the MC events with weights depending on $Q^2$ and $\nu $. The weights
%are tuned such that the $Q^2$ and $\nu $ distributions from the experimental data
%match those from the reweighted simulated data.
%The effect of this reweighting on the extracted SDMEs is small
%in most cases,
%and the difference between final SDMEs and
%those obtained with reweighted MC events is taken as systematic uncertainty.
\end{itemize}
The effect of different non-exclusive backgrounds on the extracted values of SDMEs was studied in the earlier COMPASS analysis of exclusive $\omega$ production~\cite{COMPASS-omega}. Two event samples were used for the extraction of SDMEs. The first one was obtained by applying selections similar to those described in the present $\rho ^0$ analysis. For the second one the more restrictive selections using the information from the RPD were added, which lead to a reduction of the
non-exclusive background by a factor of about 10. As a limited $p_{\rm {T}}^2$-range is covered by the RPD, the same limited kinematic region was used to compare the SDMEs obtained with and without RPD. The SDME values extracted from the two data samples were found to be consistent within statistical uncertainties. This observation confirms the correctness of the method to extract the SIDIS-background-corrected SDMEs as explained in Sec.~\Ref{sec:bg-corr-in}.

The contributions from the aforementioned sources i)-v) of systematic uncertainties are shown individually in the Appendix in Table~\ref{atab1}. The largest source is typically from group i) (the difference between the $\mu ^+$ and $\mu ^-$ beams), followed by group ii) (shift in the $E_{\rm miss}$ peak position) and group iii) (dependence on the background angular distribution).
The systematic uncertainties arising from the above discussed sources are added in quadrature to obtain the total systematic uncertainty.
The 23 SDMEs measured over the entire COMPASS kinematic region are given in Table 1 together with their statistical and total systematic uncertainties.
For most SDMEs the total systematic uncertainty is larger than the statistical uncertainty.

The systematic uncertainties of the observables that are discussed in Sec.~\ref{sec:discus} were estimated using a procedure analogous to that used for SDMEs. This means that for a given observable the contribution to its systematic uncertainty from each of the five sources indicated in points i) - v) is calculated directly as difference of the values of the concerned observable obtained when using the corresponding two sets of SDME values. For a given observable, the systematic uncertainties arising from the above discussed sources are added in quadrature to obtain the total systematic uncertainty.

%The systematic uncertainties are combined by adding in quadrature to obtain the total systematic uncertainty. The results of the 23 SDMEs for the entire COMPASS kinematic region with the total systematic uncertainties are given in Table~\ref{tab1}. When compared to statistical uncertainty, the total systematic uncertainty is larger for most SDMEs.

%The total systematic uncertainties are obtained by adding the above described components in quadrature. Table~\ref{tab1} gives the values for the total kinematic region.
%The individual contributions i) - v) to the systematic uncertainty
%for each SDME are compiled in Table \ref{atab1} in the Appendix.
%When averaged over all SDMEs it appears that the group i) systematics dominates by contributing almost half of the systematic uncertainties, while about one-fifth contributions arise from both group ii) and group iii) systematics.
% In most cases, the total systematic uncertainty larger than the statistical one.

\section{Results}
\label{Re}
\subsection{SDMEs for the entire kinematic region}
\begin{figure*}[hbt!]\centering
\includegraphics[width=12cm]{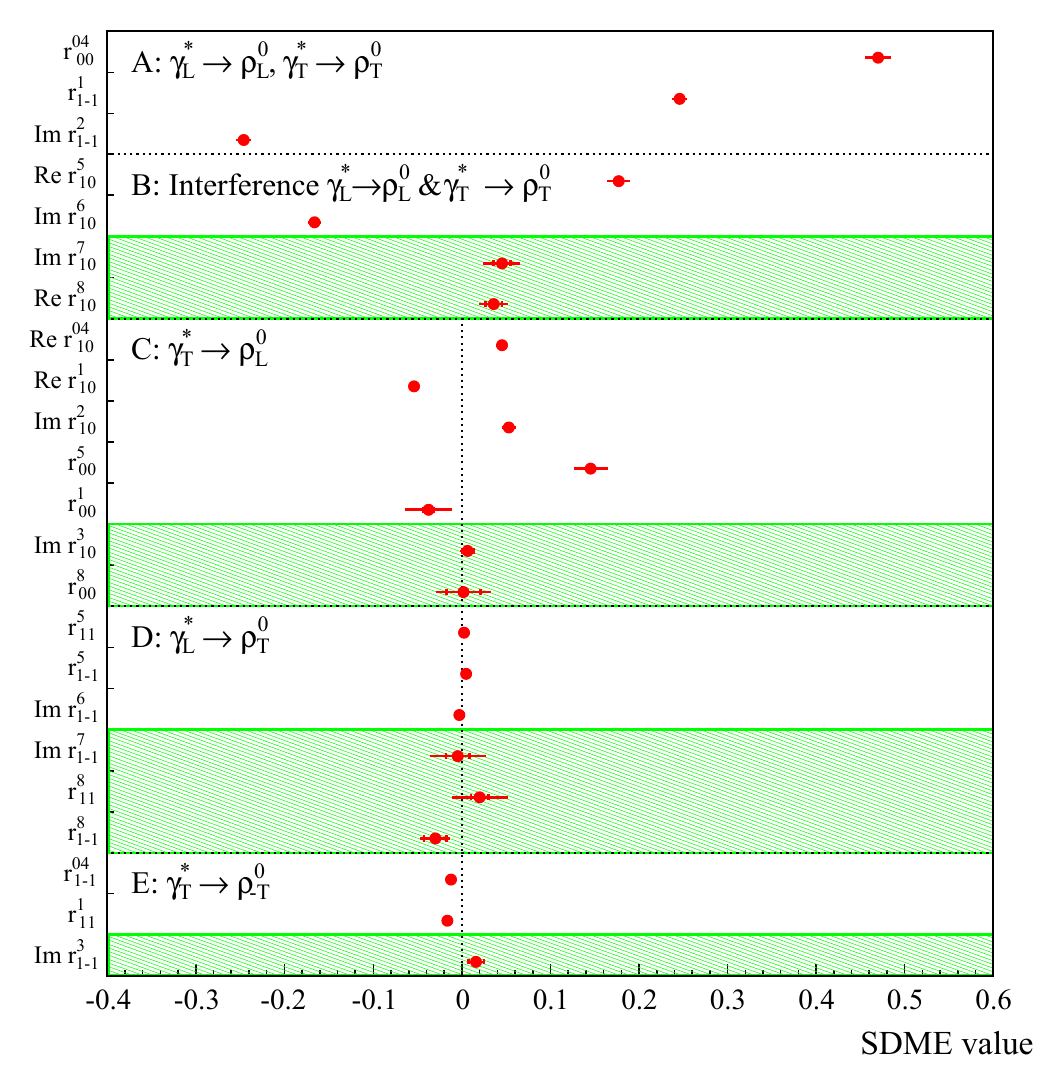}
\caption{
The 23 SDMEs for exclusive $\rho ^0$ leptoproduction extracted in the entire COMPASS kinematic
region with $ \langle Q^2 \rangle = 2.40$~(GeV/$c$)$^2$, $\langle W\rangle =9.9$~GeV/$c^2$, $ \langle p^{2}_{\rm T}\rangle = 0.18$
(GeV/$c$)$^2$.
Inner error bars represent statistical uncertainties and
outer ones statistical and systematic uncertainties added in
quadrature. Unpolarised (polarised) SDMEs are displayed in %the
unshaded (shaded) areas.
}
\label{results}
\end{figure*}

The kinematic region is defined as:
 1.0~(GeV/$c$)$^2$ $< Q^{2} < 10.0$~(GeV/$c$)$^2$,  $\;$ 5.0~GeV/$c^2$  $< W < 17.0$~GeV/$c^2$ $\;$ and
 0.01~(GeV/$c$)$^2$ $<p^{2}_{\rm T} < 0.5$~(GeV/$c$)$^2$,
with mean values
$\langle Q^{2} \rangle= 2.40$~(GeV/$c$)$^2$, $\langle W \rangle= 9.9$~GeV/$c^2$ and
$\langle p^{2}_{\rm T} \rangle = 0.18$~(GeV/$c$)$^2$. The SDMEs extracted in this region are presented in
Fig.~\ref{results} and %in
Table~\ref{tab1}. Following Refs.~\cite{DC-24,COMPASS-omega} they are assembled in five classes corresponding
to different helicity transitions.
In Fig.~\ref{results},  polarised SDMEs are shown in shaded areas.
%\textcolor{red}{check paper requirements for Tab./Fig. etc. and unify in whole paper}
%In the calculation of the statistical uncertainties the correlations between the various SDMEs are taken into account.}

The dominant contributions to the SDMEs in class A are related to the squared amplitudes for
 transitions from
longitudinally polarised virtual photons to  longitudinally polarised vector mesons, $\gamma^*_L \to
V^{ }_L$, and from transversely polarised virtual photons to transversely polarised vector mesons, $\gamma^*_T
\to V^{ }_T$.
 %The SDMEs extracted in the total kinematic region
 %1.0~(GeV/$c$)$^2$ $< Q^{2} < 10.0$~(GeV/$c$)$^2$, 5.0~GeV/$c^2$  $< W < 17.0$~GeV/$c^2$ and
 %0.01~(GeV/$c$)$^2$ $<p^{2}_{\rm T} < 0.5$~(GeV/$c$)$^2$,
%with mean values
%$\langle Q^{2} \rangle= 2.40$~(GeV/$c$)$^2$, $\langle W \rangle= 9.9$~GeV/$c^2$ and
%These SDMEs are presented in five classes corresponding
%to different helicity transitions. For the SDMEs in class A,
%the dominant contributions are related to the squared amplitudes for
% transitions from
%longitudinal virtual photons to  longitudinal vector mesons, $\gamma^*_L \to
%V^{ }_L$, and from transverse virtual photons to transverse vector mesons, $\gamma^*_T
%\to V^{ }_T$.
The former ones appear in the SDME $r^{04}_{00}$, and the latter ones in the SDMEs $r^1_{1-1}$ and Im $r^2_{1-1}$, which approximately mirror each other value (see Fig.~\ref{results} and Table~\ref{tab1}).
The dominant  terms in class B correspond to the interference
between amplitudes for the two aforementioned transitions.
The SDMEs of this class allow the determination of the phase difference between the amplitude $T_{11} $ for $\gamma^*_T
\to V^{ }_T$ transitions and the amplitude $T_{00}$ for $\gamma^*_L \to
V^{ }_L$ transition ({\it cf} Sec.~\ref{phase}).
 In class C, the main terms in most of the SDMEs are
 proportional to the interference between the helicity-flip amplitude $T_{01}$, describing $\gamma^*_T \to V^{ }_L$ transitions, and
 the large helicity-conserving amplitudes, either $T_{11}$ (for Re $r^{04}_{10}$, Re $r^1_{10}$,
Im $r^2_{10}$, Im $r^3_{10}$) or $T_{00}$ (for $r^5_{00}$,
$r^8_{00}$). The dominant terms in the SDMEs of classes D and E are proportional to the interference between the amplitude $T_{11}$ and small amplitudes describing $\gamma^*_L \to V^{ }_T$ and $\gamma^*_T \to V^{ }_{-T}$ transitions, respectively.

The experimental uncertainties of the polarised SDMEs are in most of the cases larger
than those of the unpolarised
ones
because the lepton-beam
polarisation is smaller than unity ($|P_{\text{b}}| \approx 80\%$), and in the
expressions for the
angular distributions (see Eq.~(\ref{eqang3}))
they are multiplied by the small factor
$|P_{\text{b}}|\sqrt{1- \epsilon} $,
where $ \epsilon \approx 0.90$.
\begin{table}[hbt!]
\begin{center}
\caption{\label{tab1} The 23 unpolarised and polarised SDMEs %values
 for the entire COMPASS kinematic region,
shown in the same order as in Fig.~\ref{results} for classes A to E.
The first uncertainties are statistical, the second  systematic.}
\renewcommand{\arraystretch}{1.2}
\begin{tabular}{|c|r@{\,$\pm$\,}r@{\,$\pm$\,}r|}
\hline
 SDME &\multicolumn{3}{c|}{}     \\
\hline
$r^{04}_{00}$    &$  0.4698$&$ 0.0035$&$ 0.0220$ \\
$r^1_{1-1}$      &$  0.2457$&$ 0.0037$&$ 0.0064$ \\
Im $r^2_{1-1}$   &$ -0.2459$&$ 0.0038$&$ 0.0049$ \\
\hline
Re $r^5_{10}$    &$  0.1769$&$ 0.0015$&$ 0.0041$ \\
Im $r^6_{10}$    &$ -0.1662$&$ 0.0014$&$ 0.0040$ \\
Im $r^7_{10}$    &$  0.0453$&$ 0.0096$&$ 0.0156$ \\
Re $r^8_{10}$    &$  0.0362$&$ 0.0095$&$ 0.0121$ \\
\hline
Re $r^{04}_{10}$ &$  0.0454$&$ 0.0021$&$ 0.0058$ \\
Re $r^1_{10}$    &$ -0.0539$&$ 0.0029$&$ 0.0040$ \\
Im $r^2_{10}$    &$  0.0532$&$ 0.0028$&$ 0.0043$ \\
$r^5_{00}$       &$  0.1456$&$ 0.0033$&$ 0.0129$ \\
$r^1_{00}$       &$ -0.0376$&$ 0.0062$&$ 0.0114$ \\
Im $r^3_{10}$    &$  0.0067$&$ 0.0067$&$ 0.0045$ \\
$r^8_{00}$       &$  0.0019$&$ 0.0194$&$ 0.0253$ \\
\hline
$r^5_{11}$       &$  0.0027$&$ 0.0016$&$ 0.0025$ \\
$r^5_{1-1}$      &$  0.0050$&$ 0.0020$&$ 0.0025$ \\
Im $r^6_{1-1}$   &$ -0.0028$&$ 0.0020$&$ 0.0019$ \\
Im $r^7_{1-1}$   &$ -0.0045$&$ 0.0134$&$ 0.0224$ \\
$r^8_{11}$       &$  0.0203$&$ 0.0101$&$ 0.0305$ \\
$r^8_{1-1}$      &$ -0.0300$&$ 0.0128$&$ 0.0091$ \\
\hline
$r^{04}_{1-1}$   &$ -0.0120$&$ 0.0027$&$ 0.0032$ \\
$r^1_{11}$       &$ -0.0162$&$ 0.0032$&$ 0.0037$ \\
Im $r^3_{1-1}$   &$  0.0163$&$ 0.0085$&$ 0.0043$ \\
\hline
\end{tabular}
\end{center}
\end{table}
\subsection{Dependences of SDMEs on $Q^{2}$, $p^{2}_{\rm{T}}$
and $W $}
\begin{figure*}[hbt!]\centering
\includegraphics[width=13cm]{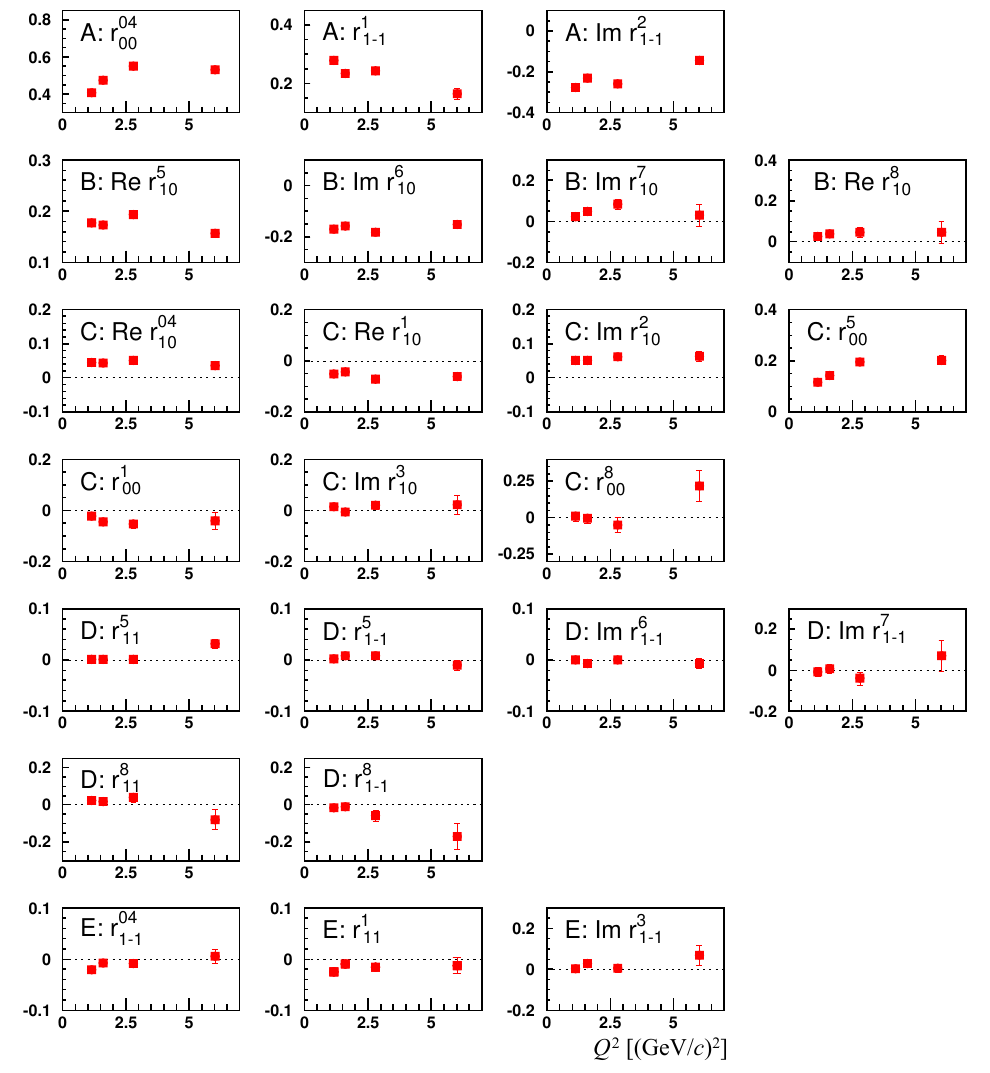}
\caption{
 $Q^2$ dependence of the measured 23 SDMEs.
 The capital letters A to E denote the class, to which the SDME belongs.
Inner error bars represent statistical uncertainties and
outer
ones statistical and systematic uncertainties added in
quadrature.
}
\label{q1_testnsys}
\end{figure*}

\begin{figure*}[hbt!]\centering
\includegraphics[width=13cm]{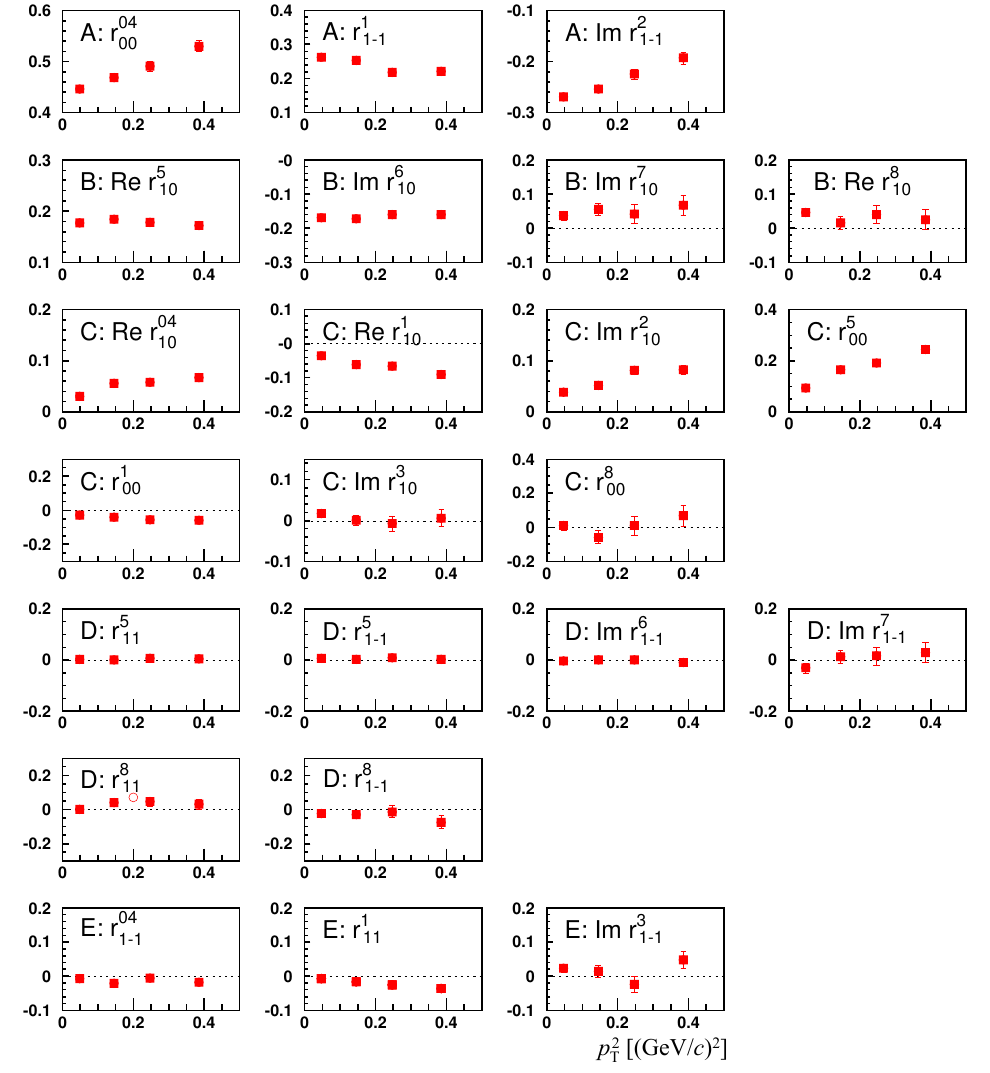}
\caption{
$p^{2}_{\rm{T}}$ dependence of the measured 23 SDMEs.
The capital letters A to E denote the class, to which the SDME belongs.
%The SDMEs are divided into five classes A, B, C, D, and E defined  in Sec.~\ref{Re}.
Inner error bars represent statistical uncertainties and
outer ones statistical and systematic uncertainties added in
quadrature.
}
\label{pt1_testnsys}
\end{figure*}

\begin{figure*}[hbt!]\centering
\includegraphics[width=13cm]{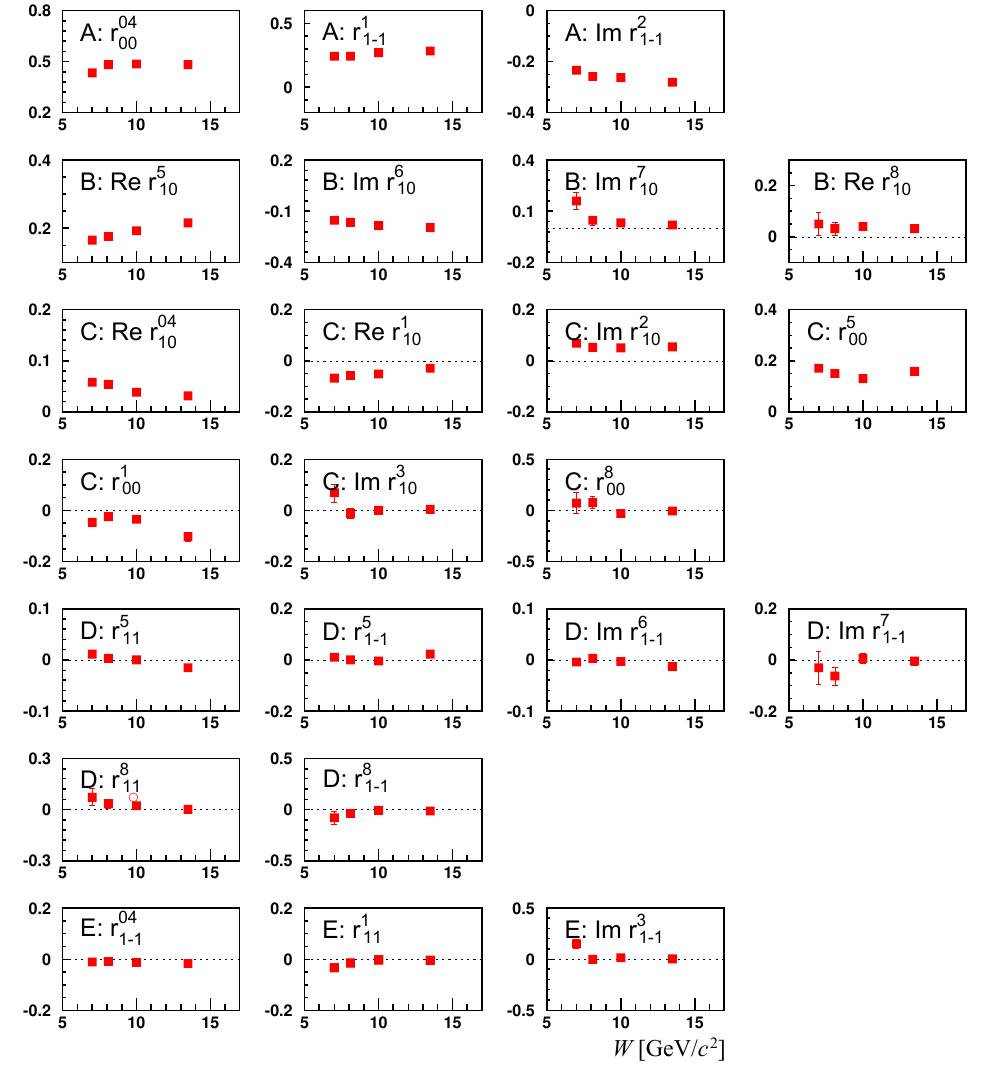}
\caption{
 $W$ dependence of the measured 23 SDMEs.
 The capital letters A to E denote the class, to which the SDME belongs.
% The SDMEs are divided into five classes A,  B, C, D, and  E defined  in Sec.~\ref{Re}.
Inner error bars represent statistical uncertainties and
outer ones statistical and systematic uncertainties added in
quadrature.
}
\label{w1_testnsys}
\end{figure*}

The SDMEs values extracted in four kinematic bins of $Q^{2}$, $p^{2}_{\rm{T}}$, or $ W $
are shown in Figs.~\ref{q1_testnsys}, \ref{pt1_testnsys} and \ref{w1_testnsys}.
The limits of the kinematic bins and the mean values of kinematic variables in each bin are given in Table~\ref{kinem-h}.

%The kinematic dependences of the SDMEs on  $Q^{2}$, $p^{2}_{\rm{T}}$ and $ W $,
%which have been determined in four bins for each of the variables, are shown in Figs.~\ref{q1_testnsys}, \ref{pt1_testnsys} and \ref{w1_testnsys}.
%In Table~\ref{kinem-h}, the limits of the kinematic bins and the mean values of kinematic variables in the bins are given.
%The values of SDMEs in bins of $Q^{2}$, $p^{2}_{\rm{T}}$ and
%$ W $ are given in Table~\ref{atab2}, \ref{atab3} and %\ref{atab4}, respectively, in the Appendix.\\
The value of the SDME $r^{04}_{00}$, which corresponds to the fractional contribution of $|T_{00}|^2$ from longitudinally polarised virtual photons to the cross section, increases with $Q^{2}$ and $p^{2}_{\rm{T}}$, while the opposite trend is observed for the absolute values of the SDMEs $r^1_{1-1}$ and Im $r^2_{1-1}$, which represent the fractional contribution of $|T_{11}|^2$ from transversely polarised virtual photons. In class C a sizeable increase of $r^5_{00}$ with $Q^{2}$ is observed. As a consequence of  angular-momentum conservation the helicity single and double-flip amplitudes should vanish as $p^{2}_{\rm{T}}\to 0$, which is consistent with the measured $p^{2}_{\rm{T}}$-dependence of SDMEs in classes C, D and E.
No clear $W$-dependence is observed for any of 23 SDMEs.

\begin{table*}

\begin{center}
\caption{ \label{kinem-h}
Kinematic binning and mean values for kinematic variables.}

\renewcommand{\arraystretch}{1.2}
\begin{tabular}{|l@{\,}l@{\,<\,}c@{\,<\,}l@{\,}l|r@{\,}l|r@{\,}l|r@{\,}l|}
\hline
\multicolumn{5}{|c|}{bin}&\multicolumn{2}{c|}{$ \langle Q^{2} \rangle$}&\multicolumn{2}{c|}{$ \langle p^{2}_{\rm{T}} \rangle$}&\multicolumn{2}{c|}{$ \langle  W  \rangle$} \\
\hline
~1.0\phantom{0}&(GeV/$c$)$^2$ & $Q^{2} $&\phantom{1}1.3& (GeV/$c$)$^2$  & 1.14&(GeV/$c$)$^2$  & 0.192 &(GeV/$c$)$^2$  & 8.8 & GeV/$c$$^2$ \\
~1.3          &(GeV/$c$)$^2$ & $Q^{2} $&\phantom{1}2.0& (GeV/$c$)$^2$  & 1.60&(GeV/$c$)$^2$  & 0.198 &(GeV/$c$)$^2$  & 8.8& GeV/$c$$^2$ \\
~2.0          &(GeV/$c$)$^2$ & $Q^{2} $&\phantom{1}4.0& (GeV/$c$)$^2$  & 2.80&(GeV/$c$)$^2$  & 0.200 &(GeV/$c$)$^2$  & 8.7 &GeV/$c$$^2$ \\
~4.0          &(GeV/$c$)$^2$ & $Q^{2} $&10.0& (GeV/$c$)$^2$  & 6.02&(GeV/$c$)$^2$  & 0.206 &(GeV/$c$)$^2$  & 8.9 &GeV/$c$$^2$ \\
\hline
\multicolumn{5}{|c|}{bin}&\multicolumn{2}{c|}{$ \langle p^2_{\rm{T}}\rangle $}&\multicolumn{2}{c|}{$ \langle Q^{2}  \rangle$}&\multicolumn{2}{c|}{$ \langle  W  \rangle$} \\
\hline
~0.01          &(GeV/$c$)$^2$&$p^{2}_{\rm{T}}$&\phantom{2}0.1
&(GeV/$c$)$^2$ & 0.053&(GeV/$c$)$^2$ & 2.56 &(GeV/$c$)$^2$  & 8.8 &GeV/$c$$^2$ \\
~0.1          &(GeV/$c$)$^2$&$p^{2}_{\rm{T}}$&\phantom{2}0.2
&(GeV/$c$)$^2$ & 0.147&(GeV/$c$)$^2$  & 2.61 &(GeV/$c$)$^2$  & 8.7 &GeV/$c$$^2$ \\
~0.2          &(GeV/$c$)$^2$ &$p^{2}_{\rm{T}}$&\phantom{2}0.3 &(GeV/$c$)$^2$ & 0.248&(GeV/$c$)$^2$   & 2.66 &(GeV/$c$)$^2$  & 8.7 &GeV/$c$$^2$ \\
~0.3          &(GeV/$c$)$^2$ &$p^{2}_{\rm{T}}$&\phantom{2}0.5 &(GeV/$c$)$^2$ & 0.391&(GeV/$c$)$^2$  & 2.70 &(GeV/$c$)$^2$  & 8.7 &GeV/$c$$^2$ \\
\hline
\multicolumn{5}{|c|}{bin}&\multicolumn{2}{c|}{$ \langle W \rangle $}&\multicolumn{2}{c|}{$ \langle Q^{2} \rangle$}&\multicolumn{2}{c|}{$ \langle p^{2}_{\rm{T}} \rangle$} \\
\hline
~5.0           &GeV/$c^2$&$ W $&\phantom{2}7.3 & GeV/$c^2$ & 7.0&GeV/$c^2$  & 2.90&(GeV/$c$)$^2$  & 0.196&(GeV/$c$)$^2$ \\
~7.3           &GeV/$c^2$&$ W $&\phantom{2}9.0 & GeV/$c^2$ & 8.1&GeV/$c^2$   & 2.65&(GeV/$c$)$^2$  & 0.201&(GeV/$c$)$^2$ \\
~9.0           &GeV/$c^2$&$ W $&12.0 & GeV/$c^2$ & 10.0&GeV/$c^2$   & 2.51&(GeV/$c$)$^2$  & 0.199&(GeV/$c$)$^2$ \\
12.0           &GeV/$c^2$&$ W $&17.0 & GeV/$c^2$ & 13.5&GeV/$c^2$   & 2.13&(GeV/$c$)$^2$  & 0.180&(GeV/$c$)$^2$ \\
\hline
\end{tabular}
\end{center}
\end{table*}

\section{Discussion}
\label{sec:discus}

\subsection {Test of the SCHC hypothesis}
In the case of SCHC only three amplitudes, $T_{00}$, $T_{11}$ and $U_{11}$, may be different from zero. As a consequence all SDMEs of classes A and B may not vanish, while SDMEs from classes C, D, and E should be equal to
zero. Six of the SDMEs in classes A and B have to fulfil the following
relations
\cite{Schill}
%\indent   In case of SCHC, only the seven SDMEs of classes A and B are not
%restricted to vanish, while all SDMEs from classes C, D, and E should be equal to
%zero. Six of the SDMEs in classes A and B have to fulfil the following
%relations
%\cite{Schill}
%~~~~~~~~~~~~~~~~~~~~~~~~\\
\begin{eqnarray}
r_{1-1}^1 &=&-\mathrm{Im} \{r_{1-1}^2\},\nonumber
\\
\mathrm{Re}\{r_{10}^5\} &=&-\mathrm{Im}\{r_{10}^{6}\},\nonumber\\
\mathrm{Im} \{r_{10}^{7}\} &=&~~\: \mathrm{Re} \{r_{10}^{8}\}.%\nonumber
\label{schc_relations}
\end{eqnarray}
Using the extracted SDMEs one obtains:
%Within uncertainties, the extracted SDMEs are consistent with these
%relations:
\begin{eqnarray}
r^{1}_{1-1}+\mathrm{Im}\{r^{2}_{1-1}\}&=&~~\, 0.000 \pm 0.006,\nonumber\\
 \mathrm{Re}\{r^{5}_{10}\}+  \mathrm{Im}\{r^{6}_{10}\}&=&~~\, 0.011 \pm
 0.003,\nonumber\\
 \mathrm{Im}\{r^{7}_{10}\} -\mathrm{Re}\{r^{8}_{10}\}&=&~~\, 0.009 \pm 0.031,\nonumber
\end{eqnarray}
where total uncertainties are quoted.
While the measurements of the first and the third relation in Eq.~(\ref{schc_relations}) are consistent with the expectation, a tension is observed for the
second relation, which may indicate a contribution
of single-helicity-flip amplitudes.
%The sensitivity of the relations in Eq.~(\ref{schc_relations}) to SCHC is low.
In the case of the first relation only the contributions from squared small double-helicity-flip amplitudes violate SCHC.
For the two other relations the contributions that violate SCHC are related to small terms corresponding to the interference of two single-helicity-flip amplitudes as well as the interference of the helicity-conserving amplitude $T_{00}$ and the double-helicity-flip amplitude $T_{1-1}$.

However, for the transitions $\gamma^*_T \to V^{}_L$ of class C the non-zero values of five unpolarised SDMEs indicate a clear SCHC violation.
%$r^5_{00}$
%and $\mathrm{Re}\{r^1_{10}\}$ show SCHC violation at the level of
%three standard deviations of the statistical uncertainty.
In the GK model
\cite{Goloskokov:2009},
these SDMEs are related to  the chiral-odd GPDs
$H_{\rm T}$ and ${\bar E}_{\rm T}$ coupled to the higher-twist wave function
of the meson.
%\textcolor{red}{Note: Check text of this paragraph vs. theory}
The kinematic dependences of these  SDMEs, as presented in
Section \ref{Re}, may
help to further constrain  the model.

 %}
 \subsection{Contribution of the helicity-flip NPE amplitudes}

 The contributions of non-zero helicity-single-flip and helicity-double-flip amplitudes to the cross section can be quantified by the ratios $\tau_{ij}$ of the helicity-flip amplitudes $T_{ij}$ to the square root of the sum of all amplitudes squared
 \begin{equation}
\tau_{ij} = \frac{|T_{ij}|}{\sqrt{\mathcal{N}}}.
\label{tau_ij}
\end{equation}
Here, the normalisation factor $\mathcal{N}$ is given by $\mathcal{N} = \mathcal{N}_T + \epsilon \mathcal{N}_L$
with
\begin{eqnarray}
\mathcal{N}_T &=&
\widetilde{\sum}(|T_{11}|^2+|T_{01}|^2+|T_{-11}|^2 \nonumber \\
& & \;\;\;+  |U_{11}|^2+|U_{01}|^2+|U_{-11}|^2),
\label{N_T}  \\
\mathcal{N}_L &=&
\widetilde{\sum}(|T_{00}|^2+2|T_{10}|^2+2|U_{10}|^2).
\label{N_L}
\end{eqnarray}

The ratios $\tau_{ij}$ can be expressed in terms of SDMEs as shown in Ref.~\cite{DC-24}.

\begin{figure*}[hbt!]\centering
\includegraphics[width=13cm]{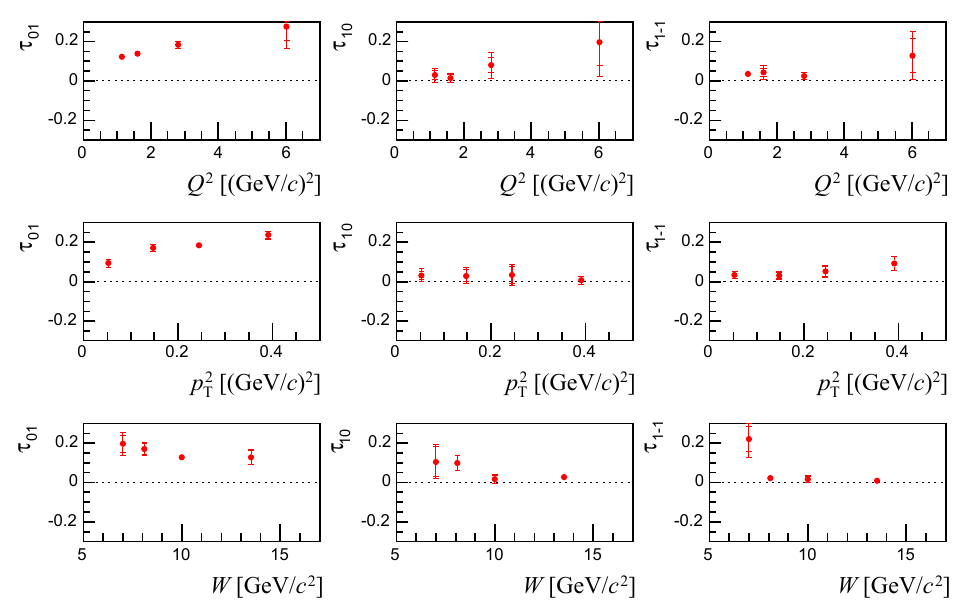}
\caption{
$Q^{2}$, $p^{2}_{\rm{T}}$ and $W$ dependences of $\tau_{01}$, $\tau_{10}$, $\tau_{1-1}$.
%The open symbols represent the values over the total kinematic region.
Inner error bars represent statistical uncertainties and
outer ones statistical and systematic uncertainties added in
quadrature.}
\label{t1t2t3}
\end{figure*}

For the amplitude $T_{01}$ describing the transition $\gamma^{*}_{T} \to \rho^{0}_{L}$ the quantity $\tau_{01}$ is given by
\begin{equation}
\tau_{01} \approx
  \sqrt{\epsilon}\frac{\sqrt{(r^{5}_{00})^{2} +(r^{8}_{00})^{2}}}{\sqrt{2r^{04}_{00}}}.
\label{tau01}
\end{equation}
The quantity $\tau_{10}$, which is related to the amplitude $T_{10}$ describing the transition $\gamma^{*}_{L} \to \rho^{0}_{T}$, is approximated by
\begin{equation}
\tau_{10}
 \approx \frac{\sqrt{(r^{5}_{11} + \mathrm{Im}\{ r^{6}_{1-1}\})^{2}+ (\mathrm{Im}\{r^{7}_{1-1}\} -r^{8}_{11})^{2}}}
{\sqrt{2(r^{1}_{1-1}-\mathrm{Im}\{r^{2}_{1-1}\})}}.
\label{tau10}
\end{equation}
For the helicity-double-flip amplitude $T_{1-1}$ describing the transition $\gamma^{*}_{-T} \to \rho^{0}_{T}$ the quantity $\tau_{1-1}$ is given by
\begin{equation}
\tau_{1-1}
 \approx \frac{\sqrt{(r^{1}_{11})^{2} + (\mathrm{Im}\{r^{3}_{1-1}\})^{2}}}
{\sqrt{r^{1}_{1-1}-\mathrm{Im}\{r^{2}_{1-1}\}}}.
\label{tau1m1}
\end{equation}
In Fig.~\ref{t1t2t3} the dependence of the quantities $\tau_{01}$, $\tau_{10}$ and $\tau_{1-1}$ on $Q^{2}$, $p^{2}_{\rm {T}}$ and $W$ is  presented.
For $\tau_{01}$, values significantly different from zero are observed, while for $\tau_{10}$ and $\tau_{1-1}$ they are much smaller. % The observation of values, which are significantly different from zero for $\tau_{01}$ and much smaller for $\tau_{10}$ and $\tau_{1-1}$.
This observation is consistent with the different degrees of SCHC violation seen for SDMEs in classes C, D and E.

A squared ratio $\tau_{ij}^2$ represents the fractional
contribution from amplitude $T_{ij}$ to the full cross section. Therefore the quantity
\begin{eqnarray}
\tau_{\rm NPE}^2 &=& (2\epsilon|T_{10}|^2 + |T_{01}|^2 +  |T_{1-1}|^2)/\mathcal{N}
\nonumber \\
&\approx& 2\epsilon\tau_{10}^2 + \tau_{01}^2 +  \tau_{1-1}^2
\label{tau2NPE}
\end{eqnarray}
represents the fractional contribution of helicity-flip
NPE amplitudes to the cross section.
%The COMPASS result for
The value of
$\tau_{\rm NPE}^2$ for the COMPASS entire kinematic range is small, equal to 0.023 $\pm$ 0.002 $\pm$ 0.004.
%{\color{red} ??? +/- ??? +/- ???}.
\subsection{UPE contribution in exclusive $\rho ^0$ meson production}
\label{UPE-contrib}

\begin{figure*}[hbt!]\centering
\includegraphics[width=13cm]{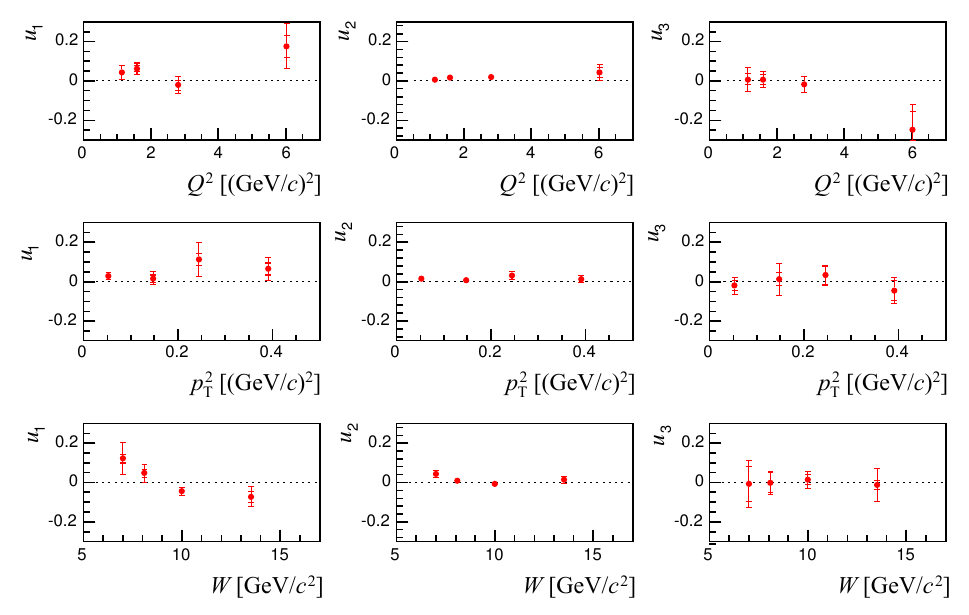}
\caption{
$Q^{2}$, $p^{2}_{\rm{T}}$, and $W$ dependences of $u_{1}$, $u_{2}$, $u_{3}$.
%The open symbols represent the values over the total kinematic region.
Inner error bars represent statistical uncertainties and
outer ones statistical and systematic uncertainties added in
quadrature.}
\label{depq2ptw}
\end{figure*}

By examining a linear combination of SDMEs
 %\magenta{[WDN 6/12 13h30: I propose to simply omit the text that follows up the the formula, as it's repeated anyway in a correct way just after.]} , which \red{'only' not precise because NPE enter in denominator of Eq. (30)} contain UPE amplitudes,
such as
%The existence of UPE transitions in
%exclusive   $\rho ^0$ production
% can  be tested by examining linear combination of SDMEs
 %\magenta{[WDN 6/12 13h30: I propose to simply omit the text that follows up the the formula, as it's repeated anyway in a correct way just after.]} , which \red{'only' not precise because NPE enter in denominator of Eq. (30)} contain UPE amplitudes,
%such as
\begin{equation}
u_1=1-r^{04}_{00}+2r^{04}_{1-1}-2r^{1}_{11}-2r^{1}_{1-1}\;,
\label{uu1}
\end{equation}
the presence of a UPE contribution can be tested.
The quantity $u_1$ is expressed in terms of helicity amplitudes  as
\begin{equation}
u_1=\widetilde{\sum}\frac{4\epsilon|U_{10}|^2+2|U_{11}+U_{-11}|^2}{\mathcal{N}},
\label{u1u}
\end{equation}
%Since the numerator depends only on UPE amplitudes,
thus a positive value of $u_1$ would indicate a non-zero contribution from UPE transitions.
For the entire kinematic
region of COMPASS $u_1$ is equal to 0.047 $\pm$ 0.010 $\pm$ 0.029, which indicates a small UPE contribution. Additional information on UPE amplitudes
can be obtained from the SDME combinations
\begin{equation}
u_2=r^{5}_{11}+r^{5}_{1-1},
\label{uu2}
\end{equation}
%and
\begin{equation}
u_3=r^{8}_{11}+r^{8}_{1-1},
\label{uu3}
\end{equation}
which in terms of
helicity amplitudes can be combined into
\begin{equation}
u_2 + \rm{i}u_3
=\sqrt2\widetilde{\sum}\frac{(U_{11}+U_{-11})U^{*}_{10}}{\mathcal{N}}.
\label{u2u3n}
\end{equation}
The value of $u_2 + \rm{i}u_3$ can vanish despite of the existence of UPE contributions.
For COMPASS $u_2$ = $-$0.008 $\pm$ 0.002 $\pm$ 0.013 and
$u_3$ = $-$0.010 $\pm$ 0.018 $\pm$ 0.037 are obtained, which are
consistent with zero at the present accuracy of the data. In  Fig.~\ref{depq2ptw} the dependence of the quantities  $u_{1}$, $u_{2}$ and $u_{3}$
 on $Q^{2}$, $p^{2}_{\rm {T}}$, and $W$ is  presented.
 The quantities $u_{1}$, $u_{2}$ and $u_{3}$ are small and
 compatible with zero within experimental uncertainties.
%The quantity
% $u_{1}$ tends to decrease with increasing $W$, which indicates that the UPE contribution becomes {\color{red} the conclusion on u1 unclear; two-sigma effect? Needed values of uncertainties}
 %smaller, while $u_{2}$, $u_{3}$ fluctuate around zero.

 The UPE fractional contribution to the cross section is given as
 \begin{eqnarray}
 \Delta_{\rm UPE} &=& (2\epsilon|U_{10}|^2 + |U_{01}|^2 +  |U_{1-1}|^2 + |U_{11}|^2)/\mathcal{N}
 \nonumber \\
 &\approx& u_1/2,
\label{tauUPE}
\end{eqnarray}
where the contributions of the amplitudes $U_{01}$ and $U_{1-1}$ was neglected for the approximate relation to $u_1$. % the contributions of amplitudes $U_{01}$ and $U_{1-1}$ has been neglected.
The value of $\Delta_{\rm UPE}(\rho^0)$ for the entire
kinematic range is 0.024 $\pm$ 0.005 $\pm$ 0.014.

Altogether, for exclusive $\rho ^0$ production at COMPASS the contribution of UPE is very small. This is in sharp contrast to the significant UPE contribution observed by COMPASS~\cite{COMPASS-omega} for exclusive $\omega$ production
in a similar kinematic range. There, this contribution is large over the entire kinematic range, $\Delta_{\rm UPE} (\omega)$ = 0.415 $\pm$ 0.037 $\pm$ 0.025. The UPE dominates $\omega$ production at small $W$ values and its contribution decreases with increasing $W$ without vanishing towards large $W$ values accessible in COMPASS. In the GK model, UPE is described by the GPDs $\widetilde{H}^{f}$ and $\widetilde{E}^{f}$ (non-pole), and by the pion-pole contribution treated as a one-boson exchange~\cite{GK:epjA-2014}. The large difference in size of the UPE contributions for $\omega$ and $\rho^0 $ production is mostly explained by the difference between  $\pi-\omega$ and $\pi-\rho^0$ transition form factors, with the former one being about three times larger than the latter~\cite{GK:epjA-2014}.

 %\red{ [AS] I do not think that we can conclude like following text; NPE %in fact grows that is consistent with growth of exclusive omega %production cross section with increasing W as observed at ZEUS; I %propose to stay with previous formulation of conclusions (starting from %"It is also consistent..."}
 %\magenta{A possible interpretation is that the NPE amplitude $T_{11}$
 %shows no strong $W$ dependence, while the UPE amplitude $U_{11}$,
 %which is mostly fed by pion-pole exchange, decreases but not vanishes %towards the high-$W$ limit of the kinematic range.}
%~~~~~~\\
\subsection{The NPE-to-UPE asymmetry of the transverse cross section for the transition $\gamma^*_T \rightarrow V^{}_T$}

Another observable that is sensitive to the relative contributions of UPE and
NPE amplitudes is the
NPE-to-UPE asymmetry of the transverse differential cross section for the transition $\gamma^*_T \rightarrow V^{}_T$. It is
defined~\cite{GK:epjA-2014} as
%as\footnote{In Ref.~\cite{HERMES:2014}
%a different
%definition of the asymmetry is used.}
\begin{eqnarray}
P &= &
\frac{d\sigma^N_T(\gamma^*_T \rightarrow V_T)
- d\sigma^U_T  (\gamma^*_T \rightarrow V_T)}
{d\sigma^N_T (\gamma^*_T \rightarrow V_T) +
d\sigma^U_T  (\gamma^*_T \rightarrow V_T)}
\nonumber \\
&\approx & \frac{2 r^1_{1-1}}{1-r^{04}_{00}-2 r^{04}_{1-1}},
%\\
%=\frac{2 r^1_{1-1}}{1-r^{04}_{00}-2 r^{04}_{1-1}},
\label{asymmGK}
\end{eqnarray}
where the superscript %s
$N$ and $U$
denotes the part of the cross section that is
fed by NPE and UPE transitions,
respectively.

\begin{figure*}
 %[hbtc!]
 \centering
 \includegraphics[width=13cm]{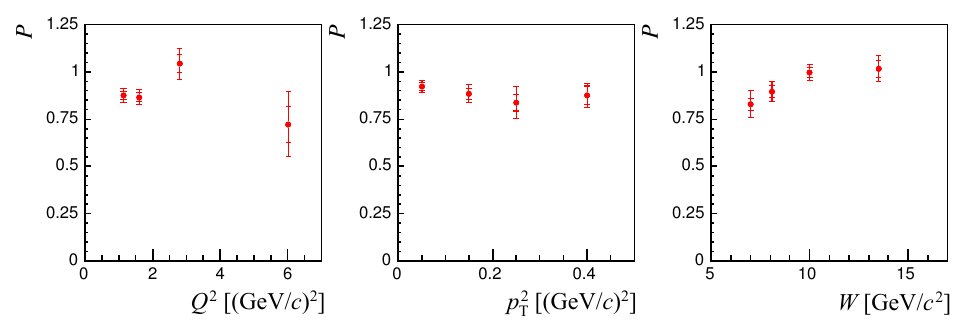}
 \caption{ $Q^{2}$, $p^{2}_{\rm{T}}$ and $W$ dependences of the NPE-to-UPE
 asymmetry of
 the transverse cross section for the transition $\gamma^*_T \rightarrow V_T$.
 %The open symbol represents the value over the total kinematic region.
 Inner error bars represent statistical uncertainties and outer
 ones statistical and systematic uncertainties added in
 quadrature. }
 \label{asyupenpenew}
 \end{figure*}

The value of $P$ obtained in the entire kinematic region is 0.887
$\pm$ 0.016 $\pm$ 0.029, which indicates that the NPE contribution dominates when averaged over the whole kinematic range of COMPASS.
The kinematic dependences of the asymmetry $P$ are shown in Fig.~\ref{asyupenpenew}. A small UPE contribution
is observed only at small values of $W$ and it becomes compatible with zero at larger $W$.
No significant $Q^2$ and $p_{\rm{T}}^2$ dependences
of the asymmetry are observed.

 The COMPASS results for exclusive $\omega$ production~\cite{COMPASS-omega} on the asymmetry $P$ and its kinematic dependences exhibit a different behaviour.
 For the whole kinematic region the value is compatible with zero, $P (\omega) = -$0.007 $\pm$ 0.076 $\pm$
 0.125, which indicates that the UPE and NPE contributions averaged over the whole kinematic range are of similar size.
 The UPE  contribution dominates at small values of $W$ and
 decreases with increasing $W$. At large values of $W$ the NPE
 contribution for $\omega$ production becomes dominant, while a non-negligible UPE
 contribution still remains.

A semi-quantitative explanation of the difference between the values of
asymmetry $P$  measured
 for $\rho ^0$ and $\omega$ production is possible by considering only the dominant contributions to the UPE
and NPE cross sections for each process.
In the framework of GK model such a contribution to the UPE cross sections is due to pion-pole exchange.
Due to the
 difference between  $\pi-\rho^0$ and $\pi-\omega$ transition form factors, which was mentioned in Sec.~\ref{UPE-contrib}, the UPE cross section for $\rho ^0$ is about 9 times
 smaller than that for $\omega$ production. For the NPE cross section, the dominant contribution
 is related to the gluon and sea-quark GPDs $H$ and their relative contributions given in
Ref.~\cite{GK2007}
%[GK EPJC50 (2007) 829-842 formula 44]
%or [Markus Diehl Habilitation]
imply that the cross section of exclusive  $\rho^0$ production is about 9 times larger than that for $\omega$. Taken together, UPE contributions are close to zero for $\rho^0$ and the $P$ value approaches unity.

%\sout{For $\rho ^0$ production this contribution is about 9 times larger than for $\omega$, which results from the SU(3) flavour symmetry that holds approximately. Hence, using the above approximations one obtains $P(\rho ^0) \approx$ 1 and $P(\omega) \approx$ 0~\cite{GK:epjA-2014}.}

\subsection{Longitudinal-to-transverse cross-section ratio}

\begin{figure*}
\centering
\includegraphics[width=13cm]{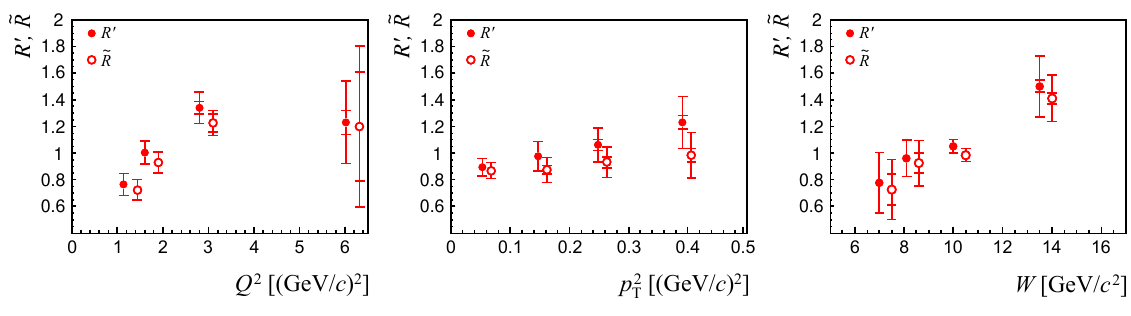}
\caption{ $Q^{2}$, $p^{2}_{\rm{T}}$ and $W$ dependences of two estimates, $R'$ and $\widetilde{R}$, of the longitudinal-to-transverse cross-section ratio $R$.
%The open symbol represents the value %over
%obtained for the total
%kinematic region.
Inner error bars represent statistical uncertainties and
outer ones statistical and systematic uncertainties added in
quadrature. For better visibility the data points for $\widetilde{R}$
are presented with a small horizontal off-set.
%Note that the additional positive uni-directional systematic uncertainty due to the approximation
%$R \approx R'$ is not shown here, see text for details.}
}
\label{depq2ptwa}
\end{figure*}

The longitudinal-to-transverse virtual-photon
cross-section ratio
\begin{equation}
R= \frac{\sigma_{L}(\gamma^{*}_{L} \to V)}{\sigma_{T}(\gamma^{*}_{T} \to
V)},
\label{trueR}
\end{equation}
is one of the most important observables in the study of light VM production since
it is sensitive to the interaction dynamics.
In order to evaluate $R$ the quantity $R'$ is commonly used:
\begin{equation}
R'  =
\frac{1}{\epsilon}\frac{r^{04}_{00}}{1-r^{04}_{00}}.
\label{sigto}
\end{equation}
Using expressions defining $r^{04}_{00}$ and $1-r^{04}_{00}$
in terms of helicity amplitudes~\cite{Schill,DC-24},
%one
%obtains
%\begin{eqnarray}
%R'=
%\frac{1}{\epsilon}
%\dfrac{
%\widetilde{\sum}(\epsilon|T_{00}|^2+|T_{01}|^2+|U_{01}|^2)}
%/
% ~~~~~~~~~~~~~~~~~~~~~~~~~~
% \nonumber \\
%{
%\widetilde{\sum}\{|T_{11}|^2+|U_{11}|^2+|T_{1-1}|^2+|U_{1-1}|^2
% ~~~~~~~~~~~~~
% \nonumber \\
%+2\epsilon (|T_{10}|^2+|U_{10}|^2)\}}.
%~~~~~~~~~~~~
%\label{Rprimhelamp}
%\end{eqnarray}
the quantity $R'$ may be interpreted~\cite{COMPASS-omega} as
longitudinal-to-transverse
ratio of ``effective''
cross sections for the production of vector mesons that are polarised
longitudinally or
transversely irrespective of the virtual-photon polarisation. In case of
SCHC, $R'$ is equal to $R$.
In spite of the observed
%clear
violation of SCHC, the % at COMPASS, we use the
approximate relation
$R \approx R'$ was used in most of the previous measurements. % is used.
%The accuracy of this approximation
%is discussed in the following.
%~\cite{lastGK}
%using
%the GK model~\cite{GK:epjC-2014,GK:epjA-2014} and
%the resulting uni-directional systematic uncertainty is
%found to be about $+15\%$ on average, while its magnitude ranges
%between 3\% and 47\% with increasing $W$ and between 6\% and 28\% with increasing $p_T^2$.
For the entire kinematic region, the ratio $R'$ is found to
be 0.980 $\pm$ 0.014 $\pm$ 0.089.
%Here, the third uncertainty is
%the systematic one
%due to the approximation  $R \approx R'$.
The kinematic dependences of $R'$ are shown in Fig.~\ref{depq2ptwa}.

In order to evaluate the effect of helicity-changing amplitudes on the estimate of the longitudinal-to-transverse cross-section
ratio one can use a quantity $\widetilde{R}$ that is defined by following
relation:~\cite{DC-24}
\begin{equation}
\widetilde{R} = R' - \frac{\eta (1+\epsilon R')}{\epsilon (1+\eta)},
\label{R-corr}
\end{equation}
where
\begin{equation}
\eta = \frac{(1+\epsilon R')}{N}\widetilde{\sum}
\{|T_{01}|^2+|U_{01}|^2-2\epsilon(|T_{10}|^2+|U_{10}|^2)\} .
\label{eta-def}
\end{equation}
The quantity $\eta$ can be approximately estimated as
\begin{equation}
\eta \approx (1+\epsilon R') (\tau^2_{01} - 2\epsilon \tau^2_{10}) .
\label{eta-approx}
\end{equation}
Here $\tau_{01}$ and $\tau_{10}$ are evaluated using
Eqs.~(\ref{tau01}) and~(\ref{tau10}), and
the small
contributions of the helicity-flip UPE amplitudes $U_{01}$ and $U_{10}$
in Eq.~(\ref{eta-def}) are neglected.

For the entire kinematic
region the ratio $\widetilde{R}$ is found to be 0.907 $\pm$ 0.014
$\pm$ 0.074. The values of
$\widetilde{R}$ as functions of kinematic variables are shown in Fig.~\ref{depq2ptwa} for comparison with $R'$.
The values of $\widetilde{R}$ and $R'$ with their statistical and total uncertainties are also given in Table~\ref{TableforRs} for the kinematic bins and for the entire kinematic region.
\begin{table*}
\begin{center}
\caption{ \label{TableforRs}
Values of the ratios $\widetilde {R}$ and $R'$ in kinematic bins and for the entire kinematic region. The first uncertainty values correspond to the statistical errors, and
the second ones to the statistical and systematic
uncertainties added in quadrature.}

\renewcommand{\arraystretch}{1.2}
\setlength{\tabcolsep}{10pt}

\begin{tabular}{|r@{\,}l|c|c|}   %c@{\,}|}
\hline
\multicolumn{2}{|c|}{ }&\multicolumn{1}{c|}{$ \widetilde{R} $} &\multicolumn{1}{c|}{$ R'$}\\%&\multicolumn{1}{c|}{$ \widetilde{R} - R'$} \\
\hline
\multicolumn{2}{|c}{ $\langle Q^2 \rangle$}&\multicolumn{2}{c|}{} \\
\hline
 1.14& (GeV/$c$)$^2$ & 0.724 $\pm$ 0.019 $\pm$ 0.076 &
 0.765 $\pm$ 0.018 $\pm$ 0.082 \\%& $-0.041$ \\
 1.60& (GeV/$c$)$^2$ & 0.930 $\pm$ 0.021 $\pm$ 0.078 &
 1.003 $\pm$ 0.022 $\pm$ 0.086 \\%& $-0.073$ \\
 2.80& (GeV/$c$)$^2$ & 1.227 $\pm$ 0.067 $\pm$ 0.094 &
 1.340 $\pm$ 0.046 $\pm$ 0.116 \\%& $-0.113$ \\
 6.02& (GeV/$c$)$^2$ & 1.200 $\pm$ 0.409 $\pm$ 0.603 &
 1.230 $\pm$ 0.091 $\pm$ 0.310 \\%& $-0.023$ \\
\hline
\multicolumn{2}{|c}{ $\langle p_{\rm T}^2\rangle$}&\multicolumn{2}{c|}{} \\
\hline
 0.053& (GeV/$c$)$^2$ & 0.868 $\pm$ 0.020 $\pm$ 0.061&
 0.894 $\pm$ 0.018 $\pm$ 0.067 \\%& $-0.056$ \\
 0.147& (GeV/$c$)$^2$ & 0.874 $\pm$ 0.030 $\pm$ 0.094&
 0.977 $\pm$ 0.028 $\pm$ 0.110 \\%& $-0.103$ \\
 0.248& (GeV/$c$)$^2$ & 0.932 $\pm$ 0.042 $\pm$ 0.114&
 1.061 $\pm$ 0.041 $\pm$ 0.127 \\%& $-0.129$ \\
 0.391& (GeV/$c$)$^2$ & 0.984 $\pm$ 0.051 $\pm$ 0.173&                  1.230 $\pm$ 0.052 $\pm$ 0.195 \\%& $-0.246$ \\
\hline
\multicolumn{2}{|c}{ $\langle W \rangle$}&\multicolumn{2}{c|}{} \\
\hline
 7.0&  GeV/$c^2$ & 0.728 $\pm$ 0.118 $\pm$ 0.225&
 0.779 $\pm$ 0.026 $\pm$ 0.227 \\%& $-0.051$ \\
 8.1&  GeV/$c^2$ & 0.924 $\pm$ 0.075 $\pm$ 0.171&
 0.961 $\pm$ 0.027 $\pm$ 0.138 \\%& $-0.037$ \\
 10.0& GeV/$c^2$ & 0.986 $\pm$ 0.025 $\pm$ 0.049&
 1.051 $\pm$ 0.025 $\pm$ 0.049 \\%& $-0.065$ \\
 13.5& GeV/$c^2$ & 1.411 $\pm$ 0.043 $\pm$ 0.175&
 1.501 $\pm$ 0.045 $\pm$ 0.232 \\%& $-0.090$ \\
\hline
\multicolumn{2}{|c|}{Entire range} & 0.907 $\pm$ 0.014 $\pm$ 0.076 &
 0.980 $\pm$ 0.014 $\pm$ 0.090 \\%& $-0.073$ \\
 \hline
\end{tabular}

\end{center}
\end{table*}
%For the majority of $\widetilde{R}$ values the corresponding statistical errors are larger than these for $R'$,
%and for three kinematic bins they are larger by a factor of 3 - 5. This difference is related to different numbers of SDMEs used for estimates of $R'$ and $\widetilde{R}$. While for the former one only the $r^{04}_{00}$ SDME is
%needed, for $\widetilde{R}$ eight other SDMES are used in addition,
%in particular 3 ``polarised'' SDMEs with large statistical errors.
%{\color{red} However,
The total uncertainties for both observables are
dominated by the systematic ones, which are similar for both cases. Using $R'$ as an estimate of $R$ introduces an additional uni-directional systematic uncertainty on $R$, which is due to the assumption of SCHC. It is estimated from  $\widetilde{R} - R'$ and found to be about $-$0.07 on average. % given in the last column of the table.
Thus for an estimate of the ratio $R$ (defined by Eq.~(\ref{trueR})) it is preferable to use the quantity $\widetilde{R}$, \textit{i.e.} $R \approx \widetilde{R}$, while $R'$ values may be used when comparing to earlier measurements that assumed SCHC.
%{\sout{This uncertainty is unidirectional and is equal to about $-$0.08 on average.}}

A strong increase of the $\sigma _{L}/\sigma _{T}$ ratio with increasing $Q^{2}$ is observed, while
the  $p^2_{\rm T}$ and $W$ dependences are weaker (see Fig.~\ref{depq2ptwa}). The $Q^2$ dependence of $R$ indicates that $\sigma_{L}$ becomes dominant at $Q^2$ larger than about 2 $({\rm GeV}/c)^2$.
%~~~~~~\\
 %\begin{figure*}
 %\centering
 %\includegraphics[width=20cm]{Fig/asyupenpeNew_pub.pdf}
 %\caption{
 % {\color{red} New Caption to be written by Andrzej }}
 %\label{asyupenpeNew}
 %\end{figure*}

\subsection{Phase difference between amplitudes $T_{11}$ and $T_{00}$}
\label{phase}

The phase difference between the amplitudes $T_{11}$ and $T_{00}$ can be evaluated as follows~\cite{DC-24}:
\begin{equation}
 \mathrm{cos} \delta
 = \frac{2\sqrt{\epsilon}(\mathrm{Re}\{r^{5}_{10}\} -\mathrm{Im}\{r^{6}_{10}\})}
{\sqrt{r^{04}_{00}(1- r^{04}_{00}+r^{1}_{1-1} -\mathrm{Im}\{r^{2}_{1-1})\}}}.
\label{cos-del}
\end{equation}
The result is $|\delta|$ = 19.6 $\pm$ 0.9 $\pm$ 3.9 degrees. The sign of $\delta$ can
be obtained from $\mathrm{sin} \delta$~\cite{DC-24} that depends on the polarised SDMEs:
\begin{equation}
 \mathrm{sin} \delta
 = \frac{2\sqrt{\epsilon}(\mathrm{Re}\{r^{8}_{10}\} +\mathrm{Im}\{r^{7}_{10}\})}
{\sqrt{r^{04}_{00}(1- r^{04}_{00}+r^{1}_{1-1} -\mathrm{Im}\{r^{2}_{1-1})\}}}.
\label{sin-del}
\end{equation}
Using Eq.~(\ref{sin-del}) one determines
%{\color{blue} one determines with a high significance that
%$\delta$ is positive.}
that the sign of
$\delta$ is positive and $\delta =
12.9 \pm 2.1 \pm 0.8$ degrees. Both results on $\delta$ are compatible within large total uncertainties.
For comparison to other experiments, which used $\cos \delta$ to estimate the phase difference, we use the positive value of $\delta =
19.6 \pm 0.9 \pm 3.9$ degrees.

\section{Comparison to other experiments}
\label{other-exp}

\begin{figure*}[hbt!]\centering
\includegraphics[width=12cm]{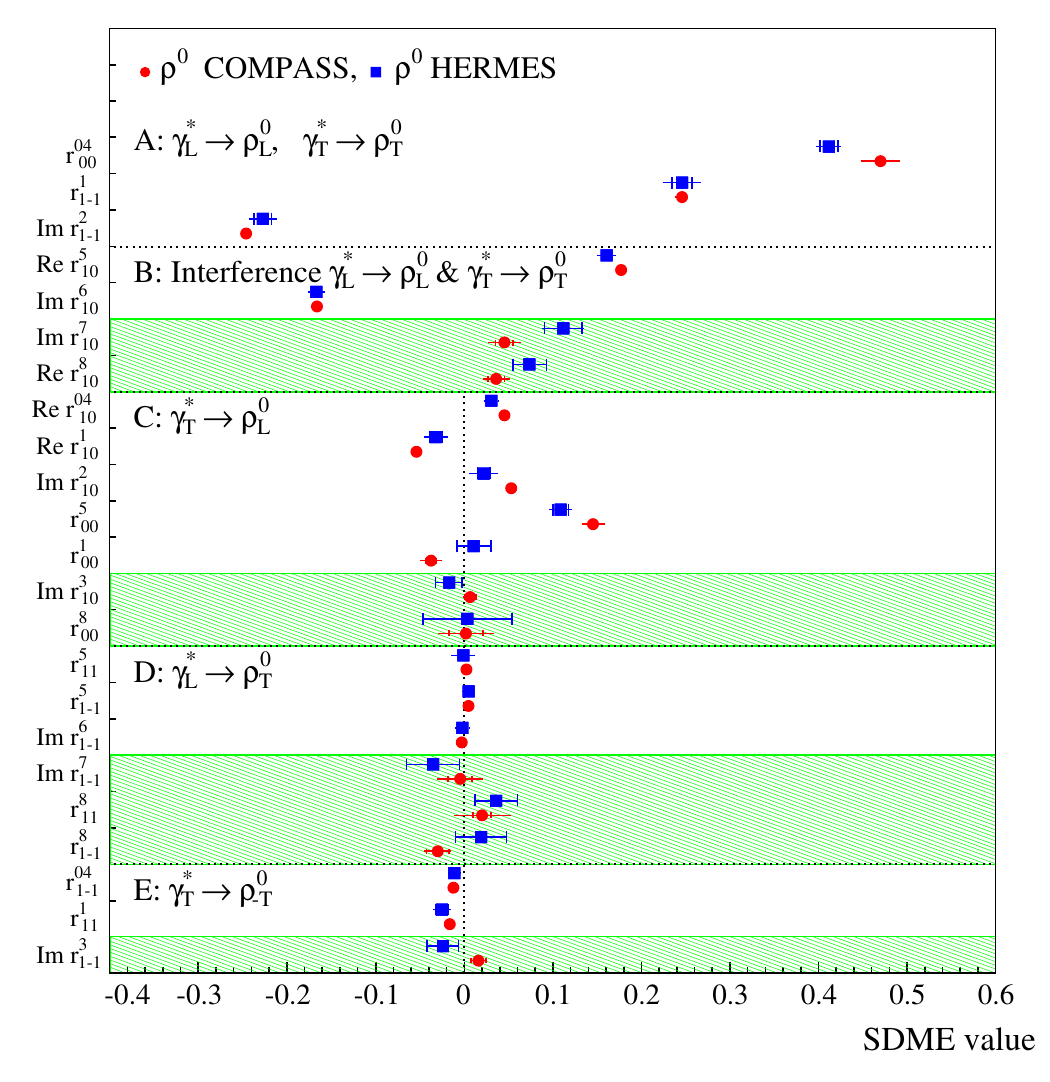}
\caption{
Comparison of the 23 SDMEs for exclusive $\rho ^0$ leptoproduction on the proton extracted in the entire kinematic
regions of the HERMES and COMPASS experiments. For HERMES the average kinematic values are $ \langle Q^2 \rangle = 1.96$~(GeV/$c$)$^2$, $\langle W\rangle =4.8$~GeV/$c^2$, $ \langle |t'| \rangle = 0.13$,
while those for COMPASS are
$ \langle Q^2 \rangle = 2.40$~(GeV/$c$)$^2$, $\langle W\rangle =9.9$~GeV/$c^2$, $ \langle p^{2}_{\rm T}\rangle = 0.18$
(GeV/$c$)$^2$.
Inner error bars represent statistical uncertainties and
outer ones statistical and systematic uncertainties added in
quadrature. Unpolarised (polarised) SDMEs are displayed in %the
unshaded (shaded) areas.
}
\label{results_H+C}
\end{figure*}

We compare the COMPASS results on SDMEs and related observables to those from the experiments that measured extensive sets of SDMEs for exclusive $\rho ^0$ electroproduction: the HERMES~\cite{DC-24}, H1~\cite{h1-2010,h1-2000} and ZEUS~\cite{zeus-2000,zeus-2007} experiments.
Only the HERMES experiment measured the complete set of 15 unpolarised and 8 polarised SDMEs, while H1 and ZEUS measured 15 unpolarised SDMEs. Compilations of selected SDMEs-related observables, including other experiments, can be found ${\it{e.g.}}$ in Refs.~\cite{DC-24,h1-2010}.

The complete sets of SDMEs from HERMES and COMPASS for their entire kinematic ranges are compared in Fig.~\ref{results_H+C}. The kinematic range for HERMES is  1.0~(${\rm GeV}/c)^2 < Q^{2} < 7.0$ $({\rm GeV}/c)^2$, 3.0~GeV/$c^2< W < 6.3 $~GeV/$c^2$, $|t'| < 0.4$~(GeV/$c$)$^2$, while that for COMPASS 1.0~(${\rm GeV}/c)^2 < Q^{2} < 10.0$ $({\rm GeV}/c)^2$, 5.0~GeV/$c^2< W <17.0$~GeV/$c^2$, 0.01~(GeV/$c$)$^2$ $<p^{2}_{\rm T} < 0.5$~(GeV/$c$)$^2$.
The ranges of $Q^2$ and the momentum transferred to the recoil proton are similar, but the $W$ ranges overlap only marginally and the COMPASS range extends significantly towards large $W$ values. In consequence, the contribution of gluons and sea quarks involved in exclusive meson production is higher by a factor of about 2.5 in COMPASS, while  the remaining contribution (from valence quarks and interference terms) is a little smaller than in HERMES (see ${\it{e.g.}}$ Ref.~\cite{Favart}).
%{\color{red} Check with Peter Kroll}.
Significant differences are observed for the SDME $r_{00}^{04}$, which is proportional to the
square of the leading helicity-conserving amplitude $T_{00}$, and for the unpolarised SDMEs from class C.

A more detailed comparison of selected observables in similar kinematic ranges for both experiments is presented in Table~\ref{comparison_H+C}. The published HERMES results for the entire kinematic range ($\langle W\rangle =4.8$~GeV/$c^2$) are compared to the COMPASS results for the lowest $W$ bin ($\langle W\rangle =7.0$~GeV/$c^2$). The quoted uncertainties are the total ones.
For most of the observables the results from both experiments are compatible within one standard deviation, except $\tau_{01}$ and $\tau_{1-1}$\,, for which the agreement is within two standard deviations.

\begin{table*}[hbt]
\renewcommand{\arraystretch}{1.2}
\begin{center}
\caption{\label{comparison_H+C} Comparison of selected observables measured by HERMES and COMPASS in similar kinematic regions. The HERMES results for the proton target~\cite{DC-24} are integrated over the entire kinematic region. The COMPASS results are given for $5.0 < W <  7.3~\mathrm{ GeV}/c^2$. The total uncertainties are given.}
%\begin{tabular}{|c|r@{$\,\pm\,$}l@{$\,\pm\,$}l|r@{$\,\pm\,$}l@{$\,\pm\,$}l|}
%\begin{tabular}{|c|r@{$\,\pm\,$}l|r@{$\,\pm\,$}@{$\,\pm\,$}l|}
\begin{tabular}{|c|>{\hskip10pt}r@{$\,\pm\,$}l|>{\hskip 10pt}r@{$\,\pm\,$}l|}
\hline
 Observable &\multicolumn{2}{c|}{HERMES}
         &\multicolumn{2}{c|}{COMPASS }\\
         &\multicolumn{2}{c|}{$\langle W\rangle$ = 4.8 GeV/$c^2$}
         &\multicolumn{2}{c|}{$\langle W\rangle$ = 7.0 GeV/$c^2$}\\
         &\multicolumn{2}{c|}{$\langle Q^2 \rangle$ = 1.98 (GeV/$c$)$^2$}
         &\multicolumn{2}{c|}{$\langle Q^2 \rangle$ = 2.90 (GeV/$c$)$^2$}\\
\hline
$r^{04}_{00}$    &$  0.412 $&$ 0.014 $&$ 0.435 $&$ 0.064$  \\
\hline
$\tau_{01}$      &$  0.114 $&$ 0.012 $&$ 0.196 $&$ 0.059$ \\
$\tau_{10}$      &$  0.075 $&$ 0.030 $&$ 0.105 $&$ 0.085$ \\
$\tau_{1-1}$     &$  0.051 $&$ 0.031 $&$ 0.222 $&$ 0.092$ \\
\hline
$u_1$            &$  0.125 $&$ 0.054 $&$ 0.122 $&$ 0.085$ \\
$u_2$            &$-$$0.011 $&$ 0.013 $&$ 0.022 $&$ 0.063$ \\
$u_3$            &$  0.055 $&$ 0.045 $&$-$$0.008 $&$ 0.116$ \\
\hline
\end{tabular}
\end{center}
\end{table*}

\begin{figure*}[hbt!]\centering
\includegraphics[width=10cm]{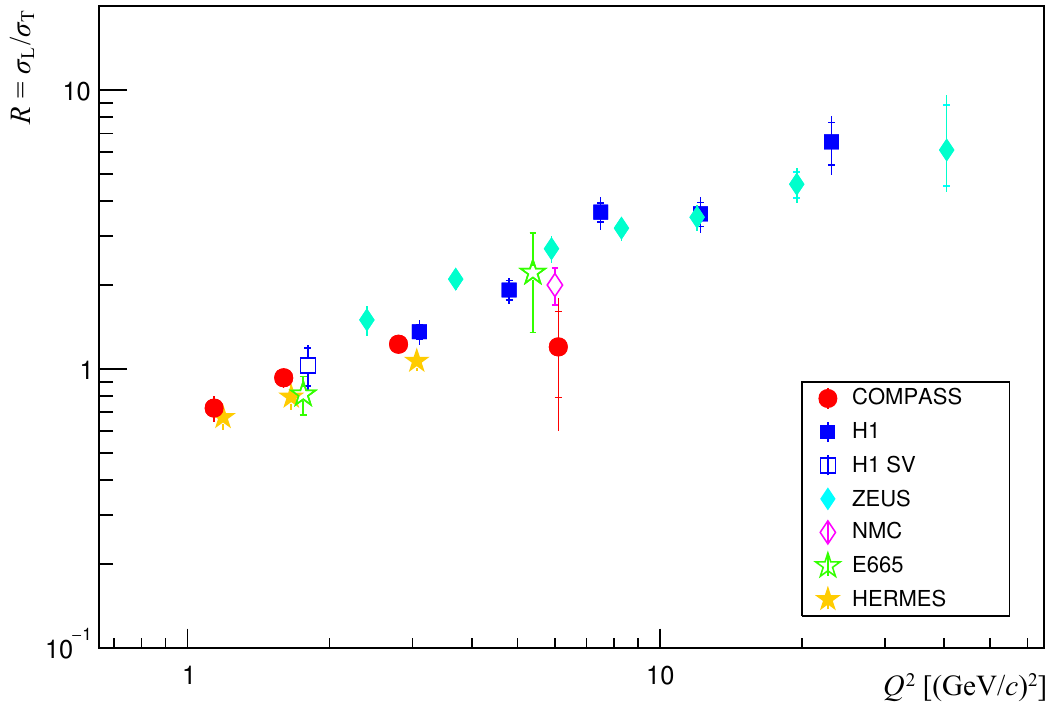}
\caption{The ratio $R = \sigma_L/\sigma_T$ as a function of $Q^2$. For comparison measurements of exclusive $\rho ^0$ leptoproduction by fixed target experiments (HERMES~\cite{DC-24}, NMC~\cite{NMC}, E665~\cite{E665})
and by HERA collider experiments (ZEUS~\cite{zeus-2007}, H1~\cite{h1-2010},
H1 SV~\cite{h1-2000}) are also shown.
}
\label{RvsQ2}
\end{figure*}

The comparison of the $Q^2$ dependence of $R$ between COMPASS and HERMES is not straightforward because the results are integrated over different $W$ ranges for each experiment. Despite the moderate $W$ dependence of $R$ observed by COMPASS ({\it{cf}} Fig.~\ref{depq2ptwa}), the estimates of $R$ from both experiments are compatible within experimental uncertainties as shown in Fig.~\ref{RvsQ2}.

The measurements of SDMEs for exclusive $\rho^0$ electroproduction by the ZEUS and H1 experiments were obtained in wide ranges of $Q^2$ and $W$ at the highest available energies. Here, for comparison with COMPASS we use the ZEUS and H1 results from Refs.~\cite{zeus-2000, h1-2000}.
The covered kinematic range for the DIS sample presented by the ZEUS experiment is 3~(${\rm GeV}/c)^2 < Q^{2} < 30$ $({\rm GeV}/c)^2$, 40~GeV/$c^2< W < 120$~GeV/$c^2$ and  $|t| < 0.6$~(GeV/$c$)$^2$, while for H1 it is
1~(${\rm GeV}/c)^2 < Q^{2} < 60$ $({\rm GeV}/c)^2$, 30~GeV/$c^2< W < 140$~GeV/$c^2$ and $|t| < 0.5$~(GeV/$c$)$^2$. In this kinematic range the value of the virtual-photon polarisation parameter~$\epsilon$
is close to 1 and the angular distribution for vector meson production and decay has a limited sensitivity to the
polarised SDMEs ({\it cf.} Eq.~\ref{eqang3}). Thus the HERA experiments could measure only the 15 unpolarised SDMEs.

Deviations from zero
are observed for five unpolarised SDMEs from classes A and B, which depend on
the helicity-conserving amplitudes $T_{00}$ and $T_{11}$. All other SDMEs are compatible with zero except $r^5_{00}$, which indicates the violation of SCHC for
$\gamma^*_T \to \rho ^0_L$ transitions. In order to quantify the size of SCHC violation, the ratios
 \begin{equation}
\tilde{\tau}_{ij} = \frac{|T_{ij}|}{\sqrt{|T_{00}|^2 + |T_{11}|^2}}
\label{tilde_tau_ij}
\end{equation}
were evaluated.
The approximate expressions for $\tilde{\tau}_{ij}$ are given in  Ref.~\cite{zeus-2000}.
In contrast to Eqs.~(\ref{tau01}, \ref{tau10}, \ref{tau1m1}) those expressions rely on the assumptions of zero phase difference between the considered amplitude $T_{ij}$ and the corresponding dominant amplitude ($T_{00}$ or $T_{11}$), neglecting UPE contributions and assuming $\epsilon = 1$.
For the kinematic ranges of the HERA experiments the values of
$\tilde{\tau}_{01}$ are equal to $0.079 \pm 0.026$ for ZEUS
and $0.08 \pm 0.03$ for H1, while the values of $\tilde{\tau}_{10}$ and $\tilde{\tau}_{1-1}$ are
compatible with zero within experimental uncertainties.
These results indicate that the helicity-flip amplitude $T_{01}$
does not vanish even at the highest available energies.
The comparison to the COMPASS result for
the entire kinematic region, $\tau_{01} = 0.143 \pm 0.011$, indicates that the relative contribution of the amplitude $T_{01}$ to the cross section becomes smaller for HERA kinematics.

Using Eqs.~(\ref{u1u}, \ref{uu2}) and the published ZEUS and H1
results on SDMEs for the entire kinematic range of each experiment~\cite{zeus-2000, h1-2000},
one obtains the values of the $u_1$ and $u_2$ observables ($u_3$ is not available at HERA), which are related to the UPE contributions.
The values of $u_1$ are equal to $0.091 \pm  0.078$ for ZEUS
and $0.058 \pm  0.125$ for H1.
The quoted uncertainties correspond to the quadratic sum of statistical and systematic uncertainties of individual SDMEs in Eq.~(\ref{u1u}) without taking into account the covariance matrix.
Both values are consistent with zero within less than 1.5 standard deviations, similar to the COMPASS result (see Sec.~\ref{UPE-contrib}).
The HERMES result for $u_1$ indicates a slightly larger UPE contribution at small $W$ values within 2.5 standard deviations from zero.
The values of $u_2$ are very small (0.015 $\pm$ 0.016 for ZEUS and 0.004 $\pm$ 0.022 for H1) and compatible with zero. In addition, the H1 measurements of the NPE-to-UPE asymmetry $P$~\cite{h1-2010} as a function of $Q^2$ and $|t|$ are compatible with unity, which supports NPE for transversely
polarised virtual-photons.
The HERA results on both $u_1$ and $P$ are consistent with the dominance of two-gluon exchange at high energy, which implies NPE.
%{\color{red} AS comment - not sure if the following should stay. -> H1 also reported on the kinematic dependences of the ratios of $T_{11}$ and the helicity-flip amplitudes $T_{01}$, $T_{10}$ and $T_{1-1}$ with respect to the dominant amplitude $T_{00}$. However, the ratios were estimated using a rather strong assumption that all  amplitudes are imaginary and cannot be directly compared to the ratios $\tau_{ij}$ from HERMES and COMPASS.}

The ZEUS and H1 results obtained from the large data sets of the 1996 -- 2000 data-taking period were published in Refs.~\cite{zeus-2007, h1-2010}, in which the $Q^2$, $W$, $|t|$ and $M_{\pi \pi}$ dependences of the cross section and SDMEs are discussed.
The detailed discussion of SDME-related quantities focuses mainly on $r_{00}^{04}$ and $R = \sigma_L/\sigma_T$.
The values of both quantities increase with increasing $Q^2$ in the whole covered range up to $Q^2 \approx 40~({\rm GeV}/c)^2$. The strong increase at small $Q^2$  becomes milder at large $Q^2$. At $Q^2$ values larger than $\approx 10~({\rm GeV}/c)^2$ the $r_{00}^{04}$ values are larger than 0.75, which indicates the predominant contribution to the cross section from longitudinal virtual photons. Within experimental uncertainties no $W$ dependence of $r_{00}^{04}$ and $R$ is observed by the two experiments, and in the case of ZEUS no $|t|$ dependence is seen.

In Fig.~\ref{RvsQ2} the COMPASS results on the $Q^2$ dependence of $R$ are compared
to the previous experiments using results with
$Q^{2} > 1.0$~(GeV/$c$)$^2$ and with moderate to large values of $W$.
The HERMES and COMPASS results are corrected for contributions of the spin-flip amplitudes
$T_{01}$ and $T_{10}$. For those from H1 the contribution of $T_{01}$ is taken into account,
whereas the SCHC approximation is used for the other data. Despite small differences due to different treatments of
small contributions of spin-flip amplitudes, and also due to a possible weak $W$ dependence,
all the results consistently show a main characteristic feature, {\it{i.e.}} the fast increase of $R$ as a function of $Q^2$ within the wide energy
range, from the fixed target experiments to the HERA collider measurements.

In leading-order pQCD and for $t = 0$
the ratio $R$ is predicted to be $R=Q^2/M_V^2$~\cite{Brodsky}, where
$M_V$ is the mass of the produced vector meson. The experimental data on $R$ for exclusive $\rho ^0$, $\phi$ and $J/\psi$ production confirm the scaling with $M_V$, but they lie systematically below this prediction
(see, e.g., Fig.~38 from Ref.~\cite{h1-2010}).
Deviations from this dependence, which become more pronounced as $Q^2$ increases, are due to effects of QCD evolution and quark transverse momentum~\cite{FKS,GK2007}.

In the framework of the colour dipole model, different transverse sizes are predicted for
virtual $q\bar{q}$ pair fluctuations originating from longitudinally and transversely polarised virtual photons, which leads to
different kinematic dependences of $\sigma_{L}$, $\sigma_{T}$ and $R$.
The transverse size of these colour dipoles is on average smaller for longitudinal photons than for transverse ones. This results in a more shallow $t$ (or $p_{\rm T}^2$) dependence of the cross section for the longitudinal photons.
In the unseparated cross section this effect
leads to a decrease of the value of the $t$-slope parameter with increasing $Q^2$ (see {\it e.g.} Refs.~\cite{h1-2010,zeus-2007}).

\section{Summary}
Using exclusive  $\rho ^0$ meson muoproduction on the proton, we have
measured 23
SDMEs at
the average COMPASS kinematics,
$\langle Q^{2} \rangle= 2.4~(\mathrm{GeV}/c)^2$, $\langle W \rangle = 9.9~\mathrm{GeV}/c^2$
and $\langle p^{2}_{\rm T} \rangle = 0.18~(\mathrm{GeV}/c)^2$.
The SDMEs are extracted in the kinematic region
 $1.0~({\rm GeV}/c)^2 < Q^{2} < 10.0$ $({\rm GeV}/c)^2$, $5.0~{\rm GeV}/c^2< W <17.0~{\rm GeV}/c^2$ and $0.01~({\rm GeV}/c)^2$, $<p^{2}_{\rm T} < 0.5~({\rm GeV}/c)^2$,
which allows us to study %ed studies of
their $Q^2$, $p^2_{\rm T}$ and $W$ dependences.

Several SDMEs that are depending on amplitudes %for
describing %transitions
$\gamma^*_T \to \rho ^0_L$ transitions
indicate a considerable violation of the SCHC hypothesis. These SDMEs are expected to be sensitive to the chiral-odd GPDs
$H_{\rm T}$ and ${\bar E}_{\rm T}$, which are coupled to the higher-twist wave function
of the meson. A particularly prominent effect is observed for
the SDME $r^5_{00}$, which strongly increases with increasing $Q^2$ and $p^2_{\rm T}$.

Using %the
specific observables that are constructed to be sensitive to the relative contributions from transitions with unnatural-parity exchanges, such as $u_1$ and the NPE-to-UPE asymmetry for the transverse cross section, we
observe a dominance of NPE exchanges. The UPE contribution is
very small and compatible with zero within experimental uncertainties.

The COMPASS results presented in this paper are obtained in the kinematic range that partially overlaps with the kinematic range of HERMES
experimental data on SDMEs for exclusive $\rho ^0$ electroproduction, but extends considerably
towards higher $W$ values. In the overlap region the results from both experiments are
compatible. Important data on
the behaviour of helicity amplitudes at very large values of $Q^2$ and $W$ are provided by ZEUS and H1.
Characteristic features are the dominant contribution of the amplitude describing the transition
$\gamma^*_L \to \rho ^0_L$ , which increases with increasing $Q^2$, as well as
negligible contributions of spin-flip and UPE
amplitudes. These results allow one to better constrain extrapolations of trends observed at fixed target experiments.

The present results
provide important input for modelling GPDs, in particular they may
help to better constrain the chiral-odd GPDs and the amplitudes for UPE
transitions in exclusive $\rho ^0$ leptoproduction.

\section*{Acknowledgements}
We are indebted to Sergey Goloskokov and Peter Kroll for numerous fruitful discussions on the interpretation of
our results
%{\color{red} and for providing us with predictions of their model (?)}.
We gratefully acknowledge the support of CERN management
and staff and the skill and effort of the technicians of our
collaborating institutions.

%\clearpage
%\onecolumn
\section*{Appendix}
\setcounter{table}{0}
\renewcommand{\thetable}{A.\arabic{table}}

%Tables \ref{atab1}, \ref{tab2}, \ref{tab3}, and %\ref{tab4} with kinematic dependences of 23 SDMEs.
Table \ref{atab1} gives the various contributions to the systematic uncertainty of the 23 SDMEs and
Tables \ref{atab2}, \ref{atab3} and \ref{atab4} list their kinematic dependences.
Table \ref{corrtab2} contains the correlation matrix for measured $\rho ^0$ SDMEs in the entire kinematic range.

\begin{table*}[hbt!]
\renewcommand{\arraystretch}{1.2}
\begin{center}
\caption{\label{atab1} Uncertainties for each SDME value: in column 3 the statistical
uncertainty (``stat.''), in columns 4--8 the values of differences for each source of systematic uncertainty as defined in Section~\ref{sec:sys}, in column
9 the total systematic uncertainty (``tot. sys.''), and in column 10 the
total uncertainty (``tot.'').}
\scriptsize
\begin{tabular}{|
>{\centering}m{0.08\textwidth}|
r@{}l|
>{\centering}m{0.05\textwidth}|
r@{}l|
r@{}l|
r@{}l|
r@{}l|
r@{}l|
>{\centering}m{0.05\textwidth}|
>{\centering\arraybackslash}m{0.05\textwidth}|
}
\hline
 SDME &
 \multicolumn{2}{m{0.06\textwidth}|}{value}  &
 stat. &
 \multicolumn{2}{>{\centering}m{0.07\textwidth}|}{beam charge} &
 \multicolumn{2}{>{\centering}m{0.08\textwidth}|}{$E_{\texttt{miss}}$} &
 %\multicolumn{2}{>{\centering}m{0.08\textwidth}|}{back\-ground} &
 \multicolumn{2}{>{\centering}m{0.08\textwidth}|}{bg SDMEs} &
 \multicolumn{2}{>{\centering}m{0.07\textwidth}|}{$f_{\texttt{bg}}$} &
 \multicolumn{2}{>{\centering}m{0.07\textwidth}|}{sim\-u\-la\-tion} &
 tot. sys. &
 tot. \\
  & & & &
 \multicolumn{2}{>{\centering}m{0.07\textwidth}|}{(i)} &
 \multicolumn{2}{>{\centering}m{0.08\textwidth}|}{(ii)} &
 \multicolumn{2}{>{\centering}m{0.08\textwidth}|}{(iii)} &
 \multicolumn{2}{>{\centering}m{0.07\textwidth}|}{(iv)} &
 \multicolumn{2}{>{\centering}m{0.07\textwidth}|}{(v)} &
 & \\
\hline
$r^{04}_{00}$ & $ $ & $0.4698$ & $0.0035$ & $ $ & $0.0071$ & $ $ & $0.0085$ & $ $ & $0.0186$ & $-$ & $0.0003$ & $ $ & $0.0040$ & $0.0220$ & $ 0.0223$ \\
$r^{1}_{1-1}$ & $ $ & $0.2457$ & $0.0037$ & $-$ & $0.0026$ & $ $ & $0.0029$ & $-$ & $0.0040$ & $ $ & $0.0031$ & $-$ & $0.0010$ & $0.0064$ & $ 0.0074$ \\
Im $r^{2}_{1-1}$ & $-$ & $0.2459$ & $0.0038$ & $ $ & $0.0009$ & $-$ & $0.0005$ & $ $ & $0.0036$ & $-$ & $0.0031$ & $ $ & $0.0007$ & $0.0049$ & $ 0.0062$ \\
\hline
Re $r^{5}_{10}$  & $ $ & $0.1769$ & $0.0015$ & $-$ & $0.0015$ & $ $ & $0.0024$ & $-$ & $0.0020$ & $ $ & $0.0020$ & $-$ & $0.0009$ & $0.0041$ & $ 0.0044$ \\
Im $r^{6}_{10}$  & $-$ & $0.1662$ & $0.0014$ & $ $ & $0.0012$ & $-$ & $0.0008$ & $ $ & $0.0031$ & $-$ & $0.0018$ & $ $ & $0.0007$ & $0.004$0 & $ 0.0042$ \\
Im $r^{7}_{10}$  & $ $ & $0.0453$ & $0.0096$ & $-$ & $0.0155$ & $-$ & $0.0007$ & $-$ & $0.0006$ & $ $ & $0.0005$ & $ $ & $0.0003$ & $0.0156$ & $ 0.0183$ \\
Re $r^{8}_{10}$  & $ $ & $0.0362$ & $0.0095$ & $ $ & $0.0120$ & $-$ & $0.0000$ & $-$ & $0.0014$ & $ $ & $0.0004$ & $ $ & $0.0002$ & $0.0121$ &  $ 0.0154$ \\
\hline
Re $r^{04}_{10}$ & $ $ & $0.0454$ & $0.0021$ & $ $ & $0.0011$ & $ $ & $0.0027$ & $ $ & $0.0050$ & $ $ & $0.0002$ & $ $ & $0.0006$ & $0.0058$ & $ 0.0062$ \\
Re $r^{1}_{10}$  & $-$ & $0.0539$ & $0.0029$ & $-$ & $0.0002$ & $-$ & $0.0039$ & $-$ & $0.0003$ & $-$ & $0.0007$ & $-$ & $0.0006$ & $0.0040$ & $ 0.0049$ \\
Im $r^{2}_{10}$  & $ $ & $0.0532$ & $0.0028$ & $ $ & $0.0028$ & $ $ & $0.0031$ & $-$ & $0.0002$ & $ $ & $0.0006$ & $ $ & $0.0004$ & $0.0043$ & $ 0.0051$ \\
$r^{5}_{00}$  & $ $ & $0.1456$ & $0.0033$ & $ $ & $0.0009$ & $ $ & $0.0124$ & $ $ & $0.0030$ & $ $ & $0.0015$ & $ $ & $0.0006$ & $0.0129$ & $ 0.0133$ \\
$r^{1}_{00}$  & $-$ & $0.0376$ & $0.0062$ & $-$ & $0.0021$ & $-$ & $0.0112$ & $ $ & $0.0009$ & $-$ & $0.0004$ & $ $ & $0.0004$ & $0.0114$ & $ 0.0130$ \\
Im $r^{3}_{10}$  & $ $ & $0.0067$ & $0.0067$ & $-$ & $0.0043$ & $ $ & $0.0005$ & $-$ & $0.0011$ & $ $ & $0.0001$ & $ $ & $0.0000$ & $0.0045$ & $ 0.0081$ \\
$r^{8}_{00}$  & $ $ & $0.0019$ & $0.0194$ & $ $ & $0.0237$ & $ $ & $0.0034$ & $-$ & $0.0082$ & $ $ & $0.0003$ & $-$ & $0.0000$ & $0.0253$ & $ 0.0319$ \\
\hline
$r^{5}_{11}$  & $ $ & $0.0027$ & $0.0016$ & $-$ & $0.0018$ & $ $ & $0.0014$ &  $ $ & $0.0011$ & $-$ & $0.0001$ & $-$ & $0.0004$ & $0.0025$ & $ 0.0030$ \\
$r^{5}_{1-1}$ & $ $ & $0.0050$ & $0.0020$ & $ $ & $0.0023$ & $-$ & $0.0004$ & $-$ & $0.0007$ & $ $ & $0.0001$ & $-$ & $0.0002$ & $0.0025$ & $ 0.0032$ \\
Im $r^{6}_{1-1}$ & $-$ & $0.0028$ & $0.0020$ & $-$ & $0.0018$ & $ $ & $0.0000$ & $-$ & $0.0007$ & $-$ & $0.0000$ & $ $ & $0.0001$ & $0.0019$ & $ 0.0028$ \\
Im $r^{7}_{1-1}$ & $-$ & $0.0045$ & $0.0134$ & $ $ & $0.0212$ & $-$ & $0.0058$ & $ $ & $0.0041$ & $-$ & $0.0002$ & $-$ & $0.0004$ & $0.0224$ & $ 0.0261$ \\
$r^{8}_{11}$  & $ $ & $0.0203$ & $0.0101$ & $-$ & $0.0304$ & $-$ & $0.0014$ & $ -$ & $0.0014$ & $ $ & $0.0002$ & $-$ & $0.0001$ & $0.0305$ & $ 0.0321$ \\
$r^{8}_{1-1}$ & $-$ & $0.0300$ & $0.0128$ & $-$ & $0.0062$ & $ $ & $0.0066$ & $ -$ & $0.0006$ & $-$ & $0.0003$ & $ $ & $0.0003$ & $0.0091$ & $ 0.0157$ \\
\hline
$r^{04}_{1-1}$& $-$ & $0.0120$ & $0.0027$ & $-$ & $0.0019$ & $ $ & $0.0002$ & $-$ & $0.0026$ & $-$ & $0.0000$ & $-$ & $0.0002$ & $0.0032$ & $ 0.0041$ \\
$r^{1}_{11}$  & $-$ & $0.0162$ & $0.0032$ & $ $ & $0.0027$ & $ $ & $0.0011$ & $ $ & $0.0023$ & $-$ & $0.0002$ & $-$ & $0.0000$ & $0.0037$ & $ 0.0049$ \\
Im $r^{3}_{1-1}$ & $ $ & $0.0163$ & $0.0085$ & $-$ & $0.0036$ & $ $ & $0.0005$ & $ $ & $0.0023$ & $ $ & $0.0001$ & $-$ & $0.0003$ & $0.0043$ & $ 0.0099 $ \\
\hline
\end{tabular}
\end{center}
\end{table*}

\begin{table*}[htp]
\renewcommand{\arraystretch}{1.2}
\begin{center}
\caption{\label{atab2} The measured 23 unpolarised and polarised $\rho ^0$ SDMEs
 in bins of $Q^2$: $1.0 - 1.3 - 2.0 - 4.0 - 10.0$~(GeV/$c$)$^2$.
The first uncertainties are statistical, the second  systematic.}
\scriptsize
\begin{tabular}{|c|r@{$\,\pm\,$}l@{$\,\pm\,$}l|r@{$\,\pm\,$}l@{$\,\pm\,$}l
                  |r@{$\,\pm\,$}l@{$\,\pm\,$}l|r@{$\,\pm\,$}l@{$\,\pm\,$}l|}

\hline
 SDME &\multicolumn{3}{c|}{$\langle Q^{2}\rangle$ = 1.14 (GeV/$c$)$^2$}
      &\multicolumn{3}{c|}{$\langle Q^{2}\rangle$ = 1.60 (GeV/$c$)$^2$}
      &\multicolumn{3}{c|}{$\langle Q^{2}\rangle$ = 2.80 (GeV/$c$)$^2$}
      &\multicolumn{3}{c|}{$\langle Q^{2}\rangle$ = 6.02 (GeV/$c$)$^2$}\\
\hline
$r^{04}_{00}$    &$  0.4080 $&$  0.0056 $&$ 0.0243$&$  0.4749 $&$ 0.0055 $&$ 0.0201$
                 &$  0.5490 $&$  0.0085 $&$ 0.0193$&$  0.5319 $&$ 0.0183 $&$ 0.0555$ \\
$r^1_{1-1}$      &$  0.2781 $&$  0.0058 $&$ 0.0088$&$  0.2337 $&$ 0.0057 $&$ 0.0074$
                 &$  0.2437 $&$  0.0089 $&$ 0.0061$&$  0.1647 $&$ 0.0193 $&$ 0.0220$ \\
Im $r^2_{1-1}$   &$ -0.2763 $&$  0.0060 $&$ 0.0083$&$ -0.2300 $&$ 0.0059 $&$ 0.0045$
                 &$ -0.2586 $&$  0.0089 $&$ 0.0165$&$ -0.1450 $&$ 0.0199 $&$ 0.0228$ \\
\hline
Re $r^5_{10}$    &$  0.1774 $&$  0.0023 $&$ 0.0042$&$  0.1726 $&$ 0.0023 $&$ 0.0025$
                 &$  0.1938 $&$  0.0036 $&$ 0.0083$&$  0.1562 $&$ 0.0078 $&$ 0.0145$ \\
Im $r^6_{10}$    &$ -0.1695 $&$  0.0021 $&$ 0.0033$&$ -0.1591 $&$ 0.0023 $&$ 0.0038$
                 &$ -0.1829 $&$  0.0035 $&$ 0.0067$&$ -0.1513 $&$ 0.0077 $&$ 0.0077$ \\
Im $r^7_{10}$    &$  0.0230 $&$  0.0148 $&$ 0.0158$&$  0.0482 $&$ 0.0152 $&$ 0.0082$
                 &$  0.0851 $&$  0.0230 $&$ 0.0343$&$  0.0296 $&$ 0.0553 $&$ 0.0073$ \\
Re $r^8_{10}$    &$  0.0253 $&$  0.0147 $&$ 0.0111$&$  0.0400 $&$ 0.0146 $&$ 0.0059$
                 &$  0.0466 $&$  0.0231 $&$ 0.0365$&$  0.0471 $&$ 0.0548 $&$ 0.0625$ \\
\hline
Re $r^{04}_{10}$ &$  0.0452 $&$  0.0034 $&$ 0.0058$&$  0.0431 $&$ 0.0034 $&$ 0.0060$
                 &$  0.0508 $&$  0.0052 $&$ 0.0069$&$  0.0358 $&$ 0.0110 $&$ 0.0089$ \\
Re $r^1_{10}$    &$ -0.0521 $&$  0.0044 $&$ 0.0049$&$ -0.0439 $&$ 0.0045 $&$ 0.0055$
                 &$ -0.0713 $&$  0.0070 $&$ 0.0014$&$ -0.0613 $&$ 0.0150 $&$ 0.0234$ \\
Im $r^2_{10}$    &$  0.0505 $&$  0.0043 $&$ 0.0031$&$  0.0508 $&$ 0.0045 $&$ 0.0041$
                 &$  0.0612 $&$  0.0069 $&$ 0.0060$&$  0.0628 $&$ 0.0151 $&$ 0.0284$ \\
$r^5_{00}$       &$  0.1150 $&$  0.0050 $&$ 0.0080$&$  0.1419 $&$ 0.0052 $&$ 0.0122$
                 &$  0.1950 $&$  0.0081 $&$ 0.0213$&$  0.2021 $&$ 0.0167 $&$ 0.0406$ \\
$r^1_{00}$       &$ -0.0217 $&$  0.0092 $&$ 0.0109$&$ -0.0441 $&$ 0.0097 $&$ 0.0103$
                 &$ -0.0532 $&$  0.0156 $&$ 0.0420$&$ -0.0419 $&$ 0.0326 $&$ 0.0254$ \\
Im $r^3_{10}$    &$  0.0144 $&$  0.0104 $&$ 0.0027$&$ -0.0068 $&$ 0.0105 $&$ 0.0084$
                 &$  0.0209 $&$  0.0161 $&$ 0.0114$&$  0.0212 $&$ 0.0380 $&$ 0.0694$ \\
$r^8_{00}$       &$  0.0095 $&$  0.0302 $&$ 0.0269$&$ -0.0041 $&$ 0.0304 $&$ 0.0294$
                 &$ -0.0498 $&$  0.0477 $&$ 0.0385$&$  0.2174 $&$ 0.1065 $&$ 0.1142$ \\
\hline
$r^5_{11}$       &$  0.0014 $&$  0.0026 $&$ 0.0040$&$  0.0009 $&$ 0.0025 $&$ 0.0017$
                 &$  0.0014 $&$  0.0037 $&$ 0.0055$&$  0.0316 $&$ 0.0080 $&$ 0.0230$ \\
$r^5_{1-1}$      &$  0.0017 $&$  0.0032 $&$ 0.0027$&$  0.0079 $&$ 0.0031 $&$ 0.0054$
                 &$  0.0087 $&$  0.0047 $&$ 0.0042$&$ -0.0096 $&$ 0.0100 $&$ 0.0087$ \\
Im $r^6_{1-1}$   &$  0.0006 $&$  0.0031 $&$ 0.0025$&$ -0.0074 $&$ 0.0031 $&$ 0.0027$
                 &$  0.0003 $&$  0.0046 $&$ 0.0029$&$ -0.0067 $&$ 0.0102 $&$ 0.0033$ \\
Im $r^7_{1-1}$   &$ -0.0079 $&$  0.0215 $&$ 0.0444$&$  0.0063 $&$ 0.0206 $&$ 0.0086$
                 &$ -0.0400 $&$  0.0314 $&$ 0.0156$&$  0.0716 $&$ 0.0755 $&$ 0.0571$ \\
$r^8_{11}$       &$  0.0227 $&$  0.0163 $&$ 0.0310$&$  0.0168 $&$ 0.0156 $&$ 0.0140$
                 &$  0.0397 $&$  0.0243 $&$ 0.0562$&$ -0.0800 $&$ 0.0546 $&$ 0.0492$ \\
$r^8_{1-1}$      &$ -0.0154 $&$  0.0206 $&$ 0.0209$&$ -0.0105 $&$ 0.0195 $&$ 0.0177$                                                  &$ -0.0575 $&$  0.0309 $&$ 0.0569$&$ -0.1683 $&$ 0.0698 $&$ 0.0418$ \\
\hline
$r^{04}_{1-1}$   &$ -0.0213 $&$  0.0044 $&$ 0.0055$&$ -0.0074 $&$ 0.0042 $&$ 0.0030$
                 &$ -0.0081 $&$  0.0064 $&$ 0.0073$&$  0.0059 $&$ 0.0136 $&$ 0.0057$ \\
$r^1_{11}$       &$ -0.0252 $&$  0.0051 $&$ 0.0083$&$ -0.0099 $&$ 0.0049 $&$ 0.0069$
                 &$ -0.0157 $&$  0.0074 $&$ 0.0101$&$ -0.0122 $&$ 0.0159 $&$ 0.0146$ \\
Im $r^3_{1-1}$   &$  0.0038 $&$  0.0134 $&$ 0.0110$&$  0.0279 $&$ 0.0131 $&$ 0.0192$
                 &$  0.0051 $&$  0.0205 $&$ 0.0094$&$  0.0702 $&$ 0.0495 $&$ 0.0476$ \\
\hline
\end{tabular}
\end{center}
\end{table*}

\begin{table*}[hbtp]
\renewcommand{\arraystretch}{1.2}
\begin{center}
\caption{\label{atab3} The measured 23 unpolarised and polarised $\rho ^0$ SDMEs
 in bins of $p^{2}_{\rm T} $: $0.01 - 0.1 - 0.3 - 0.5$ (GeV/$c$)$^2$.
The first uncertainties are statistical, the second  systematic.}
\scriptsize
\begin{tabular}{|c|r@{$\,\pm\,$}l@{$\,\pm\,$}l|r@{$\,\pm\,$}l@{$\,\pm\,$}l
                  |r@{$\,\pm\,$}l@{$\,\pm\,$}l|r@{$\,\pm\,$}l@{$\,\pm\,$}l|}
\hline
 SDME &\multicolumn{3}{c|}{$\langle p^{2}_{\rm T}\rangle$ = 0.053 (GeV/$c$)$^2$}
      &\multicolumn{3}{c|}{$\langle p^{2}_{\rm T}\rangle$ = 0.147 (GeV/$c$)$^2$}
      &\multicolumn{3}{c|}{$\langle p^{2}_{\rm T}\rangle$ = 0.248 (GeV/$c$)$^2$}
      &\multicolumn{3}{c|}{$\langle p^{2}_{\rm T}\rangle$ = 0.391 (GeV/$c$)$^2$}\\
\hline
$r^{04}_{00}$   &$  0.4458 $&$ 0.0051 $&$ 0.0173$&$  0.4690 $&$ 0.0070 $&$ 0.0260$
                &$  0.4906 $&$ 0.0095 $&$ 0.0273$&$  0.5300 $&$ 0.0105 $&$ 0.0368$ \\
$r^1_{1-1}$     &$  0.2626 $&$ 0.0052 $&$ 0.0080$&$  0.2536 $&$ 0.0072 $&$ 0.0077$
                &$  0.2177 $&$ 0.0101 $&$ 0.0197$&$  0.2212 $&$ 0.0109 $&$ 0.0123$ \\
Im $r^2_{1-1}$  &$ -0.2694 $&$ 0.0053 $&$ 0.0061$&$ -0.2534 $&$ 0.0075 $&$ 0.0144$
                &$ -0.2247 $&$ 0.0101 $&$ 0.0140$&$ -0.1929 $&$ 0.0115 $&$ 0.0116$ \\
\hline
Re $r^5_{10}$   &$  0.1774 $&$ 0.0021 $&$ 0.0035$&$  0.1841 $&$ 0.0029 $&$ 0.0043$
                &$  0.1777 $&$ 0.0039 $&$ 0.0059$&$  0.1719 $&$ 0.0046 $&$ 0.0064$ \\
Im $r^6_{10}$   &$ -0.1694 $&$ 0.0021 $&$ 0.0031$&$ -0.1718 $&$ 0.0028 $&$ 0.0051$
                &$ -0.1593 $&$ 0.0039 $&$ 0.0069$&$ -0.1603 $&$ 0.0042 $&$ 0.0079$ \\
Im $r^7_{10}$   &$  0.0368 $&$ 0.0137 $&$ 0.0125$&$  0.0549 $&$ 0.0181 $&$ 0.0122$
                &$  0.0420 $&$ 0.0271 $&$ 0.0425$&$  0.0672 $&$ 0.0301 $&$ 0.0365$ \\
Re $r^8_{10}$   &$  0.0467 $&$ 0.0133 $&$ 0.0094$&$  0.0148 $&$ 0.0188 $&$ 0.0163$
                &$  0.0402 $&$ 0.0250 $&$ 0.0147$&$  0.0252 $&$ 0.0298 $&$ 0.0491$ \\
\hline
Re $r^{04}_{10}$&$  0.0307 $&$ 0.0031 $&$ 0.0076$&$  0.0556 $&$ 0.0042 $&$ 0.0079$
                &$  0.0578 $&$ 0.0056 $&$ 0.0089$&$  0.0674 $&$ 0.0064 $&$ 0.0132$ \\
Re $r^1_{10}$   &$ -0.0352 $&$ 0.0041 $&$ 0.0085$&$ -0.0614 $&$ 0.0057 $&$ 0.0030$
                &$ -0.0665 $&$ 0.0074 $&$ 0.0084$&$ -0.0906 $&$ 0.0089 $&$ 0.0160$ \\
Im $r^2_{10}$   &$  0.0383 $&$ 0.0041 $&$ 0.0072$&$  0.0514 $&$ 0.0054 $&$ 0.0028$
                &$  0.0817 $&$ 0.0078 $&$ 0.0095$&$  0.0827 $&$ 0.0086 $&$ 0.0077$ \\
$r^5_{00}$      &$  0.0929 $&$ 0.0047 $&$ 0.0184$&$  0.1644 $&$ 0.0065 $&$ 0.0092$
                &$  0.1920 $&$ 0.0088 $&$ 0.0100$&$  0.2450 $&$ 0.0096 $&$ 0.0101$ \\
$r^1_{00}$      &$ -0.0289 $&$ 0.0088 $&$ 0.0199$&$ -0.0390 $&$ 0.0120 $&$ 0.0102$
                &$ -0.0560 $&$ 0.0167 $&$ 0.0388$&$ -0.0584 $&$ 0.0185 $&$ 0.0154$ \\
Im $r^3_{10}$   &$  0.0171 $&$ 0.0095 $&$ 0.0060$&$  0.0011 $&$ 0.0128 $&$ 0.0090$
                &$ -0.0072 $&$ 0.0186 $&$ 0.0231$&$  0.0060 $&$ 0.0208 $&$ 0.0041$ \\
$r^8_{00}$      &$  0.0122 $&$ 0.0274 $&$ 0.0520$&$ -0.0578 $&$ 0.0382 $&$ 0.0446$
                &$  0.0098 $&$ 0.0525 $&$ 0.0430$&$  0.0676 $&$ 0.0610 $&$ 0.0290$ \\
\hline
$r^5_{11}$      &$  0.0017 $&$ 0.0023 $&$ 0.0066$&$  0.0006 $&$ 0.0032 $&$ 0.0022$
                &$  0.0070 $&$ 0.0043 $&$ 0.0051$&$  0.0050 $&$ 0.0047 $&$ 0.0103$ \\
$r^5_{1-1}$     &$  0.0060 $&$ 0.0028 $&$ 0.0030$&$  0.0031 $&$ 0.0039 $&$ 0.0015$
                &$  0.0093 $&$ 0.0054 $&$ 0.0027$&$  0.0013 $&$ 0.0059 $&$ 0.0126$ \\
Im $r^6_{1-1}$  &$ -0.0037 $&$ 0.0028 $&$ 0.0038$&$  0.0009 $&$ 0.0039 $&$ 0.0022$
                &$  0.0006 $&$ 0.0051 $&$ 0.0087$&$ -0.0107 $&$ 0.0056 $&$ 0.0048$ \\
Im $r^7_{1-1}$  &$ -0.0309 $&$ 0.0191 $&$ 0.0568$&$  0.0121 $&$ 0.0265 $&$ 0.0373$
                &$  0.0159 $&$ 0.0362 $&$ 0.0595$&$  0.0293 $&$ 0.0403 $&$ 0.0224$ \\
$r^8_{11}$      &$  0.0020 $&$ 0.0145 $&$ 0.0364$&$  0.0418 $&$ 0.0200 $&$ 0.0332$
                &$  0.0459 $&$ 0.0277 $&$ 0.0066$&$  0.0301 $&$ 0.0301 $&$ 0.0401$ \\
$r^8_{1-1}$     &$ -0.0222 $&$ 0.0182 $&$ 0.0044$&$ -0.0296 $&$ 0.0252 $&$ 0.0409$
                &$ -0.0131 $&$ 0.0346 $&$ 0.0203$&$ -0.0750 $&$ 0.0380 $&$ 0.0204$ \\
\hline
$r^{04}_{1-1}$  &$ -0.0075 $&$ 0.0039 $&$ 0.0033$&$ -0.0210 $&$ 0.0053 $&$ 0.0041$
                &$ -0.0055 $&$ 0.0073 $&$ 0.0038$&$ -0.0173 $&$ 0.0079 $&$ 0.0105$ \\
$r^1_{11}$      &$ -0.0075 $&$ 0.0046 $&$ 0.0010$&$ -0.0170 $&$ 0.0063 $&$ 0.0045$
                &$ -0.0253 $&$ 0.0084 $&$ 0.0172$&$ -0.0360 $&$ 0.0091 $&$ 0.0059$ \\
Im $r^3_{1-1}$  &$  0.0234 $&$ 0.0121 $&$ 0.0026$&$  0.0138 $&$ 0.0168 $&$ 0.0092$
                &$ -0.0236 $&$ 0.0236 $&$ 0.0049$&$  0.0473 $&$ 0.0261 $&$ 0.0047$ \\
\hline
\end{tabular}
\end{center}
\end{table*}

\begin{table*}[hbtp]
\renewcommand{\arraystretch}{1.2}
\begin{center}
\caption{\label{atab4} The measured 23 unpolarised and polarised $\rho ^0$ SDMEs
 in bins of $W$: $5.00 - 7.3 - 9.0 - 12.0 - 17.0$~GeV/$c^2$.
The first uncertainties are statistical, the second  systematic.}
\scriptsize
\begin{tabular}{|c|r@{$\,\pm\,$}l@{$\,\pm\,$}l|r@{$\,\pm\,$}l@{$\,\pm\,$}l|r@{$\,\pm\,$}l@{$\,\pm\,$}l|r@{$\,\pm\,$}l@{$\,\pm\,$}l|}
\hline
 SDME &\multicolumn{3}{c|}{$\langle W\rangle$ =  7.0 GeV/$c^2$}
      &\multicolumn{3}{c|}{$\langle W\rangle$ =  8.1 GeV/$c^2$}
      &\multicolumn{3}{c|}{$\langle W\rangle$ = 10.0 GeV/$c^2$}
      &\multicolumn{3}{c|}{$\langle W\rangle$ = 13.5 GeV/$c^2$}\\
\hline
$r^{04}_{00}$    &$  0.4349 $&$ 0.0083 $&$ 0.0639 $&$  0.4819 $&$ 0.0070 $&$ 0.0335$
                 &$  0.4862 $&$ 0.0059 $&$ 0.0100 $&$  0.4836 $&$ 0.0074 $&$ 0.0356$ \\
$r^1_{1-1}$      &$  0.2432 $&$ 0.0080 $&$ 0.0165 $&$  0.2409 $&$ 0.0070 $&$ 0.0096$
                 &$  0.2689 $&$ 0.0060 $&$ 0.0054 $&$  0.2808 $&$ 0.0111 $&$ 0.0288$ \\
Im $r^2_{1-1}$   &$ -0.2344 $&$ 0.0083 $&$ 0.0081 $&$ -0.2587 $&$ 0.0069 $&$ 0.0081$
                 &$ -0.2627 $&$ 0.0063 $&$ 0.0045 $&$ -0.2804 $&$ 0.0117 $&$ 0.0192$ \\
\hline
Re $r^5_{10}$    &$  0.1659 $&$ 0.0035 $&$ 0.0110 $&$  0.1762 $&$ 0.0028 $&$ 0.0071$
                 &$  0.1927 $&$ 0.0024 $&$ 0.0042 $&$  0.2149 $&$ 0.0041 $&$ 0.0105$ \\
Im $r^6_{10}$    &$ -0.1539 $&$ 0.0033 $&$ 0.0128 $&$ -0.1671 $&$ 0.0028 $&$ 0.0082$
                 &$ -0.1849 $&$ 0.0023 $&$ 0.0035 $&$ -0.1978 $&$ 0.0039 $&$ 0.0134$ \\
Im $r^7_{10}$    &$  0.1599 $&$ 0.0472 $&$ 0.0535 $&$  0.0451 $&$ 0.0259 $&$ 0.0133$
                 &$  0.0302 $&$ 0.0122 $&$ 0.0264 $&$  0.0191 $&$ 0.0108 $&$ 0.0106$ \\
Re $r^8_{10}$    &$  0.0502 $&$ 0.0454 $&$ 0.0840 $&$  0.0313 $&$ 0.0254 $&$ 0.0068$
                 &$  0.0390 $&$ 0.0121 $&$ 0.0105 $&$  0.0314 $&$ 0.0104 $&$ 0.0137$ \\
\hline
Re $r^{04}_{10}$ &$  0.0584 $&$ 0.0049 $&$ 0.0176 $&$  0.0540 $&$ 0.0042 $&$ 0.0081$
                 &$  0.0388 $&$ 0.0037 $&$ 0.0059 $&$  0.0318 $&$ 0.0048 $&$ 0.0039$ \\
Re $r^1_{10}$    &$ -0.0685 $&$ 0.0065 $&$ 0.0135 $&$ -0.0572 $&$ 0.0055 $&$ 0.0041$
                 &$ -0.0521 $&$ 0.0048 $&$ 0.0063 $&$ -0.0285 $&$ 0.0090 $&$ 0.0093$ \\
Im $r^2_{10}$    &$  0.0684 $&$ 0.0064 $&$ 0.0055 $&$  0.0514 $&$ 0.0054 $&$ 0.0104$
                 &$  0.0502 $&$ 0.0045 $&$ 0.0097 $&$  0.0551 $&$ 0.0079 $&$ 0.0179$ \\
$r^5_{00}$       &$  0.1704 $&$ 0.0071 $&$ 0.0162 $&$  0.1505 $&$ 0.0062 $&$ 0.0056$
                 &$  0.1291 $&$ 0.0057 $&$ 0.0112 $&$  0.1589 $&$ 0.0089 $&$ 0.0508$ \\
$r^1_{00}$       &$ -0.0478 $&$ 0.0138 $&$ 0.0204 $&$ -0.0252 $&$ 0.0118 $&$ 0.0122$
                 &$ -0.0346 $&$ 0.0103 $&$ 0.0135 $&$ -0.1018 $&$ 0.0179 $&$ 0.0648$ \\
Im $r^3_{10}$    &$  0.0682 $&$ 0.0359 $&$ 0.0297 $&$ -0.0123 $&$ 0.0206 $&$ 0.0173$
                 &$  0.0003 $&$ 0.0103 $&$ 0.0080 $&$  0.0037 $&$ 0.0076 $&$ 0.0112$ \\
$r^8_{00}$       &$  0.0694 $&$ 0.1017 $&$ 0.0879 $&$  0.0776 $&$ 0.0570 $&$ 0.0335$
                 &$ -0.0332 $&$ 0.0278 $&$ 0.0109 $&$ -0.0033 $&$ 0.0225 $&$ 0.0269$ \\
\hline
$r^5_{11}$       &$  0.0118 $&$ 0.0035 $&$ 0.0111 $&$  0.0029 $&$ 0.0030 $&$ 0.0026$
                 &$ -0.0000 $&$ 0.0027 $&$ 0.0022 $&$ -0.0152 $&$ 0.0044 $&$ 0.0061$ \\
$r^5_{1-1}$      &$  0.0101 $&$ 0.0043 $&$ 0.0067 $&$  0.0010 $&$ 0.0037 $&$ 0.0018$
                 &$ -0.0032 $&$ 0.0032 $&$ 0.0033 $&$  0.0221 $&$ 0.0055 $&$ 0.0051$ \\
Im $r^6_{1-1}$   &$ -0.0035 $&$ 0.0042 $&$ 0.0061 $&$  0.0033 $&$ 0.0036 $&$ 0.0020$
                 &$ -0.0033 $&$ 0.0034 $&$ 0.0039 $&$ -0.0135 $&$ 0.0054 $&$ 0.0028$ \\
Im $r^7_{1-1}$   &$ -0.0308 $&$ 0.0653 $&$ 0.0425 $&$ -0.0639 $&$ 0.0354 $&$ 0.0147$
                 &$  0.0071 $&$ 0.0183 $&$ 0.0167 $&$ -0.0039 $&$ 0.0160 $&$ 0.0335$ \\
$r^8_{11}$       &$  0.0719 $&$ 0.0499 $&$ 0.0585 $&$  0.0352 $&$ 0.0283 $&$ 0.0075$
                 &$  0.0231 $&$ 0.0149 $&$ 0.0051 $&$  0.0019 $&$ 0.0122 $&$ 0.0370$ \\
$r^8_{1-1}$      &$ -0.0800 $&$ 0.0615 $&$ 0.0288 $&$ -0.0379 $&$ 0.0345 $&$ 0.0244$
                 &$ -0.0090 $&$ 0.0178 $&$ 0.0369 $&$ -0.0135 $&$ 0.0150 $&$ 0.0430$ \\
\hline
$r^{04}_{1-1}$   &$ -0.0110 $&$ 0.0061 $&$ 0.0098 $&$ -0.0097 $&$ 0.0053 $&$ 0.0059$
                 &$ -0.0125 $&$ 0.0046 $&$ 0.0058 $&$ -0.0179 $&$ 0.0059 $&$ 0.0089$ \\
$r^1_{11}$       &$ -0.0326 $&$ 0.0067 $&$ 0.0091 $&$ -0.0153 $&$ 0.0059 $&$ 0.0087$
                 &$ -0.0023 $&$ 0.0053 $&$ 0.0064 $&$ -0.0042 $&$ 0.0096 $&$ 0.0067$ \\
Im $r^3_{1-1}$   &$  0.1497 $&$ 0.0445 $&$ 0.0490 $&$ -0.0018 $&$ 0.0240 $&$ 0.0375$
                 &$  0.0117 $&$ 0.0124 $&$ 0.0038 $&$  0.0048 $&$ 0.0096 $&$ 0.0042$ \\
\hline
\end{tabular}
\end{center}
\end{table*}

%\begin{landscape}

\begin{table*}[ht]
\renewcommand{\arraystretch}{1.5}
\setlength{\tabcolsep}{4pt}
\caption{\label{corrtab2} The correlation matrix for the measured 23 $\rho ^0$ SDMEs for data in the entire kinematic range.}
%\adjustbox{scale=0.89,right}{
\rotatebox{90}{%
\resizebox{0.98\textheight}{!}{%
%\scriptsize
\begin{tabular}{c|rrrrrrrrrrrrrrrrrrrrrrr}
\hline
SDME & $r^{04}_{00}$  & Re $r^{04}_{10}$ & $r^{04}_{1-1}$  & $r^1_{11}$ & $r^1_{00}$ & Re $r^1_{10}$ & $r^1_{1-1}$ & Im $r^2_{10}$ & Im $r^2_{1-1}$ & $r^5_{11}$ & $r^5_{00}$ & Re $r^5_{10}$ & $r^5_{1-1}$  & Im $r^6_{10}$ & Im $r^6_{1-1}$ & Im $r^3_{10}$ & Im $r^3_{1-1}$  & Im $r^7_{10}$ & Im $r^7_{1-1}$ & $r^8_{11}$ & $r^8_{00}$ & Re $r^8_{10}$ & $r^8_{1-1}$ \\
\hline
$r^{04}_{00}$  & 1.000 & \\
Re $r^{04}_{10}$ & 0.006 & 1.000 & \\
$r^{04}_{1-1}$ &  0.016 & $-$0.055 & 1.000 & \\
$r^1_{11}$ & 0.020 & $-$0.036 & 0.364 & 1.000 & \\
$r^1_{00}$ & $-$0.029 & $-$0.058 & $-$0.027 & $-$0.387 & 1.000 & \\
Re $r^1_{10}$ & $-$0.007 & $-$0.187 & 0.045 & 0.051 & $-$0.020 & 1.000 & \\
$r^1_{1-1}$ & $-$0.242 & 0.012 & $-$0.034 & $-$0.024 & 0.020 & $-$0.006 & 1.000 & \\
Im $r^2_{10}$ & $-$0.005 & 0.123 & 0.068 & 0.032 & $-$0.027 & $-$0.066 & $-$0.001 & 1.000 & \\
Im $r^2_{1-1}$ & 0.231 & $-$0.012 & $-$0.014 & $-$0.005 & $-$0.010 & $-$0.018 & $-$0.058 & $-$0.026 & 1.000 & \\
$r^5_{11}$ & $-$0.075 & 0.293 & 0.022 & $-$0.043 & 0.047 & $-$0.112 & $-$0.049 & 0.109 & 0.048 & 1.000 & \\
$r^5_{00}$ & 0.227 & 0.294 & 0.006 & 0.049 & $-$0.204 & $-$0.226 & $-$0.059 & 0.239 & 0.058 & $-$0.305 & 1.000 & \\
Re $r^5_{10}$ & 0.031 & 0.143 & $-$0.289 & $-$0.186 & $-$0.169 & $-$0.048 & 0.064 & $-$0.076 & $-$0.070 & 0.072 & $-$0.022 & 1.000 & \\
$r^5_{1-1}$ & $-$0.009 & $-$0.289 & 0.081 & $-$0.047 & 0.001 & 0.132 & $-$0.063 & 0.132 & $-$0.001 & $-$0.280 & 0.022 & $-$0.088 & 1.000 & \\
Im $r^6_{10}$ & $-$0.026 & 0.072 & $-$0.309 & $-$0.220 & $-$0.230 & $-$0.024 & $-$0.089 & $-$0.107 & 0.098 & $-$0.015 & 0.074 & 0.164 & $-$0.013 & 1.000 & \\
Im $r^6_{1-1}$ & 0.017 & 0.283 & 0.018 & $-$0.025 & $-$0.000 & 0.157 & 0.020 & 0.151 & $-$0.087 & 0.251 & $-$0.009 & 0.053 & $-$0.074 & $-$0.043 & 1.000 & \\
Im $r^3_{10}$ & 0.002 & $-$0.000 & 0.001 & $-$0.002 & $-$0.003 & 0.002 & $-$0.001 & 0.004 & 0.003 & 0.000 & 0.018 & 0.001 & 0.010 & 0.013 & $-$0.011 & 1.000 & \\
Im $r^3_{1-1}$ & $-$0.012 & 0.001 & 0.004 & $-$0.005 & 0.006 & 0.001 & $-$0.008 & 0.008 & 0.005 & 0.013 & $-$0.008 & $-$0.015 & 0.016 & $-$0.013 & 0.007 & $-$0.049 & 1.000 & \\
Im $r^7_{10}$ & 0.008 & 0.001 & 0.013 & $-$0.001 & $-$0.017 & $-$0.001 & $-$0.010 & $-$0.002 & 0.011 & $-$0.002 & 0.001 & $-$0.009 & 0.004 & 0.011 & $-$0.004 & 0.169 & $-$0.141 & 1.000 & \\
Im $r^7_{1-1}$ & $-$0.005 & $-$0.013 & 0.013 & $-$0.002 & 0.005 & 0.012 & $-$0.024 & 0.010 & 0.022 & 0.019 & $-$0.009 & $-$0.002 & 0.007 & $-$0.004 & 0.001 & $-$0.171 & 0.116 & $-$0.066 & 1.000 & \\
$r^8_{11}$ & $-$0.010 & 0.001 & $-$0.019 & 0.019 & $-$0.008 & 0.007 & $-$0.018 & $-$0.007 & $-$0.007 & 0.001 & $-$0.009 & 0.002 & 0.009 & 0.004 & 0.005 & $-$0.151 & $-$0.034 & $-$0.027 & 0.161 & 1.000 & \\
$r^8_{00}$ & 0.006 & 0.015 & 0.009 & $-$0.006 & 0.016 & 0.019 & 0.003 & $-$0.006 & 0.008 & 0.001 & $-$0.001 & $-$0.001 & $-$0.010 & $-$0.003 & 0.002 & $-$0.198 & 0.009 & $-$0.025 & $-$0.014 & $-$0.391 & 1.000 & \\
Re $r^8_{10}$ & 0.007 & 0.007 & $-$0.006 & $-$0.001 & 0.026 & 0.030 & $-$0.011 & $-$0.006 & 0.014 & $-$0.004 & $-$0.003 & $-$0.007 & 0.011 & 0.004 & 0.008 & 0.025 & 0.148 & $-$0.037 & 0.016 & 0.017 & $-$0.031 & 1.000 & \\
$r^8_{1-1}$ & 0.020 & $-$0.003 & 0.024 & $-$0.004 & $-$0.003 & $-$0.009 & 0.025 & $-$0.006 & 0.002 & 0.010 & $-$0.003 & 0.010 & $-$0.001 & $-$0.005 & $-$0.005 & $-$0.179 & 0.008 & $-$0.030 & 0.052 & 0.169 & $-$0.021 & $-$0.013 & 1.000 \\
\hline
SDME & $r^{04}_{00}$  & Re $r^{04}_{10}$ & $r^{04}_{1-1}$  & $r^1_{11}$ & $r^1_{00}$ & Re $r^1_{10}$ & $r^1_{1-1}$ & Im $r^2_{10}$ & Im $r^2_{1-1}$ & $r^5_{11}$ & $r^5_{00}$ & Re $r^5_{10}$ & $r^5_{1-1}$  & Im $r^6_{10}$ & Im $r^6_{1-1}$ & Im $r^3_{10}$ & Im $r^3_{1-1}$  & Im $r^7_{10}$ & Im $r^7_{1-1}$ & $r^8_{11}$ & $r^8_{00}$ & Re $r^8_{10}$ & $r^8_{1-1}$ \\
\hline
\end{tabular}
}}
\end{table*}
%\end{landscape}

\clearpage

\bibliographystyle{apsrev4-2}
%\bibliography{biblio_COMPASS_rho0_SDME}{}

%\bibliographystyle{ws-rv-van}  % Bibliography: Author-Date system
%
%
%%%%%%%%%%%%%%%%%%%%%%%%%%%%%%%%%%%%%%%%%%%%%%
%
\clearpage
\textbf{The COMPASS Collaboration}

\vspace{10pt}
\begin{flushleft}
G.~D.~Alexeev$^\textrm{{\footnotesize\hyperlink{hl:dubna}{29}}}$\orcidlink{0009-0007-0196-8178},
M.~G.~Alexeev$^\textrm{{\footnotesize\hyperlink{hl:turin_u}{20},\hyperlink{hl:turin_i}{19}}}$\orcidlink{0000-0002-7306-8255},
C.~Alice$^\textrm{{\footnotesize\hyperlink{hl:turin_u}{20},\hyperlink{hl:turin_i}{19}}}$\orcidlink{0000-0001-6297-9857},
A.~Amoroso$^\textrm{{\footnotesize\hyperlink{hl:turin_u}{20},\hyperlink{hl:turin_i}{19}}}$\orcidlink{0000-0002-3095-8610},
V.~Andrieux$^\textrm{{\footnotesize\hyperlink{hl:illinois}{33}}}$\orcidlink{0000-0001-9957-9910},
V.~Anosov$^\textrm{{\footnotesize\hyperlink{hl:dubna}{29}}}$\orcidlink{0009-0003-3595-9561},
K.~Augsten$^\textrm{{\footnotesize\hyperlink{hl:praguectu}{4}}}$\orcidlink{0000-0001-8324-0576},
W.~Augustyniak$^\textrm{{\footnotesize\hyperlink{hl:warsaw}{24}}}$,
C.~D.~R.~Azevedo$^\textrm{{\footnotesize\hyperlink{hl:aveiro}{27}}}$\orcidlink{0000-0002-0012-9918},
B.~Badelek$^\textrm{{\footnotesize\hyperlink{hl:warsawu}{26}}}$\orcidlink{0000-0002-4082-1466}
% ...
\hypertarget{hl:calcutta}{$^\textrm{{\footnotesize 14}}$\footnotesize~Matrivani Institute of Experimental Research \& Education, Calcutta-700 030, India$^\textrm{{\tiny\hyperlink{hl:C}{C}}}$\\}
\hypertarget{hl:telaviv}{$^\textrm{{\footnotesize 15}}$\footnotesize~Tel Aviv University, School of Physics and Astronomy, 69978 Tel Aviv, Israel$^\textrm{{\tiny\hyperlink{hl:D}{D}}}$\\}
\hypertarget{hl:triest_a}{$^\textrm{{\footnotesize 16}}$\footnotesize~Abdus Salam ICTP, 34151 Trieste, Italy\\}
\hypertarget{hl:triest_i}{$^\textrm{{\footnotesize 17}}$\footnotesize~Trieste Section of INFN, 34127 Trieste, Italy\\}
\hypertarget{hl:triest_u}{$^\textrm{{\footnotesize 18}}$\footnotesize~University of Trieste, Dept.\ of Physics, 34127 Trieste, Italy\\}
\hypertarget{hl:turin_i}{$^\textrm{{\footnotesize 19}}$\footnotesize~Torino Section of INFN, 10125 Torino, Italy\\}
\hypertarget{hl:turin_u}{$^\textrm{{\footnotesize 20}}$\footnotesize~University of Torino, Dept.\ of Physics, 10125 Torino, Italy\\}
\hypertarget{hl:miyazaki}{$^\textrm{{\footnotesize 21}}$\footnotesize~University of Miyazaki, Miyazaki 889-2192, Japan$^\textrm{{\tiny\hyperlink{hl:E}{E}}}$\\}
\hypertarget{hl:nagoya}{$^\textrm{{\footnotesize 22}}$\footnotesize~Nagoya University, 464 Nagoya, Japan$^\textrm{{\tiny\hyperlink{hl:E}{E}}}$\\}
\hypertarget{hl:yamagata}{$^\textrm{{\footnotesize 23}}$\footnotesize~Yamagata University, Yamagata 992-8510, Japan$^\textrm{{\tiny\hyperlink{hl:E}{E}}}$\\}
\hypertarget{hl:warsaw}{$^\textrm{{\footnotesize 24}}$\footnotesize~National Centre for Nuclear Research, 02-093 Warsaw, Poland$^\textrm{{\tiny\hyperlink{hl:F}{F}}}$\\}
\hypertarget{hl:warsawtu}{$^\textrm{{\footnotesize 25}}$\footnotesize~Warsaw University of Technology, Institute of Radioelectronics, 00-665 Warsaw, Poland$^\textrm{{\tiny\hyperlink{hl:F}{F}}}$\\}
\hypertarget{hl:warsawu}{$^\textrm{{\footnotesize 26}}$\footnotesize~University of Warsaw, Faculty of Physics, 02-093 Warsaw, Poland$^\textrm{{\tiny\hyperlink{hl:F}{F}}}$\\}
\hypertarget{hl:aveiro}{$^\textrm{{\footnotesize 27}}$\footnotesize~University of Aveiro, I3N, Dept. of Physics, 3810-193 Aveiro, Portugal$^\textrm{{\tiny\hyperlink{hl:G}{G}}}$\\}
\hypertarget{hl:lisbon}{$^\textrm{{\footnotesize 28}}$\footnotesize~LIP, 1649-003 Lisbon, Portugal$^\textrm{{\tiny\hyperlink{hl:G}{G}}}$\\}
\hypertarget{hl:dubna}{$^\textrm{{\footnotesize 29}}$\footnotesize~Affiliated with an international laboratory covered by a cooperation agreement with CERN\\}
\hypertarget{hl:russia}{$^\textrm{{\footnotesize 30}}$\footnotesize~Affiliated with an institute covered by a cooperation agreement with CERN.\\}
\hypertarget{hl:cern}{$^\textrm{{\footnotesize 31}}$\footnotesize~CERN, 1211 Geneva 23, Switzerland\\}
\hypertarget{hl:taipei}{$^\textrm{{\footnotesize 32}}$\footnotesize~Academia Sinica, Institute of Physics, Taipei 11529, Taiwan$^\textrm{{\tiny\hyperlink{hl:H}{H}}}$\\}
\hypertarget{hl:illinois}{$^\textrm{{\footnotesize 33}}$\footnotesize~University of Illinois at Urbana-Champaign, Dept.\ of Physics, Urbana, IL 61801-3080, USA$^\textrm{{\tiny\hyperlink{hl:I}{I}}}$\\}

\vspace{10pt}
\hypertarget{hl:*}{$^\textrm{{\footnotesize *}}$\footnotesize~Corresponding author\\}
\hypertarget{hl:a}{$^\textrm{{\footnotesize a}}$\footnotesize~Supported by the European Union’s Horizon 2020 research and innovation programme under grant agreement STRONG–2020 - No 824093\\}
\hypertarget{hl:b}{$^\textrm{{\footnotesize b}}$\footnotesize~Supported by ANR, France with P2IO LabEx (ANR-10-LABX-0038)  in the framework "Investissements d'Avenir" (ANR-11-IDEX-0003-01)\\}
\hypertarget{hl:c}{$^\textrm{{\footnotesize c}}$\footnotesize~Supported by the DFG Research Training Group Programmes 1102 and 2044 (Germany)\\}
\hypertarget{hl:d}{$^\textrm{{\footnotesize d}}$\footnotesize~Retired from Ludwig-Maximilians-Universit\"at, 80539 M\"unchen, Germany\\}
\hypertarget{hl:d1}{$^\textrm{{\footnotesize d1}}$\footnotesize~Supported by the DFG cluster of excellence `Origin and Structure of the Universe' (www.universe-cluster.de) (Germany)\\}
\hypertarget{hl:e}{$^\textrm{{\footnotesize e}}$\footnotesize~Also at ORIGINS Excellence Cluster, 85748 Garching, Germany\\}
\hypertarget{hl:f}{$^\textrm{{\footnotesize f}}$\footnotesize~Also at Institut f\"ur Theoretische Physik, Universit\"at T\"ubingen, 72076 T\"ubingen, Germany\\}
\hypertarget{hl:g}{$^\textrm{{\footnotesize g}}$\footnotesize~Present address: NISER, Centre for Medical and Radiation Physics, Bubaneswar, India\\}
\hypertarget{hl:h}{$^\textrm{{\footnotesize h}}$\footnotesize~Also at University of Eastern Piedmont, 15100 Alessandria, Italy\\}
\hypertarget{hl:h1}{$^\textrm{{\footnotesize h1}}$\footnotesize~Supported by the Funds for Research 2019-22 of the University of Eastern Piedmont\\}
\hypertarget{hl:i}{$^\textrm{{\footnotesize i}}$\footnotesize~Also at Chubu University, Kasugai, Aichi 487-8501, Japan\\}
\hypertarget{hl:j}{$^\textrm{{\footnotesize j}}$\footnotesize~Also at KEK, 1-1 Oho, Tsukuba, Ibaraki 305-0801, Japan\\}
\hypertarget{hl:k}{$^\textrm{{\footnotesize k}}$\footnotesize~Also at Dept.\ of Physics, Pusan National University, Busan 609-735, Republic of Korea\\}
\hypertarget{hl:k1}{$^\textrm{{\footnotesize k1}}$\footnotesize~Also at Physics Dept., Brookhaven National Laboratory, Upton, NY 11973, USA\\}
\hypertarget{hl:l}{$^\textrm{{\footnotesize l}}$\footnotesize~Also at Dept.\ of Physics, National Central University, 300 Jhongda Road, Jhongli 32001, Taiwan\\}
\hypertarget{hl:m}{$^\textrm{{\footnotesize m}}$\footnotesize~Also at Dept.\ of Physics, National Kaohsiung Normal University, Kaohsiung County 824, Taiwan\\}
\hypertarget{hl:$\dagger$}{$^\textrm{{\footnotesize $\dagger$}}$\footnotesize~Deceased\\}

\vspace{10pt}
\hypertarget{hl:A}{$^\textrm{{\tiny A}}$\footnotesize~Supported by MEYS, Grants LM2015058, LM2018104 and LTT17018 and Charles University Grant PRIMUS/22/SCI/017 (Czech Republic)\\}
\hypertarget{hl:B}{$^\textrm{{\tiny B}}$\footnotesize~Supported by BMBF - Bundesministerium f\"ur Bildung und Forschung (Germany)\\}
\hypertarget{hl:C}{$^\textrm{{\tiny C}}$\footnotesize~Supported by B. Sen fund (India)\\}
\hypertarget{hl:D}{$^\textrm{{\tiny D}}$\footnotesize~Supported by the Israel Academy of Sciences and Humanities (Israel)\\}
\hypertarget{hl:E}{$^\textrm{{\tiny E}}$\footnotesize~Supported by MEXT and JSPS, Grants 18002006, 20540299, 18540281 and 26247032, the Daiko and Yamada Foundations (Japan)\\}
\hypertarget{hl:F}{$^\textrm{{\tiny F}}$\footnotesize~Supported by NCN, Grant 2020/37/B/ST2/01547 (Poland)\\}
\hypertarget{hl:G}{$^\textrm{{\tiny G}}$\footnotesize~Supported by FCT, Grants CERN/FIS-PAR/0022/2019 and CERN/FIS-PAR/0016/2021 (Portugal)\\}
\hypertarget{hl:H}{$^\textrm{{\tiny H}}$\footnotesize~Supported by the Ministry of Science and Technology (Taiwan)\\}
\hypertarget{hl:I}{$^\textrm{{\tiny I}}$\footnotesize~Supported by the National Science Foundation, Grant no. PHY-1506416 (USA)\\}

\end{flushleft}

% The main differences between the asymmetry $P$ observed
% for $\rho ^0$ and $\omega$ production may be explained by the
% GK model. In addition to the different transition form factors
% for the two processes, which were mentioned in the previous section, the NPE cross section for $\rho ^0$ is about 9 times
% larger than for $\omega$ production. The difference comes from the
% gluon and sea-quark GPDs, while the contribution from valence quarks is about the same~\cite{GK:epjA-2014} {\color{red} check with Peter Kroll}.

\end{document}